\documentclass[10pt,nofootinbib,aps,pra]{revtex4-1} 
\newcommand\nc{\newcommand}  \nc\rc{\renewcommand}  
\nc\PartOn[1]{#1}  \nc\PartOff[1]{}  

\usepackage{silence}                               
\usepackage{graphicx}
\graphicspath{{figs/}}

\usepackage{float}
\usepackage{multirow}

\usepackage{amsmath,amsfonts,amssymb,nicefrac,esint,bbm}
\usepackage{upgreek}

\usepackage{hyperref}  

\nc\x\hskip       \nc\y\vskip       \nc\X\kern       
\nc\xph\hphantom  \nc\yph\vphantom  \nc\ph\phantom
\nc\bi{}  \AtBeginDocument{\newdimen\bitx \bitx=.06em}  
\nc\bit{\x       \bitx}  \nc\biT{\x       -\bitx}  
\nc\bitt{\x     2\bitx}  \nc\biTT{\x     -2\bitx}  
\nc\bittt{\x    3\bitx}  \nc\biTTT{\x    -3\bitx}  
\nc\bitttt{\x   4\bitx}  \nc\biTTTT{\x   -4\bitx}  
\nc\bittttt{\x  5\bitx}  \nc\biTTTTT{\x  -5\bitx}  
\nc\bitttttt{\x 6\bitx}  \nc\biTTTTTT{\x -6\bitx}  

\nc\lhs{l.h.s.}  \nc\lhss{\lhs\ }  \nc\rhs{r.h.s.}  \nc\rhss{\rhs\ }
\nc\wrt{w.r.t.\ }  \nc\cf{cf.\ }
\nc\emp{\textit}  
\nc\lat{\textit}  
\nc\ie{\lat{i.e.,\ }}  \nc\etal{\lat{et al.}}  \nc\etall{\etal\ }  \nc\eg{\lat{e.g.,\ }}

\DeclareMathAlphabet{\mathpzc}{OT1}{pzc}{m}{it}
\nc\tensorr\mathbf
\nc\Tensor\boldsymbol

\nc\qfont{\mathsf}

\nc\FTintbypartsfont{\mathsf}  \nc\SzMmathsf{\mathtt}  

\nc\re[1]{(\ref{#1})}  
\nc\ree[2]{(\ref{#1}\text{#2})}  
\nc\m[1]{$            #1         $}  
\nc\mm[1]{$   \,      #1  \,     $}  
\nc\mmm[1]{$  \,\,    #1  \,\,   $}  
\nc\mmmm[1]{$ \,\,\,  #1  \,\,\, $}  

\nc\0[2]{\lt\1#1{{#2}\rt}}
\nc\1[2]{\ifcase #1{#2}\or(#2)\or[#2]\or\{#2\}\or\mathord
  <{#2}\mathord>\or\langle#2\rangle\or\lvert#2\rvert\or\lVert#2\rVert\fi}
\nc\2[2]{\mathinner{\ifcase #1{#2}\or\bigl(#2\bigr)\or\bigl[#2\bigr]\or
  \bigl\{#2\bigr\}\or\bigl<#2\bigr>\or\bigl\langle#2\bigr\rangle\or
  \bigl\lvert#2\bigr\rvert\or\bigl\lVert#2\bigr\rVert\fi}}
\nc\3[2]{\mathinner{\ifcase #1{#2}\or\Bigl(#2\Bigr)\or\Bigl[#2\Bigr]\or
  \Bigl\{#2\Bigr\}\or\Bigl<#2\Bigr>\or\Bigl\langle#2\Bigr\rangle\or
  \Bigl\lvert#2\Bigr\rvert\or\Bigl\lVert#2\Bigr\rVert\fi}}
\nc\4[2]{\mathinner{\ifcase #1{#2}\or\biggl(#2\biggr)\or\biggl[#2\biggr]\or
  \biggl\{#2\biggr\}\or\biggl<#2\biggr>\or\biggl\langle#2\biggr\rangle\or
  \biggl\lvert#2\biggr\rvert\or\biggl\lVert#2\biggr\rVert\fi}}
\nc\5[2]{\mathinner{\ifcase #1{#2}\or\Biggl(#2\Biggr)\or\Biggl[#2\Biggr]\or
  \Biggl\{#2\Biggr\}\or\Biggl<#2\Biggr>\or\Biggl\langle#2\Biggr\rangle\or
  \Biggl\lvert#2\Biggr\rvert\or\Biggl\lVert#2\Biggr\rVert\fi}}
\nc\9[2]{\ifcase #1{#2}\or\left(#2\right)\or\left[#2\right]\or\left
  \{#2\right\}\or\left\langle#2\right\rangle\or\left\langle#2\right
  \rangle\or\left\lvert#2\right\rvert\or\left\lVert#2\right\rVert\fi}
\nc\lt{\mathopen{}\mathclose\bgroup\left}  \nc\rt{\aftergroup\egroup\right}

\nc\f\frac
\nc\Nicefrac[2]{\nicefrac{#1}{#2}}  

\nc\scp[2]{\left \langle #1 , #2 \right \rangle}

\nc\dimless{\hat}
\nc\Appdimless{\check}

\nc\FTst[1]{#1}

\makeatletter
\nc\defeq{ \mathrel{\rlap{\raisebox{0.3ex}{$\m@th\cdot$}}\raisebox{-0.3ex}{$\m@th\cdot$}}=}
\nc\eqdef{=\mathrel{\rlap{\raisebox{0.3ex}{$\m@th\cdot$}}\raisebox{-0.3ex}{$\m@th\cdot$}} }
\makeatother

\nc\dd{\mathrm{d}}
\nc\pd{\partial}
\nc\pdt[1]{\frac{\dd #1}{\dd \hht}}
\nc\pdx[1]{\frac{\dd #1}{\dd \hx}}
\nc\ppdx[2]{\frac{\dd^#2 #1}{\dd \hx^#2}}
\nc\qpdt[1]{\frac{\pd #1}{\pd \hht}}
\nc\qpdx[1]{\frac{\pd #1}{\pd \hx}}
\nc\qppdx[2]{\frac{\pd^#2 #1}{\pd \hx^#2}}
\nc\dht[1]{\frac{\dd #1}{\dd \hht}}
\nc\pdht[1]{\frac{\pd #1}{\pd \hht}}
\nc\pdhx[1]{\frac{\pd #1}{\pd \hx}}
\nc\FTpdpdt[1]{\f{\partial #1}{\partial \qt}}
\nc\FTpdpdtdimless[1]{\f{\partial #1}{\partial \ct}}
\nc\FTpdpdx[1]{\f{\partial #1}{\partial \qx}}
\nc\FTpdpdxx[1]{\f{\partial^2 #1}{\partial \qx^2}}
\nc\FTpdpdxdimless[1]{\f{\partial #1}{\partial \hx}}
\nc\FTpdpdxxdimless[1]{\f{\partial^2 #1}{\partial \hx^2}}

\nc\zero{\tensorr 0}
\nc\ident{\tensorr 1}
\nc\ee{\mathrm{e}}
\nc\ii{\mathrm{i}}
\nc\RE{\Re}                \nc\IM{\Im}

\nc\Nm{{}}  \nc\Nn{{}}

\nc\qBi{\mathpzc{Bi}}

\nc\FTa{a}  
\nc\qc{c}  
\nc\qqc{c}  
\nc\eif{\SzMmathsf{f}}  
\nc\FTintbypartsf{\FTintbypartsfont{f}}  
\nc\eig{\SzMmathsf{g}}  
\nc\FTintbypartsg{\FTintbypartsfont{g}}  
\nc\qq{\qfont{q}}  
\nc\qqnull{\qq_0}  
\nc\qqA{Q_A}  
\nc\qqV{\qq_V}  
\nc\qt{t}  
\nc\qtdiff{\qt_\text{diff}}  
\nc\qtp{\qt_\text{p}}  
\nc\hht{\dimless{\qt}}  
\nc\ct{\Appdimless{\qt}}  
\nc\qx{x}  
\nc\qxpexp{\qx_\text{exp}}  
\nc\qxchar{\qX}  
\nc\hx{\dimless{\qx}}  
\nc\qv{{\SzMmathsf{v}}}

\nc\qA{A}  
\nc\hA{\dimless{A}}
\nc\cL{\Tensor{\mathcal{L}}}
\nc\qM{M}
\nc\qqM{\tensorr{\qM}}
\nc\qN{N}
\nc\qNnullc{\qN^{\left( 0,\qchi \right)}}
\nc\qNct{\qN^{\left( C,\qtheta \right)}}
\nc\qNst{\qN^{\left( S,\qtheta \right)}}
\nc\qNcc{\qN^{\left( C,\qchi \right)}}
\nc\qNsc{\qN^{\left( S,\qchi \right)}}
\nc\tNsc{{\tilde \qN}^{\left( S,\qchi \right)}}
\nc\tcNsc{{\tilde \qN}^{\left( S,\qchi \right)^*}}
\nc\qQ{Q}  
\nc\qT{T}  
\nc\qTinfty{\qT_\infty}  
\nc\qTnull{\qT_0}  
\nc\FTV{V}  
\nc\qX{X}  
\nc\hX{\dimless{\qX}}  
\nc\qxp{\qX_\text{p}}  
\nc\hxp{\dimless{\qX}_\text{p}}  

\nc\FTalpha{\alpha}  
\nc\eialpha{\upalpha}  \nc\eibeta{\upbeta}  
\nc\qeps{\varepsilon}
\nc\qchi{\chi}
\nc\qchinull{\qchi_0}
\nc\qtheta{\vartheta}
\nc\qthetanull{\qtheta_0}
\nc\qlambda{\lambda}  
\nc\qmu{\mu}  
\nc\qnu{\nu}  
\nc\qrho{\varrho}  
\nc\qtau{\tau}  
\nc\htau{\dimless{\qtau}}  
\nc\ctau{\Appdimless{\qtau}}  
\nc\qomega{\omega}

\nc\cPhi{\Phi}  
\nc\ccPhi{\Tensor{\cPhi}}  

\nc\qPsi{\Psi}  
\nc\qqPsi{\Tensor{\qPsi}}  
\nc\qPt{\qPsi^\qtheta}
\nc\qPc{\qPsi^\qchi}

\nc\qPhi{\qPsi}  
\nc\qqPhi{\Tensor{\qPhi}}  
\nc\qpt{\qPhi^\qtheta}  \nc\tpt{\tilde{\qPhi}^\qtheta}
\nc\qpc{\qPhi^\qchi}    \nc\tpc{\tilde{\qPhi}^\qchi}

\nc\qTheta{\Theta}

\begin{document}


\setlength{\tabcolsep}{0.45em}  

\title{Exact analytical solution for
the non-selfadjoint problem of
Maxwell--Cattaneo--Vernotte heat conduction
with
heat-transfer
boundary condition
}



\author{Mátyás Szücs}
\affiliation{Department of Energy Engineering, Faculty of Mechanical Engineering, Budapest University of Technology and Economics, Műegyetem rkp.\ 3., H-1111 Budapest, Hungary}
\affiliation{Department of Theoretical Physics, Institute for Particle and Nuclear Physics, HUN-REN Wigner Research Centre for Physics, Konkoly-Thege Miklós út 29--33., H-1121 Budapest, Hungary}
\affiliation{Montavid Thermodynamic Research Group, Society for the Unity of Science and Technology, Lovas út 18., H-1012 Budapest, Hungary}

\author{Tamás Fülöp}
\email[]{fulop.tamas@gpk.bme.hu}
\affiliation{Department of Energy Engineering, Faculty of Mechanical Engineering, Budapest University of Technology and Economics, Műegyetem rkp.\ 3., H-1111 Budapest, Hungary}
\affiliation{Montavid Thermodynamic Research Group, Society for the Unity of Science and Technology, Lovas út 18., H-1012 Budapest, Hungary}


\date{\today}

\begin{abstract}

The most well-known beyond-Fourier heat conduction model, the Maxwell--Cattaneo--Vernotte equation is solved analytically in the presence of heat transfer boundary condition.
In contrast to the corresponding Fourier problem, this boundary condition renders the underlying differential operator non-selfadjoint.
With a suitable scalar product, the adjoint eigenvalue problem is established.
The resulting left and right eigenfunctions constitute a biorthogonal system, allowing the expansion coefficients to be determined from
arbitrary
square-integrable initial conditions.
This enables
a convenient
infinite-sum analytical solution, which is presented and
thoroughly
investigated for various values of the model parameters (including near-Fourier and highly hyperbolic regimes) and for two practically important initial conditions (equilibrium and flash pulse initiated).
The spectral structure is analyzed in detail, including the occurrence of real,
imaginary, and complex-conjugate eigenvalue roots, their asymptotic distribution, and their dependence on the dimensionless
relaxation time and Biot number.
We find good agreement with corresponding finite-difference numerical solutions.
The comparison also illustrates the
effects of spectral truncation
(i.e., 
the Gibbs phenomenon) and numerical dissipation near propagating thermal-wave fronts.
Completeness of the eigenfunction set is numerically demonstrated.
The initial condition induced by the flash pulse is derived analytically.
The results provide an analytical benchmark for non-selfadjoint hyperbolic heat conduction,
unveil
how boundary heat transfer and relaxation time influence the transition between Fourier-like diffusion and thermal-wave propagation,
and open the possibility to find and investigate beyond-Fourier heat conduction via heat transfer in experiments and practical applications.

\end{abstract}


\maketitle 

\section{Introduction}

Despite the applicability of Fourier's law \cite{fourier1822theorie} for most heat conduction situations, there have long been various motivations to find alternative models for heat conduction. Both the atomic-level kinetic-theory considerations  by Maxwell
\cite{maxwell1867dynamical,grad1958principles},
and the efforts to replace the infinite effect propagation speed in Fourier's theory with a finite one
\cite{cattaneo1948sulla,vernotte1958paradoxes}
(see also thermoelasticity-related aspects, \eg \cite{chandrasekharaiah1998hyperbolic,auriault2016cattaneo,fulop2018emergence})
have led to what is now frequently called the Maxwell--Cattaneo--Vernotte (MCV) heat conduction equation. This equation is the prototype for the subsequently
emerging
collection of
beyond-Fourier models, which
 \begin{list}{\m{\bullet}}{\parsep0ex \partopsep0ex
   \leftmargin2 em \topsep0.75 ex \itemsep0.75 ex}
\item
provide
phenomenological description of
the so-called ``second sound'',
the
wave-like mode of heat propagation first
anticipated
and observed in superfluid helium~II \cite{landau1941theory,kapitza1941heat},
 \item
aimed to describe experimentally observed non-Fourier behavior in, \eg
ultrashort-pulse laser heating, proton exchange membrane fuel cell vehicles, rapid solidification phenomena, thin film physics, heterogeneous rocks, and biophysical phenomena (see, \eg \cite{mariano2022solutions,kovacs2024heat} and references therein),
 \item
followed from later statistical considerations (e.g., \cite{guyer1966solution,ruggeri2015rational}
), or
 \item
resulted from continuum thermodynamical consistency when describing rich internal material structures and phenomena (e.g., see the road from \cite{jou1988extended} through \cite{van2012universality}
).
 \end{list}

Interestingly -- but not unexpectedly -- the structure of the MCV equation, with its damped wave-propagation physical content, appears in various other fields as well. The telegrapher's equations give account of ohmically damped signal propagation in electrical transmission lines. The same form of equation describes, at much higher propagation speed, optical signal transmission
with losses.
In parallel, damped elastic vibrations (\eg of strings) are also modeled via this equation structure. Other mentionable examples are pressure waves in fluid flow 
through porous formations or with viscoelastic effects \1 1 {see, \eg \cite{zhang2025analytical} and references therein}. 

The MCV \1 1 {structured} equation describes dissipation and irreversibility in the bulk. In parallel, such effects can occur at the boundaries as well. In heat conduction, the first-kind boundary condition (BC1) prescribes temperature at the boundary (Dirichlet BC for temperature), while the second-kind one (BC2) dictates normal heat current density at the boundary (for Fourier heat conduction, this is equivalent to Neumann BC), and the third-kind one (BC3) describes heat transfer at the boundary, where normal heat current density equals boundary temperature 
deviation
from external temperature
multiplied by the heat transfer coefficient.
In fact, both BC1 and BC2 can be regarded as limiting cases of BC3: BC1
is attained
in the limit
of infinite heat transfer coefficient,
while BC2 is obtained when
the heat transfer coefficient is sent to zero while the external temperature parameter is sent to infinity in such a way that their product is finite and
provides
the desired value of normal heat current density.

Actually,
in practice, both BC1 and BC2 are
typically
chosen as mathematically simpler approximate replacements of a BC3 situation.

Analogously
to that the MCV equation structure occurs widely outside of heat conduction as well, the dissipative and irreversible nature of BC3 also has its analogs in the corresponding situations. For example, waves in cavities, resonators, and wave guides, suffer damping/losses at the walls.

All these enlight that solving BC3 problems is at least as important for practice than
dealing with
BC1 and BC2 situations.

Within the context of Fourier heat conduction, there is actually no principal difficulty regarding BC3.
Namely, the Robin BC still defines a self-adjoint problem -- the Laplacian operator residing at the heart of the problem is self-adjoint --, hence, the eigenfunctions of this operator form a complete orthogonal set so any \1 1 {reasonable} initial value can be expanded with respect to this set, the coefficients provided by scalar products \wrt the eigenfunctions; and the subsequent time dependences of the coefficients are easily obtainable exponential decreases.

The picture changes, however, drastically for the MCV equation. As it is presented below, BC3 ruins self-adjointness. Then success is not obvious any more. First, the eigenfunctions may not form a complete set. Next, assuming that completeness holds, the most favorable situation is when one is able to find a suitable scalar product and appropriate \textit{adjoint} boundary conditions with respect to which the adjoint operator can be established. Then the eigenfunctions of the adjoint operator form a dual set \1 1 {\ie have bi-orthogonal relationship to the original eigenfunction set}. If the dual set is complete, too, then taking the scalar product \wrt them provides the coefficients of the initial condition expanded \wrt the original eigenbasis.

Even if
success is achieved
at all these steps,
a problem for practice can arise: in applications, one typically determines only finitely many coefficients and hopes that the finite sum satisfactorily approximates the infinite one. However, this
property
is far from sure, as discussed in \cite{kostenbauder1997eigenmode,seshadri2006eigenmode} for a non-selfadjoint optical waveguide problem.

Unfortunately, while mathematics is very well-developed for self-adjoint problems, progress is much slower -- as being more difficult -- for the non-selfadjoint cases \1 1 {see \cite{krejcirik2014similarity,bogli2022eigenvalues,badeau2025spectral} and references therein for
certain
results
nearest
to our problem here}. For example, based on \cite{krejcirik2006closed}, Bögli et al.\ \cite{bogli2022eigenvalues} present an example where completeness of the eigenfunctions does not hold.

In the present work, we solve the MCV equation for a finite-size one space dimensional sample \1 1 {equivalent to an infinite plane wall situation as well}, with zero BC2 at one end and BC3 at the other end.
This setup is also applicable when both sides of a sample have BC3 with the same heat transfer coefficient and the initial condition is space-reflection invariant so it is enough to consider half of the sample with the reflection invariant zero BC2 at the original center.

We determine the eigenfunctions
and the eigenvalues (which, unlike in simpler settings such as the MCV + BC2 + BC2 problem \cite{nascimento2024exact}, are complex roots of a transcendental equation), identify an appropriate scalar product and dual boundary conditions, and determine the dual eigenfunctions. An asymptotic formula for the eigenvalue roots is derived analytically, and the parameter dependence of the existence of real and imaginary eigenvalue roots -- in addition to the always-present infinite series of complex conjugate pair ones -- is investigated, which results also aid in finding all the eigenvalue roots.

Completeness is
demonstrated
numerically. The MCV + BC2 + BC3 problem is solved and analyzed for various values of the free parameters, including near-Fourier \1 1 {in other words, near-parabolic} and highly hyperbolic regimes. The analytical \1 1 {truncated infinite-sum} solutions are compared to implicit Euler space-staggered finite-difference numerical outcomes, finding satisfactory agreement.

Two initial conditions are considered, both motivated by the typical practical applications: one is the equilibrized initial state and the other is the initial state induced by a short heat pulse at the BC2 side, such as a radiative pulse applied in the flash experimental method \cite{parker1961flash}. For the latter, an analytical derivation of the induced initial state is provided.

\section{Problem statement and general solution}  \label{probstatprep}

Wave-like
heat propagation within a one-dimensional spatial
region
$ 0 \le \qx \le \qX $ is investigated for the time
domain
$ \qt \ge 0 $. Assuming that the sample is at rest with respect to an inertial reference frame and
that
only heat-type interactions are to be taken into account,
the energy balance equation is
\begin{align}
    \label{eq:bal-e}
    \qrho \qc \frac{\pd \qT}{\pd \qt} &= - \frac{\pd \qq}{\pd \qx} ,
\end{align}
where $ \qT $ and $ \qq $ represent temperature and heat current density, respectively%
, and
(mass) density $ \qrho $ and
specific heat capacity $ \qc $ are both assumed to be constant.
Heat conduction with a finite propagation speed is
described
via the simplest hyperbolic model, namely, the Maxwell--Cattaneo--Vernotte equation
\begin{align}
    \label{eq:q-MCV}
    \tau \frac{\pd \qq}{\pd \qt} + \qq &= - \lambda \frac{\pd \qT}{\pd \qx} ,
\end{align}
where the thermal relaxation time $ \qtau $ and the thermal conductivity $ \qlambda $ are also considered constants.

Restricting ourselves to specific boundary conditions, we investigate an asymmetric
cooling problem; namely, an adiabatic insulation is assumed at the left boundary ($ \qx = 0 $), while convective warming or cooling to an ambient environment with constant temperature is considered at the right boundary ($ \qx = \qX $).\footnote{As expounded in the Introduction, this setup also realizes a symmetric problem.}
The corresponding boundary conditions are expressed as
\begin{align}
    \label{eq:BC-0}
    \qq ( \qt , 0 ) &= 0 , \\
    \label{eq:BC-X}
    \qq ( \qt , \qX ) &= \FTalpha \big( \qT ( \qt , \qX ) - \qTinfty \big) ,
\end{align}
where $ \FTalpha $ is the heat transfer coefficient, and $ \qTinfty $ denotes the ambient temperature. The initial conditions
are
\begin{align}
    \label{eq:IC-T}
    \qT ( 0 , \qx ) &= \qTnull ( x ) , \\
    \label{eq:IC-q}
    \qq ( 0 , \qx ) &= \qqnull ( \qx ) , 
\end{align}
with $ \qTnull ( \qx ) - \qTinfty $ and $ \qqnull ( \qx ) $ both being square-integrable functions.

Introducing the non-dimensional variables and fields
\begin{align}  \label{nondim}
  &&
    \hx &\defeq \frac{\qx}{\qX} , &
    \hht &\defeq \frac{\qt}{\frac{\qX^2}{\FTa}} , &
    \qtheta &\defeq \frac{\qT - \qTinfty}{\qTheta_{\mathrm{u}}} , &
    \qchi &\defeq \frac{\qq}{\qlambda \frac{\qTheta_{\mathrm{u}}}{\qX}} 
  &&
\end{align}
with
thermal diffusivity \mm { \FTa \defeq \f{\qlambda}{\qrho \qc} }
and an arbitrarily chosen\footnote{%
Different concrete problems provide different natural choices
for $ \qTheta_{\mathrm{u}} $, incorporating, for instance, the initial or boundary conditions, as well as the conservation of energy through the energy integral. Two specific cases are presented in Section~\ref{sec:num-ex}.} reference temperature (difference) unit $ \qTheta_{\mathrm{u}} $, the dimensionless initial-boundary value problem 
turns out to be
\begin{align}
    \label{eq:MCV-nondim-1}
    \qpdt{\qtheta} &= - \qpdx{\qchi} , \\
    \label{eq:MCV-nondim-2}
    \htau \qpdt{\qchi} + \qchi &= - \qpdx{\qtheta} , \\
    \label{eq:MCV-nondim-BC-0}
    \qtheta ( 0 , \hx ) &= \qthetanull ( \hx ) , \\
    \label{eq:MCV-nondim-BC-1}
    \qchi ( 0 , \hx ) &= \qchinull ( \hx ) , \\
    \qchi \left( \hht , 0 \right) &= 0 , \label{ftj} \\ 
    \qchi \left( \hht , 1 \right) &= \qBi \qtheta \left( \hht , 1 \right) ,\label{ftk}
\end{align}
where $ \htau \defeq
\frac{\qtau}{\nicefrac{\qX^2}{\FTa}}
> 0 $ and
$ \qBi \defeq \frac{\qX}{\nicefrac{\qlambda}{\FTalpha}} > 0 $ denote the dimensionless relaxation time and the Biot number, respectively.

\subsection{The eigenvalue--eigenfunction problem}  \label{fti}

The operator form of the dynamical set of equations \re{eq:MCV-nondim-1}--\re{eq:MCV-nondim-2} is
\begin{align}
    \qpdt{} \begin{pmatrix} \qtheta
    \vphantom{\frac{1}{\htau} \qpdx{}} \\[.7ex] \vphantom{\frac{1}{\htau} \qpdx{}}
    \qchi \end{pmatrix} =
    \begin{pmatrix}
        0 & - \qpdx{} \\[.7ex]
        - \frac{1}{\htau} \qpdx{} & - \frac{1}{\htau}
    \end{pmatrix}
    \begin{pmatrix} \qtheta
    \vphantom{\frac{1}{\htau} \qpdx{}} \\[.7ex] \vphantom{\frac{1}{\htau} \qpdx{}}
    \qchi \end{pmatrix} ,
\qquad\qquad
\text{that is,}
\qquad\qquad
    \label{eq:MCV-op-form}
    \qpdt{} \begin{pmatrix} \qtheta
    \vphantom{\frac{1}{\htau} \qpdx{}} \\[.7ex] \vphantom{\frac{1}{\htau} \qpdx{}}
    \qchi \end{pmatrix} =
    \cL \begin{pmatrix} \qtheta
    \vphantom{\frac{1}{\htau} \qpdx{}} \\[.7ex] \vphantom{\frac{1}{\htau} \qpdx{}}
    \qchi \end{pmatrix}
    \qquad\quad
    \qquad
 \end{align}
with the linear operator
 \begin{align}
    \cL
    \defeq
    \begin{pmatrix}
        0 & - \qpdx{} \\[.7ex]
        - \frac{1}{\htau} \qpdx{} & - \frac{1}{\htau}
    \end{pmatrix} .
\end{align}
The corresponding
eigenvalue--eigenfunction problem is
\begin{align}  \label{ftc}
    \cL \qPhi_n ( \hx ) = \qmu_n \qPhi_n ( \hx ) ,
\end{align}
where the eigenfunctions are two-component ones, denoted as
\begin{align}
    \qPhi_n ( \hx ) = \begin{pmatrix} \qpt_n ( \hx ) \\ \qpc_n ( \hx ) \end{pmatrix} .
\end{align}
We expect 
a solution of \eqref{eq:MCV-op-form}
in the form
\begin{align}
    \begin{pmatrix} \qtheta \\ \qchi \end{pmatrix} \left( \hht , \hx \right) = \sum_{n = 1}^\infty \qqc_n \left( \hht \right) \begin{pmatrix} \qpt_n ( \hx ) \\ \qpc_n ( \hx ) \end{pmatrix} .
\end{align}
Replacing into \re{eq:MCV-op-form}, and assuming that the
involved formal
steps are
mathematically
allowed:
\begin{align}  \label{ftf}
    \sum_{n = 1}^\infty \pdt{\qqc_n} \left( \hht \right) \qPhi_n ( \hx ) = \sum_{n = 1}^\infty \qqc_n \left( \hht \right) \cL \qPhi_n ( \hx ) = \sum_{n = 1}^\infty \qqc_n \left( \hht \right) \qmu_n \qPhi_n ( \hx ) .
\end{align}

Turning now towards the details of the eigenfunctions, writing \eqref{ftc} component-wise yields
\begin{align}
    \label{eq:eigen-1}
    - \pdx{\qpc_n} ( \hx ) &= \qmu_n \qpt_n ( \hx ) , \\
    \label{eq:eigen-2}
    -  \frac{1}{\htau} \pdx{\qpt_n} ( \hx ) - \frac{1}{\htau} \qpc_n ( \hx ) &= \qmu_n \qpc_n ( \hx ) .
\end{align}
With \m { \qmu_n = 0 },
only the identically zero solution of \eqref{eq:eigen-1}--\eqref{eq:eigen-2} satisfies
\eqref{ftj}--\eqref{ftk} so, hereafter, \m { \qmu_n \ne 0 }.

Eliminating $ \qpt_n $ [multiplying \eqref{eq:eigen-1} by \m { -1 } and differentiating it, and substituting \m { - \pdx{\qpt_n} ( \hx ) } expressed from \eqref{eq:eigen-2}]:
\begin{align}  \label{redeq}
    \ppdx{\qpc_n}{2} ( \hx ) = \qmu_n^{} \big( \qpc_n ( \hx ) + \htau \qmu_n^{} \qpc_n ( \hx ) \big) = \left( \qmu_n^{} + \htau \qmu_n^2 \right) \qpc_n ( \hx )
    = \qmu_n^{} \left( 1 + \htau \qmu_n \right) \qpc_n ( \hx ).
\end{align}

In addition to the already excluded \m { \qmu_n = 0 },
\mm { \qmu_n = - \f{1}{\htau} } also makes the coefficient of the \rhs\ of \re{redeq} zero. If \mm { \qmu_n = - \f{1}{\htau} } then \eqref{eq:eigen-2} gets simplified to \mm { \pdx{\qpt_n} ( \hx ) = 0 \,, } \mm { \qpt_n = \text{const.} \eqdef \qNnullc_n \,. } This reduces \eqref{eq:eigen-1} to \mm { \pdx{\qpc_n} ( \hx ) = \frac{1}{\htau} \qNnullc_n } which, in light of \eqref{ftj}, leads to \mm { \qpc_n ( \hx ) = \frac{1}{\htau} \qNnullc_n \hx \,. } Finally, \eqref{ftk} requires \mm { \frac{1}{\htau} = \qBi } for a nonzero \m { \qNnullc_n } \1 1 {\ie for a not-identically-zero \m { \qPhi_n }}. Summarizing: \mm { \qmu_n = - \f{1}{\htau} } is an eigenvalue
only for
\mm { \htau \qBi = 1 \,, } and then the eigenfunction is
 \begin{align}  \label{speceigenfn}
\qPhi_n ( \hx ) = \qNnullc_n \0 1 {
\begin{matrix}
1 \\ \f{1}{\htau} \hx
\end{matrix}
} \,.
 \end{align}

Let us next turn to the generic case \mm { \qmu_n^{} + \htau \qmu_n^2 \ne 0 \,. } Introducing
the nonzero \m { \qnu_n } with
\begin{align}
    \label{eq:nu-def}
    - \qnu_n^2 \defeq \qmu_n^{} \left( 1 + \htau \qmu_n^{} \right) ,
  \qquad \qquad
    \RE \1 0 { \qnu_n^{} } \ge 0, \quad \text{if} \,\, \RE \1 0 { \qnu_n^{} } = 0 \,\, \text{then} \,\, \IM \1 0 { \qnu_n^{} }
    >
    0
    ,
\end{align}
the reduced equation \re{redeq} becomes
\begin{align}  \label{redeqq}
    \ppdx{\qpc_n}{2} ( \hx ) = - \qnu_n^2 \qpc_n ( \hx ) .
\end{align}
Then the eigenfunctions are of the form
\begin{align}  \label{ftz}
    \begin{pmatrix} \qpt_n ( \hx ) \\ \qpc_n ( \hx ) \end{pmatrix} =
    \begin{pmatrix}
        \qNct_n \cos \left( \qnu_n \hx \right) + \qNst_n \sin \left( \qnu_n \hx \right) \\
        \qNcc_n \cos \left( \qnu_n \hx \right) + \qNsc_n \sin \left( \qnu_n \hx \right)
    \end{pmatrix} ,
\end{align}
with coefficients \mm { \qN^{\0 1 {\cdots}}_n } that are arbitrary as long as only \re{redeqq} is concerned. Requirements come, however, from the
boundary conditions:
\begin{align}  \label{ftq}
    \qpc_n (0) &= 0 , \\  \label{fto}
    \qpc_n (1) &= \qBi \qpt_n (1) .
\end{align}
Exploiting \re{ftq} gives\footnote
{A note on notations: for example, \m {A
\stackrel{\re{eq:MCV-nondim-2}}{=}
B} will denote that, using
\re{eq:MCV-nondim-2},
\m { A } turns out to be equal to \m { B }.}
\begin{align}  \label{ftx}
    \qpc_n (0) &= \qNcc_n = 0 , \qquad\qquad \Longrightarrow \qquad\qquad
    \qpc_n ( \hx ) = \qNsc_n \sin \left( \qnu_n \hx \right) ;
    \\  \label{fty}
    \qpt_n ( \hx ) & \stackrel{\re{eq:eigen-1}}{=} - \frac{1}{\qmu_n} \pdx{\qpc_n} ( \hx ) \stackrel{\ree{ftx}{b}}{=} - \frac{\qnu_n}{\qmu_n} \qNsc_n \cos \left( \qnu_n \hx \right) .
\end{align}

One immediate consequence is that \re{fty} also means \mm { \qNst_n = 0 }, in view of \ree{ftz}{a}.

Writing \m { \qpt_n ( 0 ) } is now possible in two ways: based on \ree{ftz}{a}, and utilizing \re{fty}. The two versions must be equal so
\begin{align}
    \label{eq:Nt-Nc}
    \qNct_n
    =
    - \frac{\qnu_n}{\qmu_n} \qNsc_n .
\end{align}
Substituting \re{ftz} into
\eqref{fto}, with all that we have learnt so far about the coefficients \m { \qN^{\0 1 {\cdots}}_n },
yields (after dividing by \m { \qNsc_n } which cannot be 0 otherwise both eigenfunction components were 0)
\begin{align}  \label{ftn}
    \sin \qnu_n
    = - \qBi \frac{\qnu_n}{\qmu_n} \cos \qnu_n .
\end{align}
If \m { \cos \qnu_n = 0 } then \m { |\sin \qnu_n| = 1 }, the \lhss is nonzero but the \rhss is zero, which is impossible, therefore, hereafter, \m { \cos \qnu_n \ne 0 }.
In parallel, if \m { \sin \qnu_n = 0 } then \m { |\cos \qnu_n| = 1 }, the \lhss is zero while the \rhss is nonzero -- again a contradiction -- so, hereafter, \m { \sin \qnu_n \ne 0 }. Then we can write
the relationship \eqref{ftn} between $ \qmu_n $ and $ \qnu_n $ as
\begin{align}
    \label{eq:mu-nu}
    \qmu_n = - \qBi \frac{\qnu_n}{\tan \qnu_n} .
\end{align}
Multiplying \re{eq:nu-def} by $ \sin \qnu_n $ and substituting \re{eq:mu-nu}, the transcendental equation on $ \qnu_n $ is
\begin{align}  \label{ftp}
    - \left( \sin^2 \qnu_n + \htau \qBi^2 \cos^2 \qnu_n \right) \qnu_n + \qBi \sin \qnu_n \cos \qnu_n = 0 .
\end{align}
If \m { \qnu_n^{} } is a solution to it then \m { \qnu_n^*} is also a solution. Via the complex conjugate of \eqref{eq:mu-nu} we see that if \m { \qmu_n^{} } corresponds to \m { \qnu_n^{} } then \m { \qmu_n^* } corresponds to \m { \qnu_n^*}.

Applying \re{eq:mu-nu}, the relationship \re{eq:Nt-Nc} is
\begin{align}  \label{NN}
    \qNct_n = \frac{\tan \qnu_n}{\qBi} \qNsc_n ,
\end{align}
hence,
the eigenfunctions are
\begin{align}  \label{NNN}
    \begin{pmatrix} \qpt_n ( \hx )
    \vphantom{\frac{\tan \qnu_n}{\qBi} \qNsc_n}
    \\
    \vphantom{\qNsc_n}
    \qpc_n ( \hx ) \end{pmatrix} =
    \begin{pmatrix}
        \frac{\tan \qnu_n}{\qBi} \qNsc_n \cos \left( \qnu_n \hx \right) \\
        \qNsc_n \sin \left( \qnu_n \hx \right)
    \end{pmatrix} .
\end{align}

We mention that the \m { \qnu_n \to 0 } limit of \re{NN}--\re{NNN} reproduces \re{speceigenfn} with \m { \qNct_n \to \qNnullc_n }, which is actually
plausible from the \m { \qnu_n \to 0 } limit \1 1 {the \m { \qmu_n \to - \f{1}{\htau} } limit} of \re{eq:nu-def}. In this sense, \m { \qnu_n = 0 } is also allowed.

Some first comments on the spectrum are as follows.
We can see that each \m { \qmu_n } is nondegenerate, since a single \m { \qnu_n } belongs to it, and a single \m { \qnu_n } describes a one-dimensional eigensubspace.
More closely, a given \m { \qmu_n } assigns a given \m { \qnu_n } via \re{eq:nu-def} \1 1 {releasing the sign convention of \m { \qnu_n } does not introduce multiplicity because the one-dimensional eigenspace given by \re{NNN} is invariant under \m { \qnu_n
\leftrightarrow
- \qnu_n }}. Conversely, a given \m { \qnu_n } assings a given \m { \qmu_n }, through \re{eq:mu-nu}. This \m { \qmu_n } is real when \m { \qnu_n } is real and also when it is imaginary \1 2 {see \re{eq:mu-nu} again}.

\subsection{The scalar product and the adjoint problem} \label{sec:adjoint}

We define the
scalar product
\begin{align}  \label{fte}
    \scp{ \rule{0pt}{2 ex} \qqPsi_1 }{ \qqPsi_2 }
    \defeq \int\limits_0^1 \qqPsi_1^\dagger \qqM \qqPsi_2^{} \dd \hx
    = \int\limits_0^1 \left( {\qPt_1}^* ( \hx ) \qPt_2 ( \hx ) + \htau {\qPc_1}^* ( \hx ) \qPc_2 ( \hx ) \right) \dd \hx
\end{align}
where $^\dagger$ denotes transpose combined with complex conjugation -- note the presence of \m { \htau } through the matrix
\begin{align}
    \qqM = \begin{pmatrix} 1 & 0 \\ 0 & \htau \end{pmatrix} .
\end{align}
This scalar product defines our Hilbert space. Its elements \m { \qqPsi } are those functions where both components \m { \qpt } and  \m { \qpc } are square integrable.

In this Hilbert space, the
adjoint problem is addressed as follows:
\begin{align}  \label{ftd}
    \scp{ \tilde{\qqPsi}_\Nm }{ \cL \qqPsi_\Nn }
    = \scp{ \cL^\dagger \tilde{\qqPsi}_\Nm }{ \qqPsi_\Nn } :
\end{align}
\begin{align}
    & \scp{ \begin{pmatrix}  \tpt_\Nm \\ \tpc_\Nm  \end{pmatrix} }{ \begin{pmatrix}
        0 & - \qpdx{} \\ - \frac{1}{\htau} \qpdx{} & - \frac{1}{\htau}
    \end{pmatrix} \begin{pmatrix} \qpt_\Nn \\ \qpc_\Nn \end{pmatrix} }
    = - \int\limits_0^1 \left( \rule{0pt}{3.7 ex}
    \mbox{\m{\tpt_\Nm}}^*
    ( \hx ) \pdx{\qpc_\Nn} ( \hx ) + \mbox{\m{\tpc_\Nm}}^* ( \hx ) \left( \pdx{\qpt_\Nn} ( \hx ) + \qpc_\Nn ( \hx ) \right) \right) \dd \hx \\
    &
    \hskip 8 em  
    = \int\limits_0^1 \left( \pdx{\mbox{\m{\tpt_\Nm}}^*} ( \hx ) \qpc_\Nn ( \hx ) + \pdx{\mbox{\m{\tpc_\Nm}}^*} ( \hx ) \qpt_\Nn ( \hx ) - \mbox{\m{\tpc_\Nm}}^* \qpc_2 ( \hx ) \right) \dd \hx - \left[ \mbox{\m{\tpt_\Nm}}^* \qpc_\Nn + \mbox{\m{\tpc_\Nm}}^* \qpt_\Nn \right]_0^1 \\
    &
    \hskip 8 em  
    =     \scp{ \begin{pmatrix}
        0 & \qpdx{} \\
        \frac{1}{\htau} \qpdx{} & - \frac{1}{\htau}
    \end{pmatrix} \begin{pmatrix} \tpt_\Nm ( \hx ) \\ \tpc_\Nm ( \hx ) \end{pmatrix} }{\begin{pmatrix} \qpt_\Nn ( \hx ) \\ \qpc_\Nn ( \hx ) \end{pmatrix} } - \left[ \mbox{\m{\tpt_\Nm}}^* \qpc_\Nn + \mbox{\m{\tpc_\Nm}}^* \qpt_\Nn \right]_0^1 ,
\end{align}
 \begin{align}  \label{ftb}
\left[ \mbox{\m{\tpt_\Nm}}^* \qpc_\Nn + \mbox{\m{\tpc_\Nm}}^* \qpt_\Nn \right]_0^1 & =
\mbox{\m{\tpt_\Nm}}^*(1) \qpc_\Nn(1) + \mbox{\m{\tpc_\Nm}}^*(1) \qpt_\Nn(1) -
\mbox{\m{\tpt_\Nm}}^*(0) \qpc_\Nn(0) - \mbox{\m{\tpc_\Nm}}^*(0) \qpt_\Nn(0)
 \\  & =
\left( \mbox{\m{\tpt_\Nm}}^*(1) \qBi + \mbox{\m{\tpc_\Nm}}^*(1) \right) \qpt_\Nn(1) -
\mbox{\m{\tpt_\Nm}}^*(0) \cdot 0 - \mbox{\m{\tpc_\Nm}}^*(0) \qpt_\Nn(0) .
 \end{align}
Therefore, the adjoint problem is identified as
\begin{align}
    \begin{pmatrix}
        0 & \qpdx{} \\
        \frac{1}{\htau} \qpdx{} & - \frac{1}{\htau}
    \end{pmatrix} \begin{pmatrix} \tpt_\Nm ( \hx ) \\ \tpc_\Nm ( \hx ) \end{pmatrix} &= \tilde{\qmu}_\Nm \begin{pmatrix} \tpt_\Nm ( \hx ) \\ \tpc_\Nm ( \hx ) \end{pmatrix}, \\[1.5 ex]  
    \mbox{\m{\tpc_\Nm}}^* (0) &= 0 , \\[0 ex]  
    \mbox{\m{\tpc_\Nm}}^* (1) &= - \qBi \mbox{\m{\tpt_\Nm}}^* (1) ;
\end{align}
since $ \qBi \in \mathbb{R} $, the adjoint boundary conditions are
\begin{align}  \label{ftg}
    {\tpc_\Nm} (0) &= 0 , \\    \label{fth}
    {\tpc_\Nm} (1) &= - \qBi {\tpt_\Nm} (1) .
\end{align}
These lead to
the same
eigenvalue roots
as of the original problem \1 2 {\ie in \eqref{eq:nu-def} and \eqref{ftn}--\eqref{ftp}, one is allowed to replace \m { \qnu_n } with \m { \tilde{\qnu}_m } and \m { \qmu_n } with \m { \tilde{\qmu}_m }}, and the corresponding
\1 1 {frequently called ``left''}
eigenfunctions are:
\begin{align}
    \begin{pmatrix} {\tpt}_m ( \hx )
    \vphantom{\frac{\tan \tilde{\qnu}_m}{\qBi} \tNsc_n}
    \\
   \vphantom{\tNsc_n}
   {\tpc}_m ( \hx ) \end{pmatrix} =
    \begin{pmatrix}
        - \frac{\tan \tilde{\qnu}_m}{\qBi} \tNsc_m \cos \left( \tilde{\qnu}_m \hx \right) \\
        \tNsc_m \sin \left( \tilde{\qnu}_m \hx \right)
    \end{pmatrix} .
\end{align}

\subsubsection{Biorthogonality of the eigenfunctions and normalization}

On the ``left'' and ``right'' eigenfunctions, \eqref{ftd} tells
\begin{align}
    \scp{\tilde{\qqPhi}_m
    }{\qmu_n \qqPhi_n
    } = \scp{\tilde{\qmu}_m
    \tilde{\qqPhi}_m
    }{\qqPhi_n
    } ,
\end{align}
\begin{align}
    \left( \qmu_n^{} - \tilde{\qmu}_m^* \right) \scp{\tilde{\qqPhi}_m^{} 
    }{\qqPhi_n^{}
    } = 0 .
\end{align}
If $ \qmu_n^{} \neq \tilde{\qmu}_m^* $ then $ \scp{\tilde{\qqPhi}_m
}{\qqPhi_n } = 0 $, \ie the functions $ \tilde{\qPhi}_m $
form
a dual
system of functions
for the $ \qqPhi_n $s.
It is expected that both the right and the left eigenfunction sets are complete.
If, on the other side,
$ \scp{\tilde{\qqPhi}_m}{\qqPhi_n} \ne 0 $
then $ \qmu_n^{} = \tilde{\qmu}_m^* $.
This enables to order the
left
eigenfunctions along the order of the
right
ones \1 1 {\ie \m { m \equiv n }, \m { \qnu_m \equiv \qnu_n } is possible}%
. Namely, by \re{eq:nu-def},
\mm { \qmu_n^{} = \tilde{\qmu}_m^* }
implies
\mm { \qnu_n^{} = \tilde{\qnu}_m^* }
\1 1 {unless \m { \tilde{\qmu}_m } is real, in which case \m { \qnu_n }
suffices%
\footnote{%
In view of \ree{eq:nu-def}{a} and its complex conjugate \mm { - \qnu_n^{2} = - \0 1 {\tilde{\qnu}_m^*}^2 }, the only question is whether \m { \tilde{\qnu}_m^* } respects the convention \ree{eq:nu-def}{b}. Only \mm { \Im \tilde{\qnu}_m^* = 0 }
is to address,
and in that case both roots \m { \tilde{\qmu}_m^* } of \mmm { - \0 1 {\tilde{\qnu}_m^*}^2 = \tilde{\qmu}_m^* \left( 1 + \htau \tilde{\qmu}_m^* \right) } are real, \m { \tilde{\qmu}_m } is real, and is characterized by \m { \qnu_n }.}
for the characterization of
\m { \tilde{\qmu}_m^{} = \qmu_n }}.
This suggests to choose the numbering (ordering) of \m { \tilde{\qnu} }s in tune with the numbering (ordering) of \m { \qnu }s: hereafter, let \mm { \tilde{\qnu}_n = \qnu_n^{(*)} } \1 2 {\m{^{(*)}} denoting that, for real eigenvalues \m { \tilde{\qmu}_n }, \mm { \tilde{\qnu}_n^{} = \qnu_n^{} \,,} otherwise \mm { \tilde{\qnu}_n^{} = \qnu_n^{*} }}.
Recall that if \m { \qnu_n^{} } is a solution to \eqref{ftp} then \m { \qnu_n^*} is also a solution of it, hence, the collection of solutions \m { \tilde{\qnu} } is merely a rearrangement of the collection of solutions \m { \qnu }.

$ \scp{\tilde{\qqPhi}_m}{\qqPhi_n} \ne 0 $ can be concretized as
$ \scp{\tilde{\qqPhi}_m}{\qqPhi_n} = \delta_{mn} $
(where $ \delta_{mn} $ denotes Kronecker delta) with appropriate normalization of the eigenfunctions as follows.
For
generic \m { m, n }, in virtue of 
elementary trigonometric identities,
\begin{align}
    \scp{ \begin{pmatrix} \tpt_m \\ \tpc_m \end{pmatrix} }
        { \begin{pmatrix} \qpt_n \\ \qpc_n \end{pmatrix} }  =
    \tcNsc_m \qNsc_n \int\limits_0^1 \left( - \frac{\tan \tilde{\qnu}_m^* \tan \qnu_n^{}}{\qBi^2} \cos \left( \tilde{\qnu}_m^* \hx \right) \cos \left( \qnu_n^{} \hx \right) + \htau \sin \left( \tilde{\qnu}_m^* \hx \right) \sin \left( \qnu_n^{} \hx \right) \right) \dd \hx ,
\end{align}
hence, $ \scp{\tilde{\qqPhi}_m}{\qqPhi_n} = \delta_{mn} $ can be achieved by, for example, $ \qNsc_n \defeq 1 $ and
\begin{align}
    \tNsc_m \defeq - \frac{2 \qBi^2}{\left( \tan^2 \tilde{\qnu}_m - \htau \qBi^2 \right) + \left(\tan^2 \tilde{\qnu}_m + \htau \qBi^2 \right) \frac{\sin \left( 2 \tilde{\qnu}_m \right)}{2 \tilde{\qnu}_m}} .
\end{align}

\subsection{Checking completeness}

Completeness means that any square integrable function -- including the test functions \m { \ccPhi } known from distribution theory,
which form a dense subset in our Hilbert space --
can be given as the linear combination of the ``right'' eigenfunctions, where the coefficients are its scalar products with the corresponding "left" eigenfunctions:
\begin{align}
    \ccPhi ( \hx ) = \sum_{n=1}^\infty \qqc_n \qqPhi_n ( \hx )
    \qquad  \text{with}  \qquad
    \qqc_n = \scp{\tilde{\qqPhi}_n}{\ccPhi}.
\end{align}

In indexed notation with $ i,j = 1,2 $ \1 1 {indexing the \m { \qtheta } and \m { \qchi } components, respectively},
\begin{align}
    \cPhi^i ( \hx )
    &=
    \sum_{n=1}^\infty \scp{\tilde \qqPhi_n}{\ccPhi} \qPhi_n^i ( \hx )
    =
    \sum_{n=1}^\infty \left( \int\limits_0^1 \tilde \qqPhi_n^\dagger ( \hx' ) \qqM \ccPhi ( \hx' ) \dd \hx' \right) \qPhi_n^i ( \hx )
    =
    \int\limits_0^1 \left( \sum_{n=1}^\infty
    {\qPhi_n \vphantom{\qqPhi_n^\dagger}\hskip-.55 em}^i\hskip.25 em
    ( \hx )
    \left( {{\tilde{\qqPhi}}_n}^\dagger ( \hx' ) \right)^j
    { \qM \vphantom{\qPhi_n^\dagger} }^{jk} \right) \cPhi^k ( \hx' ) \dd \hx' .
\end{align}
On the other side,
 \begin{align}  \label{fts}
    \cPhi^i ( \hx ) &= \int\limits_0^1 \delta^{ik} \delta \left( \hx - \hx' \right) \cPhi^k ( \hx' ) \dd \hx'
 \end{align}
with the Dirac delta \m { \delta \left( \hx - \hx' \right) }. Comparison then gives the requirement
\begin{align}
    \label{eq:dirac-mx}
    \sum_{n=1}^\infty
    \begin{pmatrix}
        \qpt_n ( \hx ) \left( {\tpt_n} ( \hx' ) \right)^* & \htau \qpt_n ( \hx ) \left( {\tpc_n} ( \hx' ) \right)^* \\
        \qpc_n ( \hx ) \left( {\tpt_n} ( \hx' ) \right)^* & \tau \qpc_n ( \hx ) \left( {\tpc_n} ( \hx' ) \right)^*
    \end{pmatrix}
    =
    \begin{pmatrix}
        \delta \left( \hx - \hx' \right) & 0 \\ 0 & \delta \left( \hx - \hx' \right)
    \end{pmatrix} .
\end{align}

\subsection{Time evolution of the eigenmodes}

Returning to \re{ftf}, repeated here as
\begin{align}
    \sum_{n = 1}^\infty \pdt{\qqc_n} \0 1 { \hht } \begin{pmatrix} \qpt_n ( \hx ) \\ \qpc_n ( \hx ) \end{pmatrix} = \sum_{n = 1}^\infty \qmu_n \qqc_n \0 1 {\hht } \begin{pmatrix} \qpt_n ( \hx ) \\ \qpc_n ( \hx )    \end{pmatrix} ,
\end{align}
both sides here provide expansion \wrt the eigenbasis.
Expansion of any square integrable function in a basis has
unique coefficients, which leads to
\begin{align}
    && && &&
    \pdt{\qqc_n} \0 1 { \hht } &= \qmu_n \qqc_n \0 1 { \hht } & \Longrightarrow &&
    \qqc_n (\hht) &= \qqc_n (0) \ee^{\qmu_n \hht} ,
    && && &&
\end{align}
\begin{align}
    \begin{pmatrix} \qtheta \\ \qchi \end{pmatrix} \left( \hht , \hx \right) = \sum_{n = 1}^\infty \qqc_n ( 0 ) \ee^{\qmu_n \hht}  \begin{pmatrix} \qpt_n ( \hx ) \\ \qpc_n ( \hx ) \end{pmatrix} .
\end{align}

The initial conditions fix the coefficients \m { \qqc_n (0) }:
\begin{align}
    \begin{pmatrix} \qtheta \\ \qchi \end{pmatrix} \left( 0 , \hx \right) = \sum_{n = 1}^\infty \qqc_n ( 0 ) \begin{pmatrix} \qpt_n ( \hx ) \\ \qpc_n ( \hx ) \end{pmatrix} = \begin{pmatrix} \qtheta_0 ( \hx ) \\ \qchi_0 ( \hx ) \end{pmatrix} ,
\end{align}
\begin{align}
    \scp{ \begin{pmatrix} \tpt_m
    \\ \tpc_m
    \end{pmatrix} }{\sum_{n = 1}^\infty \qqc_n ( 0 ) \begin{pmatrix} \qpt_n
    \\ \qpc_n
    \end{pmatrix} } = \sum_{n = 1}^\infty \qqc_n ( 0 ) \underbrace{ \scp{ \begin{pmatrix} \tpt_m
    \\ \tpc_m
    \end{pmatrix} }{\begin{pmatrix} \qpt_n
    \\ \qpc_n
    \end{pmatrix} }}_{\delta_{mn}} = \scp{ \begin{pmatrix} \tpt_m
    \\ \tpc_m
    \end{pmatrix} }{\begin{pmatrix} \qtheta_0
    \\ \qchi_0
    \end{pmatrix}} ,
\end{align}
\begin{align}
    \qqc_m ( 0 ) = \scp{ \begin{pmatrix} \tpt_m
    \\ \tpc_m
    \end{pmatrix} }{\begin{pmatrix} \qtheta_0
    \\ \qchi_0
    \end{pmatrix}} = - \tcNsc_m  \int\limits_0^1 \left( \frac{\tan \tilde{\qnu}_m^*}{\qBi} \cos \left( \tilde{\qnu}_m^* \hx \right) \qtheta_0 \left( \hx \right) - \htau \sin \left( \tilde{\qnu}_m^* \hx \right) \qchi_0 \left( \hx \right) \right) \dd \hx .
\end{align}

\subsection{Summarizing the workflow}

For practical use, the essential ingredients of the analytical solution are
collected below in a compact,
ready-to-apply recipe.
\begin{enumerate}
    \item
    Solve the transcendental equation
    \begin{align}
        \label{eq:MCV-eig-val-eq}
        - \left( \sin^2 \qnu_n + \htau \qBi^2 \cos^2 \qnu_n \right) \qnu_n + \qBi \sin \qnu_n \cos \qnu_n = 0 .
    \end{align}
The same equation is obtained for the adjoint problem. The ordering of the adjoint roots are suggested to be \m { \tilde{\qnu}_n^{} = \qnu_n^* }. Note that the underlying differential operator of the investigated problem is neither self-adjoint nor skew-adjoint; therefore, the
eigenvalue roots
are generally complex numbers. Depending on the parameters $ \htau $ and $ \qBi $, real as well as purely imaginary
eigenvalue roots
can also emerge. A detailed analysis of the transcendental equation \re{eq:MCV-eig-val-eq} is presented in Appendix~\ref{sec:eigenvalues}.

    \item
    Determine the eigenvalues $ \qmu_n $ and the normalization factors $ \tNsc_n $ via
    \begin{align}
        \label{eq:MCV:eig-vals}
        \qmu_n &= - \qBi \frac{\qnu_n}{\tan \qnu_n} , \\
        \label{eq:MCV:norm-fac}
        \tNsc_n & = - \frac{2 \qBi^2}{\left( \tan^2 \tilde{\qnu}_n - \htau \qBi^2 \right) + \left(\tan^2 \tilde{\qnu}_n + \htau \qBi^2 \right) \frac{\sin \left( 2 \tilde{\qnu}_n \right)}{2 \tilde{\qnu}_n}} .
    \end{align}
    
    \item
    Determine the expansion coefficients from the initial conditions $ \qtheta_0 \left( \hx \right) $ and $ \qchi_0 \left( \hx \right) $ using the formula
    \begin{align}
        \label{eq:c_n(0)}
        \qqc_n ( 0 ) = - \tcNsc_n  \int\limits_0^1 \left( \frac{\tan \tilde{\qnu}_n^*}{\qBi} \cos \left( \tilde{\qnu}_n^* \hx \right) \qtheta_0 \left( \hx \right) - \htau \sin \left( \tilde{\qnu}_n^* \hx \right) \qchi_0 \left( \hx \right) \right) \dd \hx .
    \end{align}
    
    \item
    The dimensionless temperature field and the dimensionless heat current density field are determined using the previously calculated quantities in the form
    \begin{align}
        \label{eq:MCV-dimless-sol}
        \begin{pmatrix} \qtheta \\ \qchi \end{pmatrix} \left( \hht , \hx \right) = \sum_{n = 1}^\infty \qqc_n ( 0 ) \ee^{\qmu_n \hht}
        \begin{pmatrix}
            \frac{\tan \qnu_n}{\qBi} \cos \left( \qnu_n \hx \right) \\
            \sin \left( \qnu_n \hx \right)
        \end{pmatrix} .
    \end{align}
    
\end{enumerate}

\section{Example applications and their analysis} \label{sec:num-ex}

Next, we present the results obtained from the analytical solutions for two
relevant problems
for practice,
evaluated by truncating the infinite series to a finite number of eigenmodes.

\subsection{One-sided convective cooling from a homogeneous initial state}

This subsection extends the classical engineering problem of the transient cooling of an infinite wall -- or a slab -- adiabatically insulated on one side to the framework of the Maxwell--Cattaneo--Vernotte equation. While traditional heat conduction solutions rely on the parabolic Fourier law, the present approach incorporates hyperbolic non-Fourier effects, accounting for a finite thermal propagation speed. Assuming a spatially uniform initial temperature field, the analytical solution is evaluated to explore the interplay between convective surface cooling and internal heat propagation appearing as damped waves.

A fundamental distinction from Fourier's heat conduction is that, while a uniform initial temperature field $ \qTnull $ inherently implies a zero heat 
current density,
the hyperbolic nature of the MCV equation allows for a non-zero initial heat
current density
even under a homogeneous temperature distribution. However, by assuming that the specimen has been in thermal equilibrium for a sufficiently long time before the cooling process initiates, the initial heat current density can be reasonably considered to be zero. Consequently, the initial conditions for this case are given by
\begin{align}
    \qT ( 0 , \qx ) &= \qTnull \neq \qTinfty , \\
    \qq ( 0 , \qx ) &= 0 .
\end{align}
In this case, it is advisable to choose the temperature unit as $ \qTheta_{\mathrm{u}} = \qTnull - \qTinfty $, resulting in the dimensionless initial conditions
\begin{align}
    \label{eq:IC-T-hom}
    \qtheta \left( 0 , \hx \right) &= 1 , \\
    \label{eq:IC-q-hom}
    \qchi \left( 0 , \hx \right) &= 0 .
\end{align}
Correspondingly, the expansion coefficients---determined from \re{eq:c_n(0)} via the initial conditions \re{eq:IC-T-hom}--\re{eq:IC-q-hom}---are
\begin{align}
    \label{eq:MCV-c_n(0)-hom}
    \qqc_n ( 0 ) = - \tcNsc_n \frac{\tan \tilde{\qnu}_n^*}{\qBi} \frac{\sin \tilde{\qnu}_n^*}{\tilde{\qnu}_n^*}
    = - \tcNsc_n \frac{\tan {\qnu}_n}{\qBi} \frac{\sin {\qnu}_n}{{\qnu}_n} .
\end{align}

\subsubsection{Numerical
testing
of biorthogonality and completeness}

In order to demonstrate the mathematical consistency and practical applicability of the derived analytical framework, a numerical evaluation of the biorthogonality and completeness relations is performed. For this illustrative analysis, we choose the dimensionless parameters $ \qBi = 0.2 $ and $ \htau = 1 $, which correspond to a physically representative regime of weakly damped thermal wave propagation. Within this regime, the non-self-adjoint nature of the operator is strongly evident, making it an ideal case to numerically illustrate and support the analytical characteristics of the eigensystem.

The numerical values characterizing the first 23 eigenmodes are summarized in Table~\ref{tab:eig-val-test}. It is worth noting
that, for these dimensionless parameters,
the spectrum features a single, purely real
eigenvalue root,
indexed as $ \qnu_0 $. Consequently, the corresponding expansion coefficient $ \qqc_0 ( 0 ) $ remains strictly real. This unique mode fundamentally reflects the classical Fourier-like heat conduction behavior, which preserves the self-adjoint nature inherent to parabolic characteristics. At this lowest-order frequency, the characteristic diffusion time is significantly
larger
than the thermal relaxation time, rendering the wave-like effects negligible for this specific mode. In contrast, all higher-order modes emerge as complex conjugate pairs, indicating that the hyperbolic non-Fourier features and the non-self-adjoint nature of the underlying differential operator dominate at higher frequencies.
\begin{table}[!ht]
    \centering
    \renewcommand{\arraystretch}{1.2}
    \centering
    \begin{tabular}{c||c|c|c}
    $ n $ & $ \qnu_n $ & $ \qmu_n $ & $ \qqc_n ( 0 ) $ \\
    \hline
    $ 0\hphantom{0} $ & $ 0.3920 $ & $ - 0.1896 $ & $ 0.5023 \hphantom{\ \, \cdot 10^{-4}} $ \\
    $ 1,2\hphantom{0} $ & $ \hphantom{1}3.1739 \pm 0.1979 \ii $ & $ - 0.7004 \pm \hphantom{1}3.1344 \ii $ & $ - 0.0621 \pm 0.0142 \ii \hphantom{\ \, \cdot 10^{-4}} $ \\
    $ 3,4\hphantom{0} $ & $ \hphantom{1}6.2997 \pm 0.2015 \ii $ & $ - 0.7021 \pm \hphantom{1}6.2798 \ii $ & $ \hphantom{-}0.0321 \mp 0.0037 \ii \hphantom{\ \, \cdot 10^{-4}} $ \\
    $ 5,6\hphantom{0} $ & $ \hphantom{1}9.4358 \pm 0.2022 \ii $ & $ - 0.7025 \pm \hphantom{1}9.4225 \ii $ & $ - 0.0215 \pm 0.0017 \ii \hphantom{\ \, \cdot 10^{-4}} $ \\
    $ 7,8\hphantom{0} $ & $ 12.5746 \pm 0.2024 \ii $ & $ - 0.7026 \pm 12.5647 \ii $ & $ \hphantom{-}0.0162 \mp 9.3100 \cdot 10^{-4} \ii $ \\
    $ 9,10 $ & $ 15.7146 \pm 0.2025 \ii $ & $ - 0.7026 \pm 15.7066 \ii $ & $ - 0.0130 \pm 5.9724 \cdot 10^{-4} \ii $ \\
    $ 11,12 $ & $ 18.8551 \pm 0.2026 \ii $ & $ - 0.7027 \pm 18.8484 \ii $ & $ \hphantom{-}0.0108 \mp 4.1503 \cdot 10^{-4} \ii $ \\
    $ 13,14 $ & $ 21.9959 \pm 0.2026 \ii $ & $ - 0.7027 \pm 21.9902 \ii $ & $ - 0.0093 \pm 3.0505 \cdot 10^{-4} \ii $ \\
    $ 15,16 $ & $ 25.1369 \pm 0.2027 \ii $ & $ - 0.7027 \pm 25.1319 \ii $ & $ \hphantom{-}0.0081 \mp 2.3362 \cdot 10^{-4} \ii $ \\
    $ 17,18 $ & $ 28.2780 \pm 0.2027 \ii $ & $ - 0.7027 \pm 28.2736 \ii $ & $ - 0.0072 \pm 1.8462 \cdot 10^{-4} \ii $ \\
    $ 19,20 $ & $ 31.4192 \pm 0.2027 \ii $ & $ - 0.7027 \pm 31.4153 \ii $ & $ \hphantom{-}0.0065 \mp 1.4956 \cdot 10^{-4} \ii $ \\
    $ 21,22 $ & $ 34.5605 \pm 0.2027 \ii $ & $ - 0.7027 \pm 34.5569 \ii $ & $ - 0.0059 \pm 1.2362 \cdot 10^{-4} \ii $
    \end{tabular}
    \caption{The first 23 roots $ \qnu_n $ of the transcendental equation \re{eq:MCV-eig-val-eq}, the corresponding eigenvalues $ \qmu_n $ and expansion coefficients $ c_n ( 0 ) $ determined from \re{eq:MCV:eig-vals} and \re{eq:MCV-c_n(0)-hom}, respectively, for the parameters $ \qBi = 0.2 $ and $ \htau = 1 $. Note that the single real eigenvalue is indexed via $ 0 $, while the complex eigenvalue pairs are listed pairwise.}
    \label{tab:eig-val-test}
\end{table}

\begin{figure}[!ht]
	\includegraphics[width=0.45\textwidth]{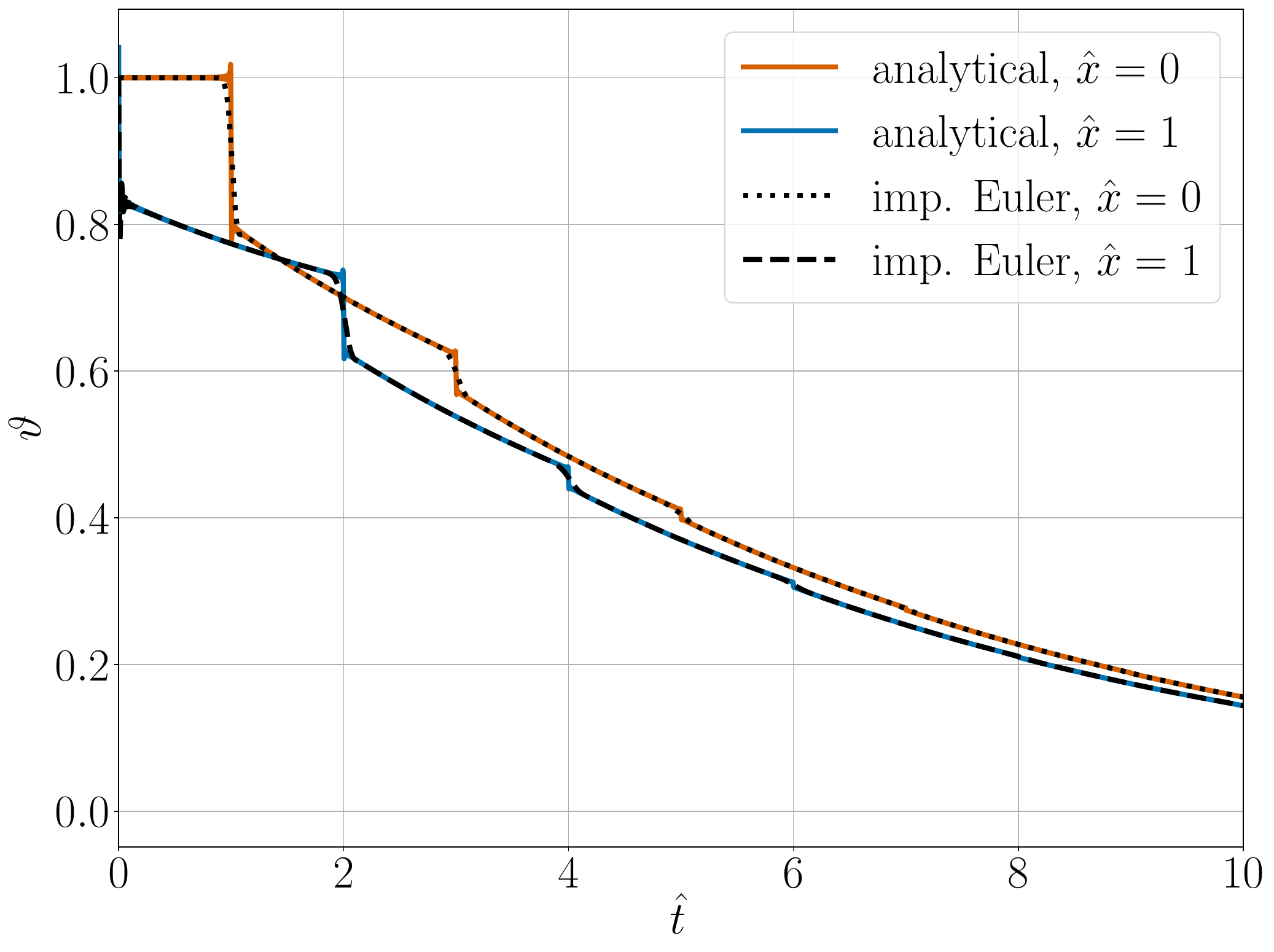} \centering
    \hfill
    \includegraphics[width=0.45\textwidth]{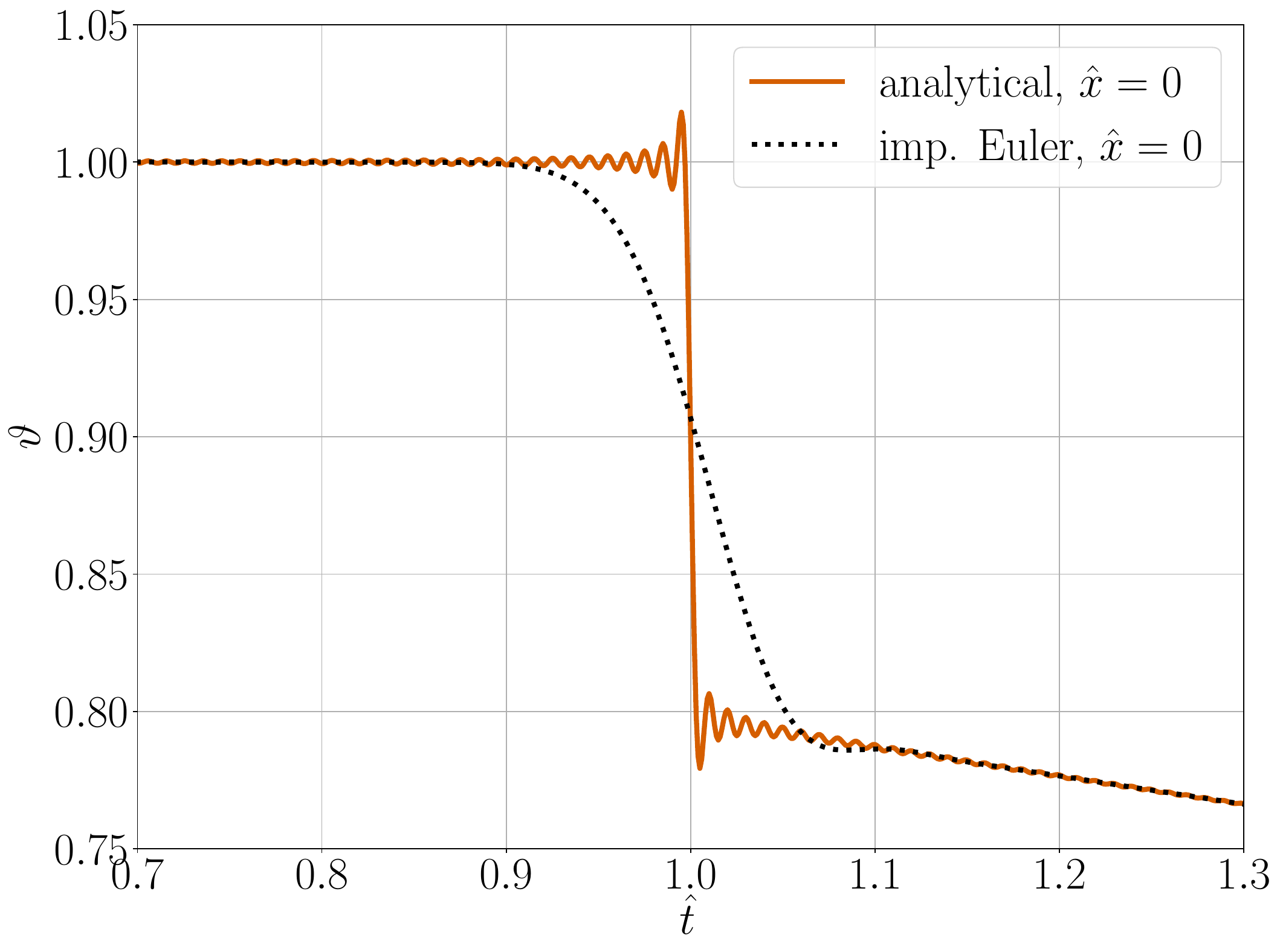} \centering 
    \caption{Dimensionless transient temperature response evaluated at the boundaries. \emph{Left:}
    long-time
    evolution showing multiple wave reflections. \emph{Right:} detailed view of the
    short-time
    behavior highlighting the Gibbs phenomenon in the analytical solution and the numerical dissipation inherent to the implicit Euler scheme.}
    \label{fig:T-t_Bi-02_tau-1}
\end{figure}

The transient temperature response evaluated at the boundaries ($ \hx = 0 $ and $ \hx = 1 $) is illustrated in the left panel of Figure~\ref{fig:T-t_Bi-02_tau-1}, comparing the derived analytical solution---computed using $ 401 $ eigenmodes (comprising the single real eigenvalue and $ 2 \times 200 $ complex conjugate pairs)---with the numerical results obtained via the implicit Euler scheme (for details, see Appendix~\ref{sec:num-meth}). For evaluating the analyitical solution, as well as for parametrizing the implicit Euler solver, the temporal and spatial steps of $ \Delta \hht = 0.001 $ and $ \Delta \hx = 0.01 $ are applied. A remarkable agreement is observed between the two independent solution methods, which reinforces the mathematical validity of the infinite series approach.

Physically, the temperature profiles clearly demonstrate the hyperbolic, wave-like nature of the heat propagation dictated by the MCV equation. At the insulated left boundary ($ \hx = 0 $), temperature remains strictly constant at its initial value ($ \qtheta = 1 $) until $ \hht = 1 $, which represents the exact
time required for the thermal wave front to traverse the domain length. Sharp steps and slope discontinuities appear periodically at $ \hht = 1 $, $ 2 $, $ 3 $, and $ 4 $, mapping the subsequent reflections of the underdamped thermal wave from the boundaries.

A detailed view of the short-time transient behavior during the first wave reflections is shown in the right panel of Figure~\ref{fig:T-t_Bi-02_tau-1}. This zoomed plot clearly highlights the distinct mathematical features of the two solution methods. On one hand, the analytical solution exhibits rapid, localized oscillations immediately preceding and following the sharp wave fronts (most prominently visible at $ \hht = 1 $ for the insulated boundary at $ \hx = 0 $). This behavior is a direct consequence of the Gibbs phenomenon, originating from the
finite
truncation of the infinite eigenmode series when reconstructing the steep gradients and mathematical discontinuities of the propagating thermal shock wave. On the other hand, the implicit Euler solution remains largely free of these non-physical artificial oscillations, yet it suffers from significant numerical dissipation. This dissipative effect subsequently manifests as a substantial artificial smoothing of the thermal wave fronts, spreading the sharp discontinuity over a wider temporal interval and underestimating the peak values immediately after the reflection. Consequently, while the numerical scheme provides a monotonic and smooth approximation in the long term, the analytical solution proves superior in preserving the strict physical location and steepness of the wave front, despite the inherent presence of the Gibbs phenomenon.

\begin{figure}[!ht]
	\includegraphics[width=0.45\textwidth]{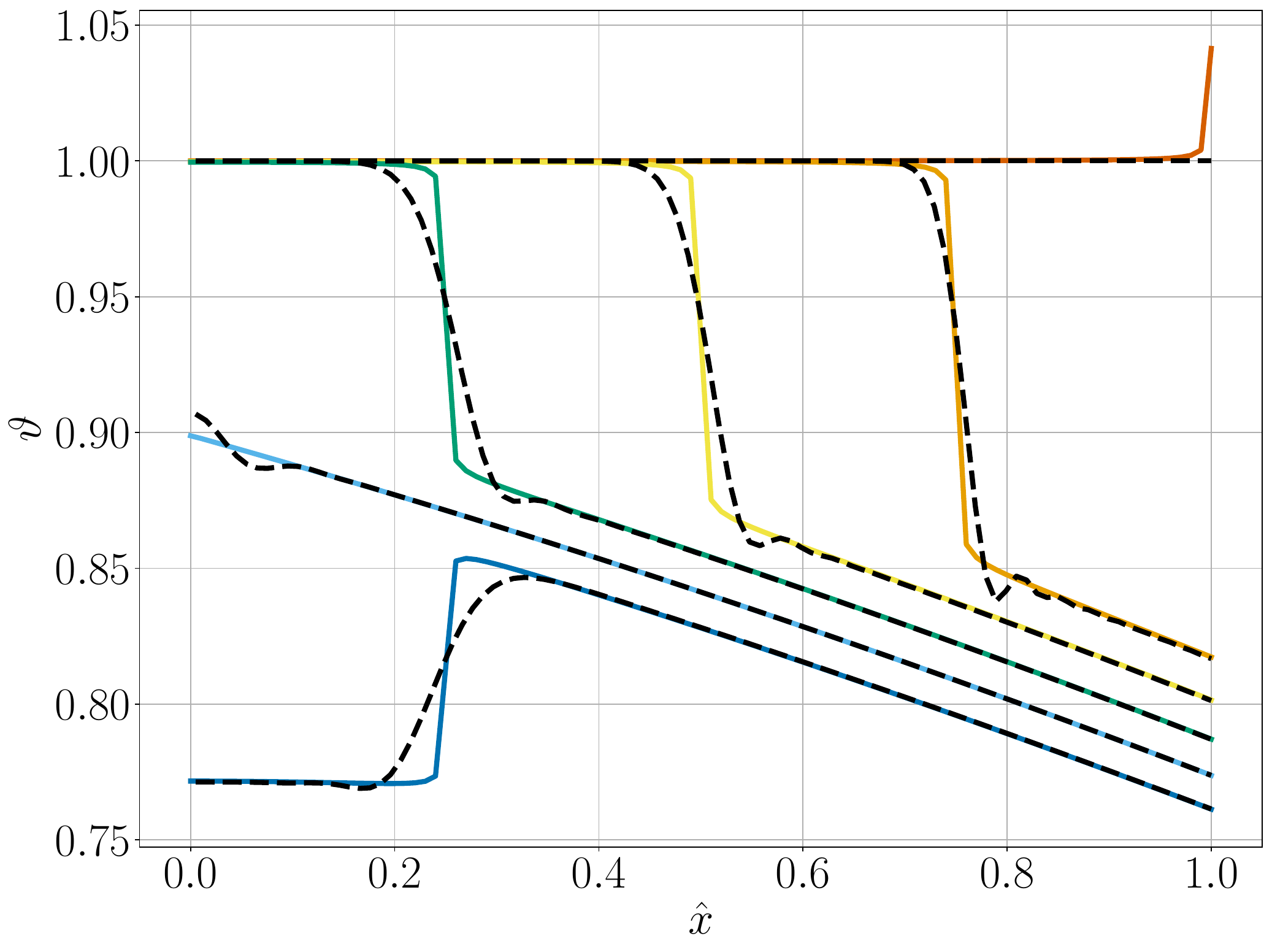} \centering
    \caption{Dimensionless spatial temperature distributions across the domain $ 0 \le \hx \le 1 $ the dimensionless time instances $ \hht = 0 $, $ 0.25 $, $ 0.5 $, $ 0.75 $, $ 1 $, and $ 1.25 $, comparing the derived analytical solution (solid colored lines, time propagates from the red to the blue line) with the implicit Euler numerical results (dashed black lines).}
    \label{fig:T-x_Bi-02_tau-1}
\end{figure}

The spatial temperature distributions $ \qtheta $ across the dimensionless domain $ 0 \le \hx \le 1 $ are presented in Figure~\ref{fig:T-x_Bi-02_tau-1} for the time instances $ \hht = 0 $, $ 0.25 $, $ 0.5 $, $ 0.75 $, $ 1 $, and $ 1.25 $. The snapshots clearly visualize the sharp, step-like thermal shock wave traveling through the medium. Since the convective cooling
\1 1 {which is in effect starting from \m { \hht = 0 }}
is applied at the right boundary, the wave front initiates from $ \hx = 1 $ and propagates to the left toward the insulated boundary at $ \hx = 0 $. This motion separates the hot, undisturbed zone (where $ \qtheta = 1 $) from the expanding cooled region behind the front. Once the wave reaches the insulated boundary at $ \hx = 0 $, it reflects and starts propagating backward to the right.

Comparing
the space dependence of the two solutions
further exposes the structural differences between the analytical and numerical approaches. The analytical solution successfully maintains the strict physical location and steepness of the moving thermal wave front. However, due to the relatively coarse spatial resolution compared to the temporal step, the localized Gibbs oscillations are not explicitly present in the analytical solution; instead, the steep mathematical discontinuity of the shock front appears slightly smoothed out. In parallel, the implicit Euler solution exhibits subtle artificial oscillations at the very beginning of the transient process, which is a direct manifestation of numerical dispersion. These spurious oscillations are quickly damped by the scheme's inherent numerical dissipation, which subsequently dominates the long-time behavior and leads to a significant broadening and diffusion of the thermal shock front over time.

\begin{figure}[!ht]
    \centering
    \includegraphics[width=0.7\textwidth]{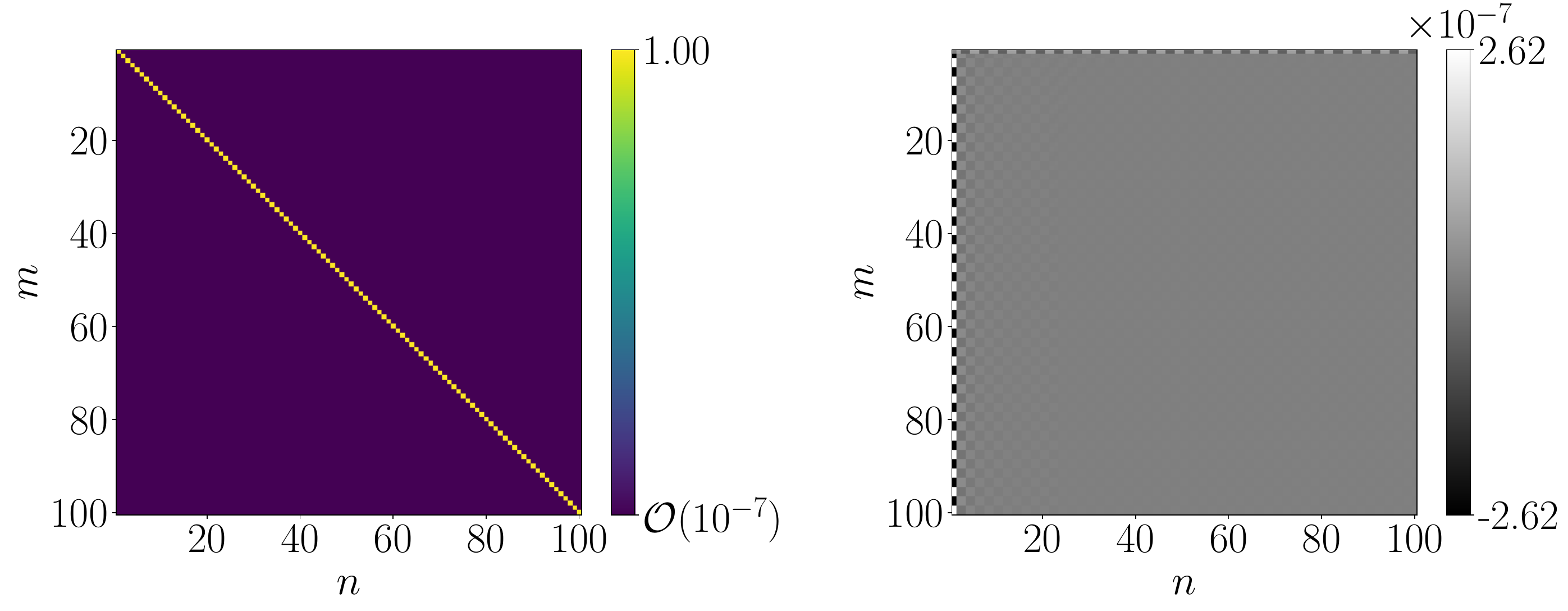}
    \caption{Graphical verification of the biorthogonality relation: the inner product matrix computed via numerical integration of the left and right eigenfunctions. \emph{Left:} the real part of the inner product matrix, yielding the identity matrix (representing proper normalization).
    \emph{Right:} the imaginary part of the inner product matrix, which is vanishing.}
    \label{fig:num-orthonorm}
\end{figure}

The normed biorthogonality of the right and left eigenfunctions is also
checked.
By numerically evaluating the scalar product defined in \re{fte}
via the trapezoidal rule, Figure~\ref{fig:num-orthonorm} graphically
demonstrates that the real part of the inner product matrix closely
approximates the identity matrix, while its imaginary part vanishes toward
zero.

\begin{figure}[!ht]
	\includegraphics[width=0.45\textwidth]{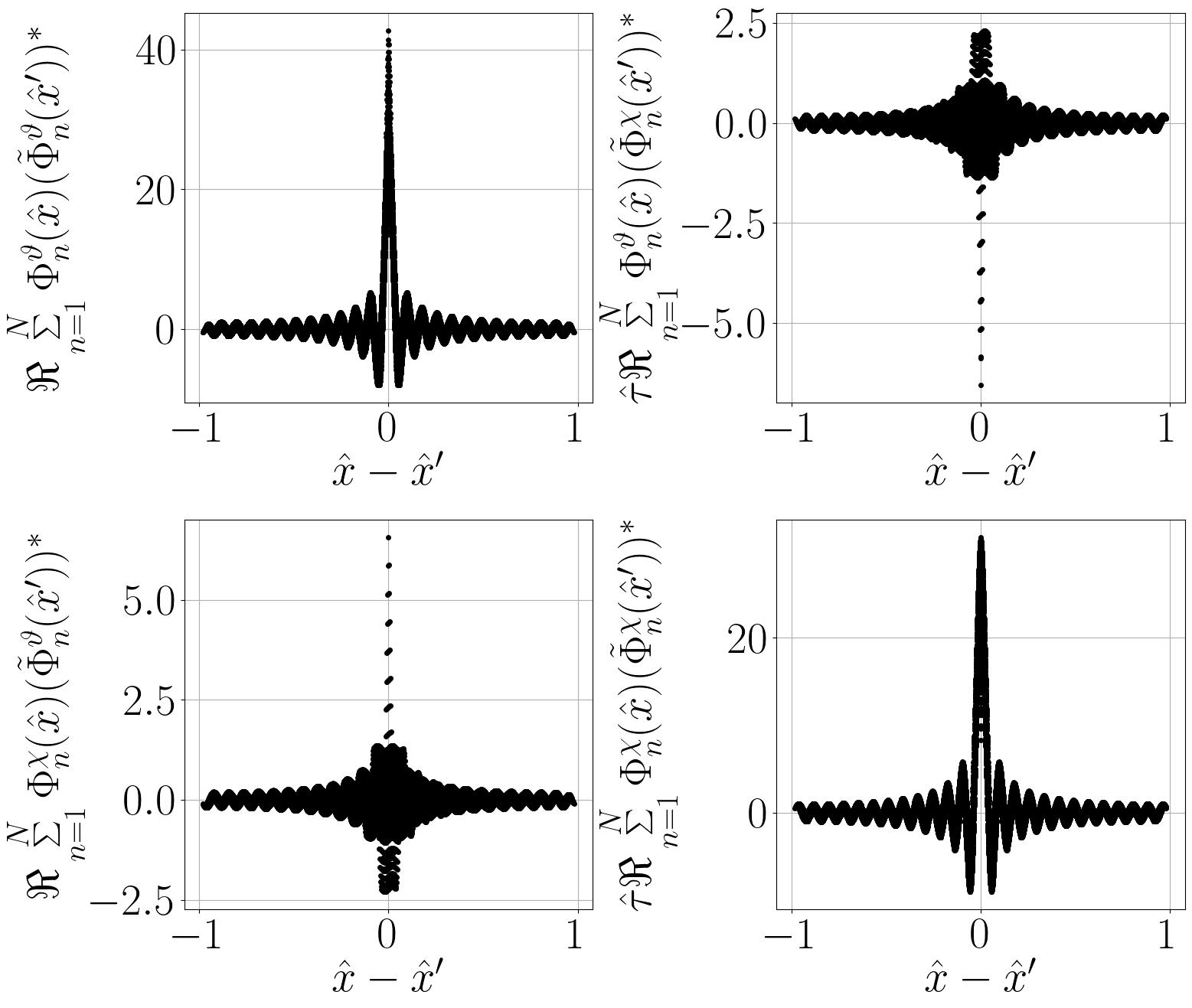} \centering
    \hfill
    \includegraphics[width=0.45\textwidth]{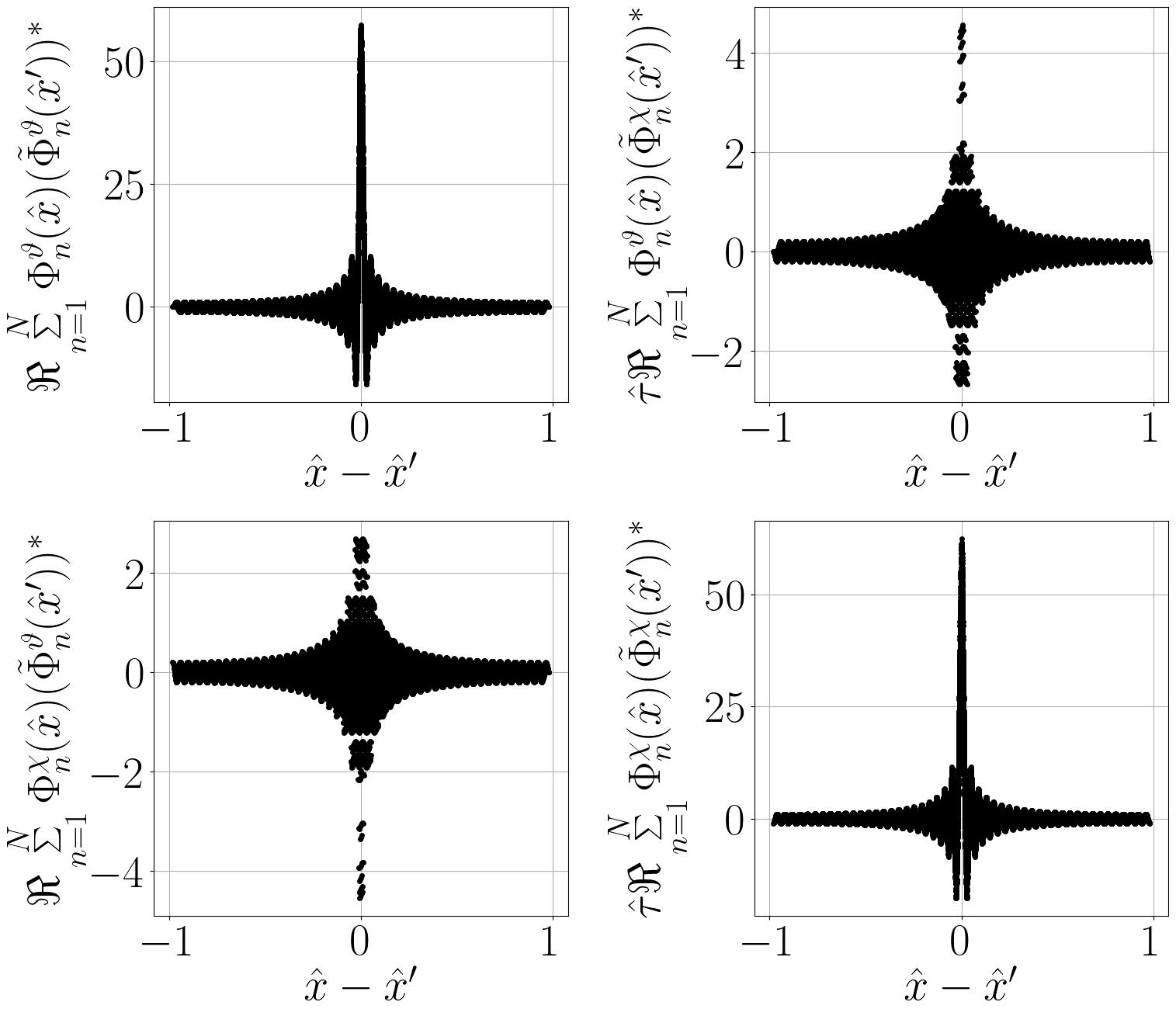} \centering 
    \\ \vspace{5ex}
    \includegraphics[width=0.45\textwidth]{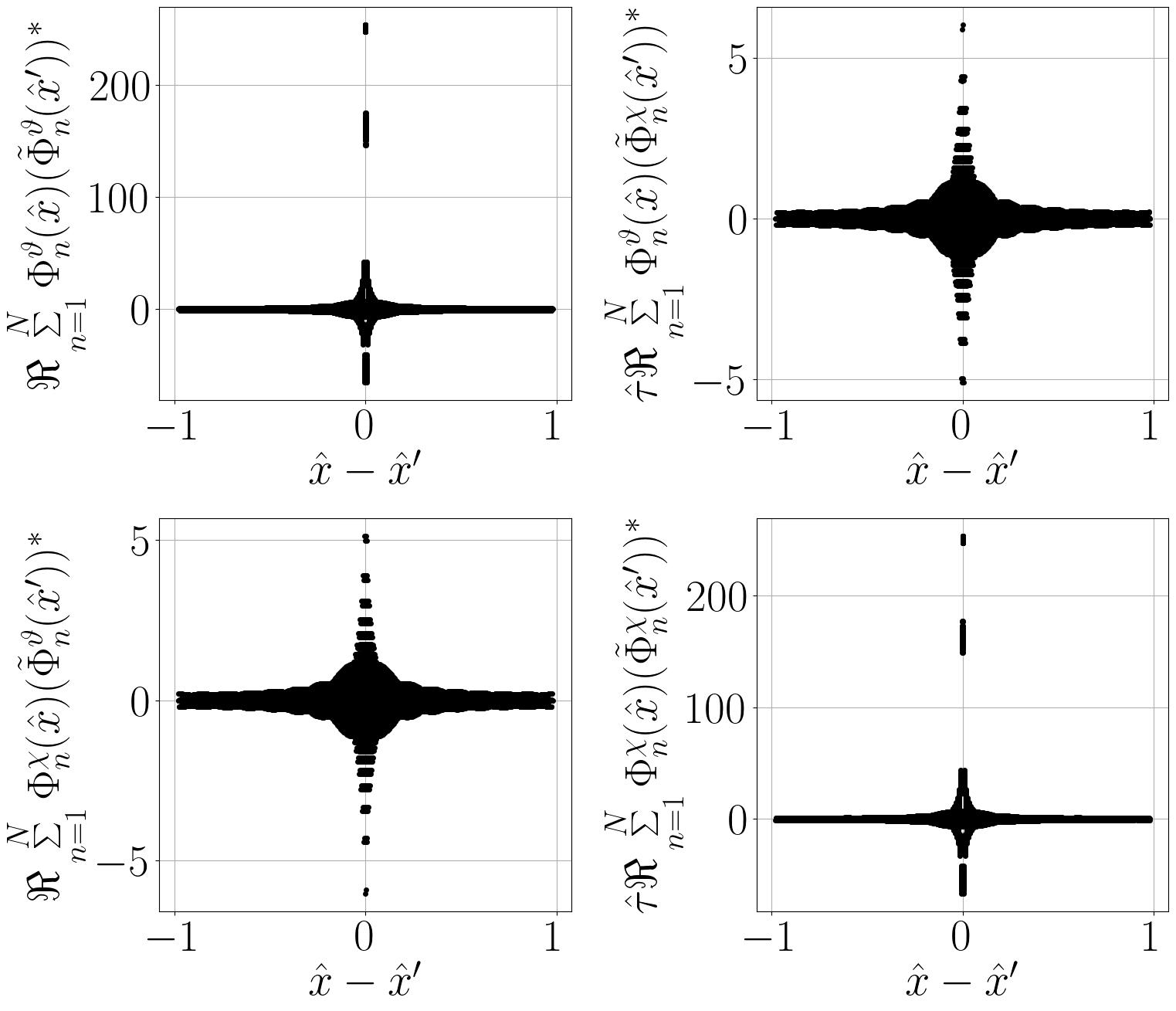} \centering
    \hfill
    \includegraphics[width=0.45\textwidth]{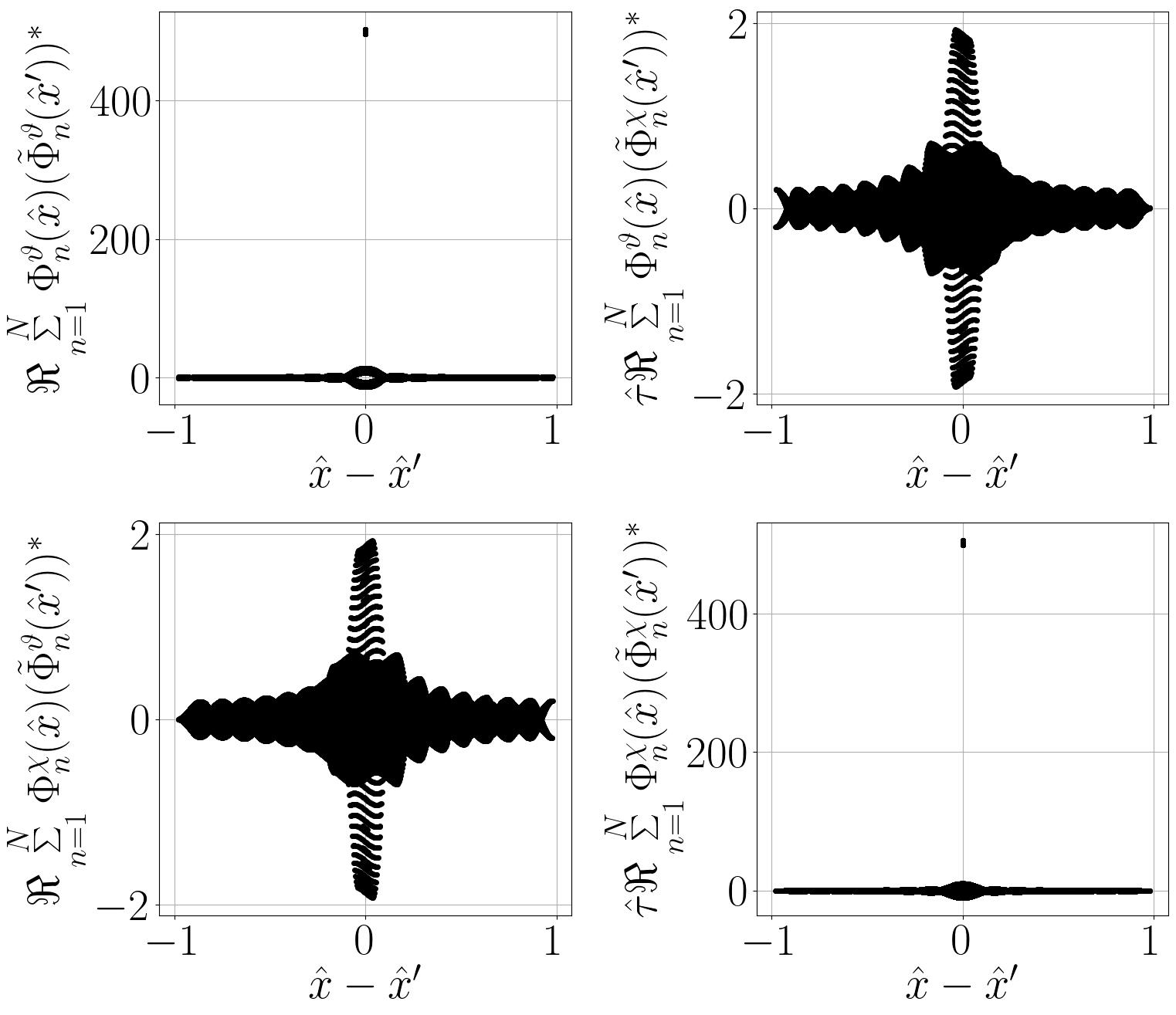} \centering
    \caption{Numerical
    check
    of the completeness property through the graphical visualization of the \m {2 \times 2} matrix \re{eq:dirac-mx}. The results are shown for an increasing number of eigenmodes: $ N = 51 $ (\emph{top left}), $ N = 101 $ (\emph{top right}), $ N = 501 $ (\emph{bottom left}), and $ N = 1001 $ (\emph{bottom right}).}
    \label{fig:num-dirac}
\end{figure}

Finally, the completeness of the eigenfunctions is investigated by examining the convergence of the eigenmode summation. Figure~\ref{fig:num-dirac} represents the four components of the matrix \re{eq:dirac-mx} graphically for an increasing number of eigenmodes. As the number of eigenmodes increases, the matrix elements $ (1,1) $ and $ (2,2) $ tend to a Dirac delta distribution, while the off-diagonal elements $ (1,2) $ and $ (2,1) $ fluctuate tightly around zero. Their outlying values observed at the right edge correspond to the fields evaluated near the
BC3
at $\hx = 1$.
It is understandable why convergence is the slowest near the boundary: completeness means to realize all \1 1 {square integrable} functions, even those not respecting the given boundary conditions. On the other side, we provide this realization with functions which all respect the prescribed boundary conditions. No wonder only the infinite sum performs this realization properly enough; any finite truncation causes strongest deviation where the eigenfunctions have the least freedom to compensate each other: at the boundaries.

\subsubsection{Parameter dependence of the solution}

To comprehensively map the physical and mathematical behavior of the investigated problem, a parametric study is conducted across the two-dimensional parameter space spanned by the relaxation time $ \htau $ and the Biot number $ \qBi $. Specifically, we analyze how $ \htau $ and $ \qBi $ affect the spectral distribution of both the spatial coefficients $ \qnu_n $ and the eigenvalues $ \qmu_n $. Furthermore, we investigate how these quantities distinguish between the hyperbolic wave-like and parabolic diffusive transport regimes. For all subsequent parametric computations, the analytical solution is evaluated by considering the real or purely imaginary
eigenvalue roots
along with $ 200 $ complex conjugate pairs, utilizing a temporal and spatial resolution of $ \Delta \hht = 0.001 $ and $ \Delta \hx = 0.01 $, respectively.

\begin{figure}[!ht]
	\includegraphics[height=0.35\textwidth]{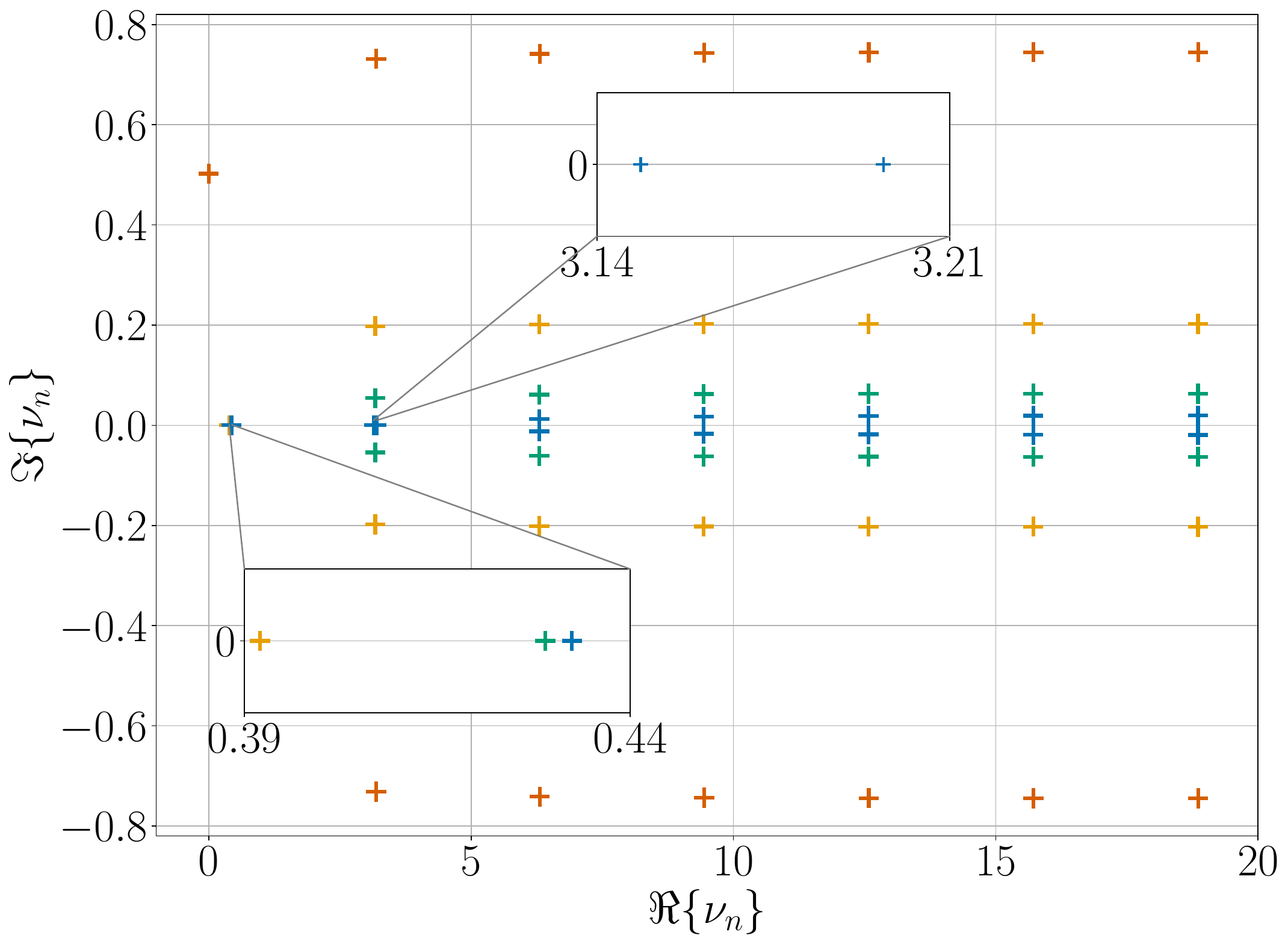} \centering
    \hfill
    \includegraphics[height=0.35\textwidth]{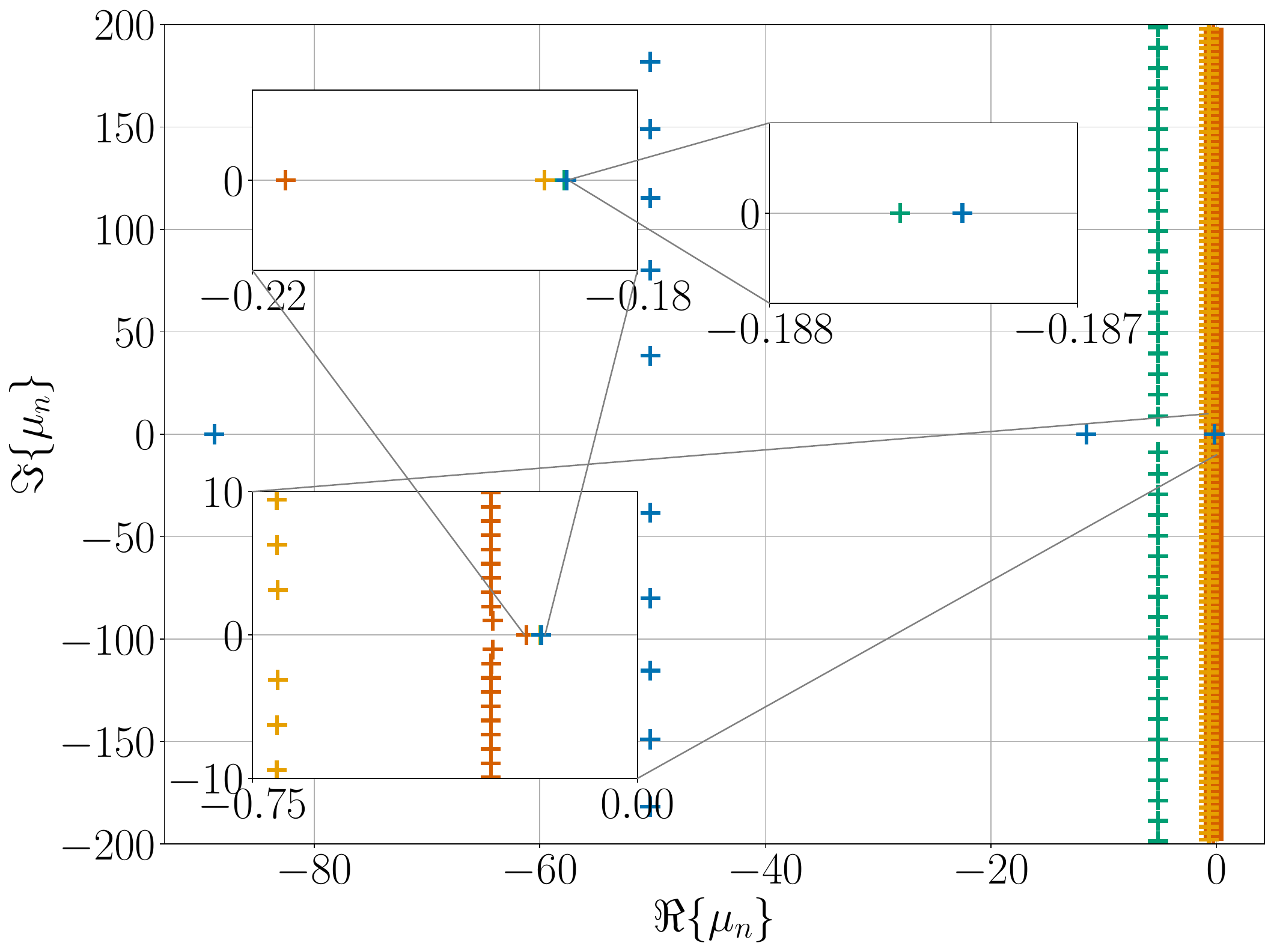} \centering
    \caption{Distribution of the spatial coefficients $ \qnu_n $ (\emph{left}) and eigenvalues $ \qmu_n $ (\emph{right}) on the complex plane under a fixed Biot number $ \qBi = 0.2 $ for the dimensionless relaxation times $ \htau = 0.01 $, $ 0.1 $, $ 1 $, and $ 10 $ (increase is visualized by the coloring from blue to red).
    }
    \label{fig:spect-tau-test}
\end{figure}

First, the dependence on the relaxation time is investigated under a fixed Biot number of $ \qBi = 0.2 $, varying the parameters as $ \htau = 0.01 $, $ 0.1 $, $ 1 $, and $ 10 $. The resulting distribution of the spatial coefficients $ \qnu_n $ and eigenvalues $ \qmu_n $ on the complex plane is illustrated in the  left panel of Figure~\ref{fig:spect-tau-test}. As $ \htau $ increases, the number of purely real spatial 
coefficients
monotonically decreases; specifically, for $ \htau = 0.01 $ there are three real roots, whereas for $ \htau = 0.1 $ only a single real root remains. Beyond a certain critical value of $ \htau $, no further real roots emerge. Interestingly, an increase in $ \htau $ does not significantly alter the real parts of $ \qnu_n $, except for the lowest-order modes. Instead, its primary effect is a pronounced expansion of their imaginary parts, shifting the
eigenvalue roots
vertically away from the real axis. This spectral behavior is directly reflected in the eigenvalues $ \qmu_n $ as well. As $ \htau $ increases, the absolute values of the real parts of the complex eigenvalues significantly decrease, shifting the characteristic vertical branch toward the imaginary axis, meaning that larger relaxation times result in substantially weaker physical damping, allowing the thermal waves to persist longer in the time domain. Concurrently, the imaginary parts of the eigenvalues become much denser along these vertical asymptotes, indicating a higher density of available oscillatory frequencies.

\begin{figure}[!ht]
	\includegraphics[width=0.45\textwidth]{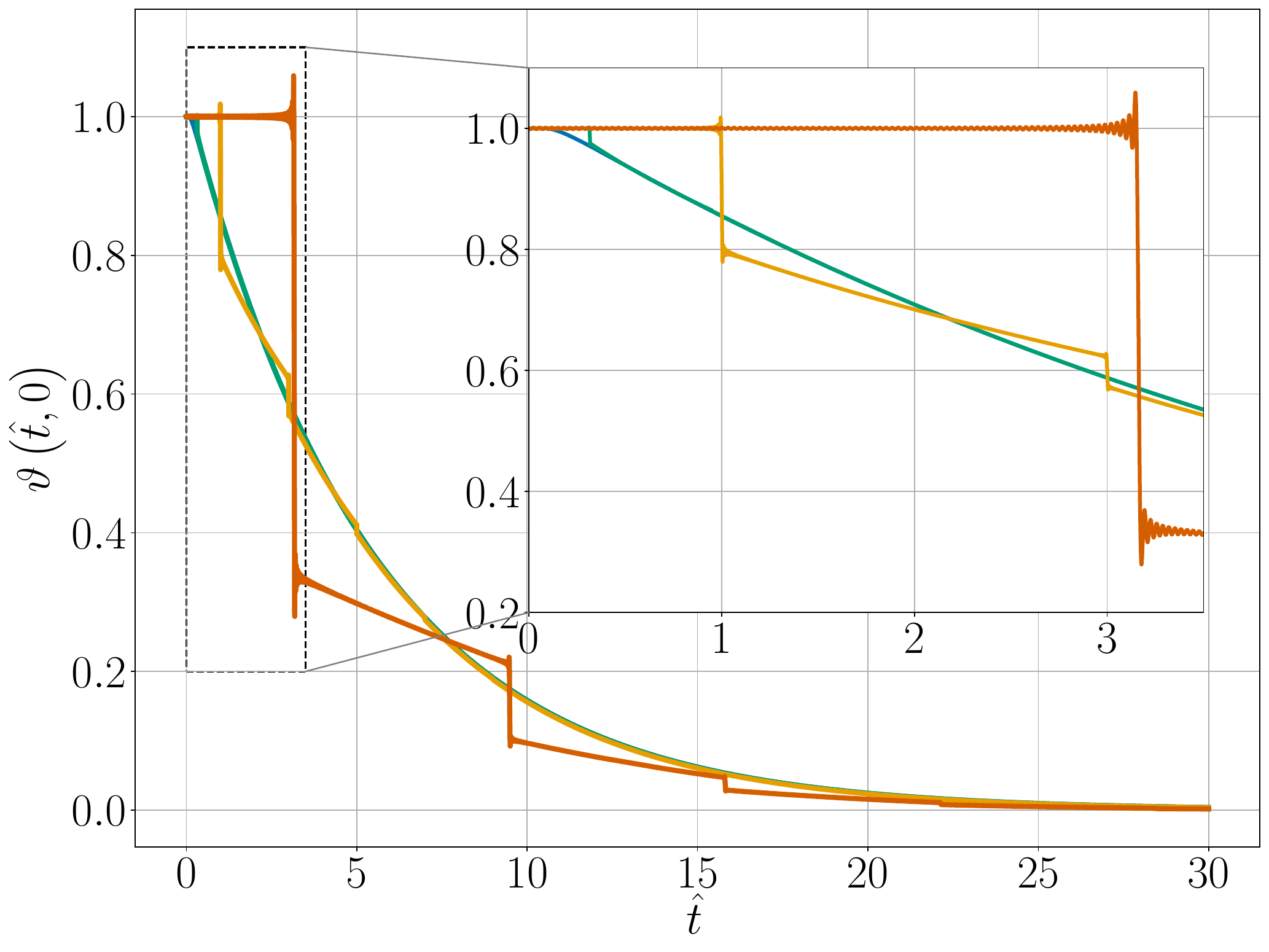} \centering
   \caption{Dimensionless transient temperature response $ \qtheta $ evaluated at the insulated boundary $ \hx = 0 $ under a fixed Biot number $ \qBi = 0.2 $ for the dimensionless relaxation times $ \htau = 0.01 $, $ 0.1 $, $ 1 $, and $ 10 $ (increase is visualized by the coloring from blue to red). The profiles demonstrate the continuous transition from highly underdamped, step-like hyperbolic wave propagation ($ \htau = 10 $) to smooth diffusive parabolic decay ($ \htau \le 0.1 $). The inset provides a close-up of the early transient, showcasing
   the Gibbs phenomenon.
   }
    \label{fig:T_t_Bi_0-2_tau_test}
\end{figure}

The physical consequence of this parametric transition is presented in Figure~\ref{fig:T_t_Bi_0-2_tau_test}, which illustrates the transient temperature response evaluated at the insulated boundary ($ \hx = 0 $) using the exact same color-coding as the spectral plots. For the highly hyperbolic regime ($ \htau = 10 $), the temperature profile exhibits sharp, periodic step-like discontinuities and flat plateaus, mapping the subsequent arrivals of the weakly damped thermal shock wave. Because the infinite series is truncated, these sharp physical gradients are inherently accompanied by pronounced Gibbs oscillations near the wave fronts. As $ \htau $ decreases ($ \htau = 1 $), the enhanced internal dissipation smooths out the wave front, reducing the amplitude of the step changes. Under strongly damped conditions ($ \htau = 0.1 $ and $ \htau = 0.01 $), the wave-like characteristics are entirely suppressed, and the temperature decay transitions into a smooth, monotonic, Fourier-like exponential curve (analyzed in detail in Subsection~\ref{sec:MCV-vs-F}). While all curves are evaluated using the same number of complex eigenmodes and fixed temporal and spatial resolutions, the Gibbs phenomenon naturally vanishes at lower relaxation times because the physical transport process itself becomes purely diffusive. Under these conditions, the system exhibits seemingly no dispersion. Consequently, the thermal wave front arrives at the boundary almost exactly at the theoretical dispersion-free arrival time given by $\hht_{\mathrm{arr}} = \sqrt{\htau} $ (resulting from the ideal dimensionless wave speed $ \dimless{\qv} = \frac{1}{\sqrt{\htau}} $). As captured in the inset panel, this is remarkably confirmed by the onset of the transient at $ \hht \approx 0.3 $ for $ \htau = 0.1 $, and the sharp steps exactly at $ \hht = 1 $ for $ \htau = 1 $ and at $ \hht \approx 3.1 $ for $ \htau = 10 $.

\begin{figure}[!ht]
	\includegraphics[height=0.35\textwidth]{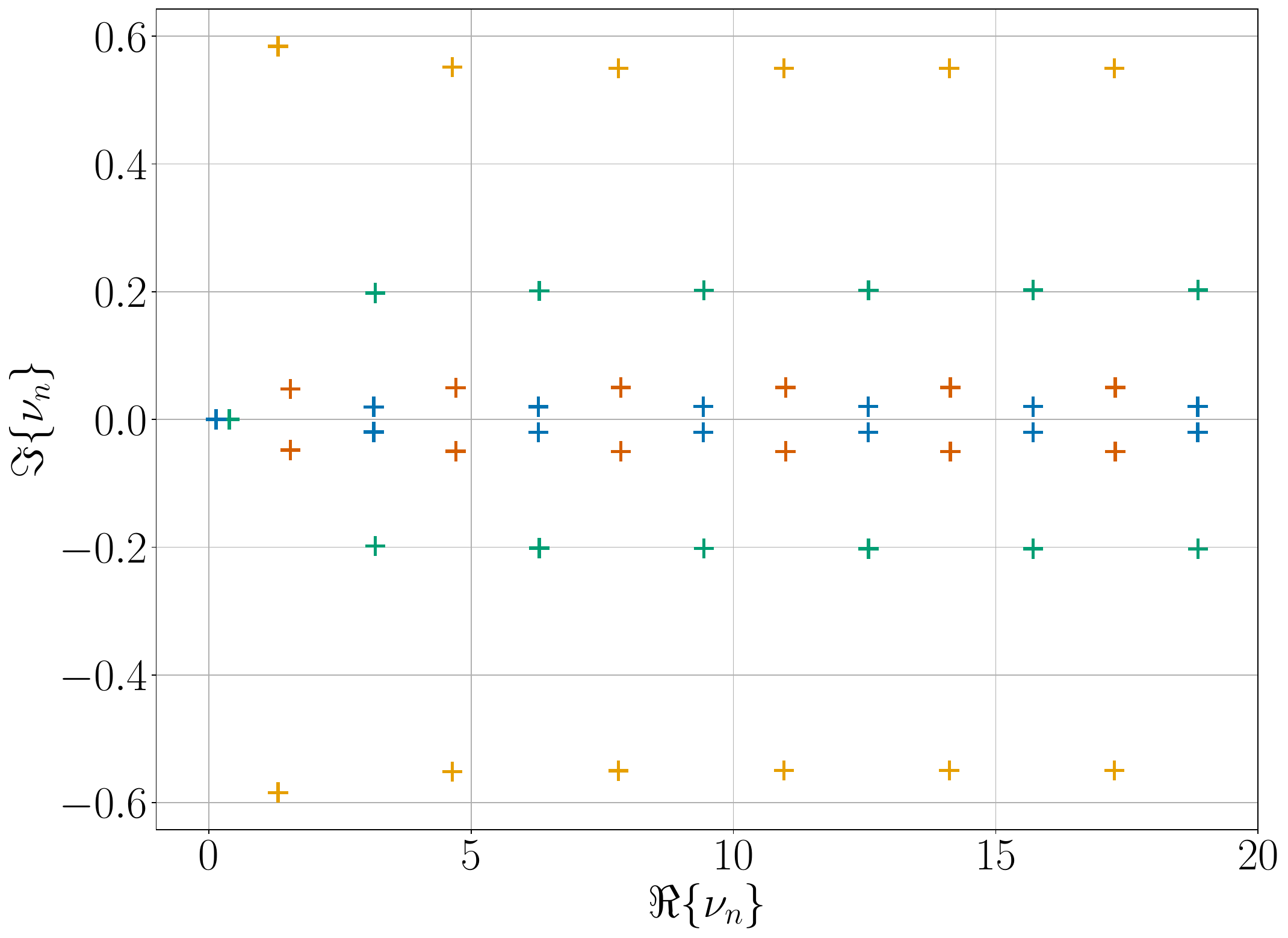} \centering
    \hfill
    \includegraphics[height=0.35\textwidth]{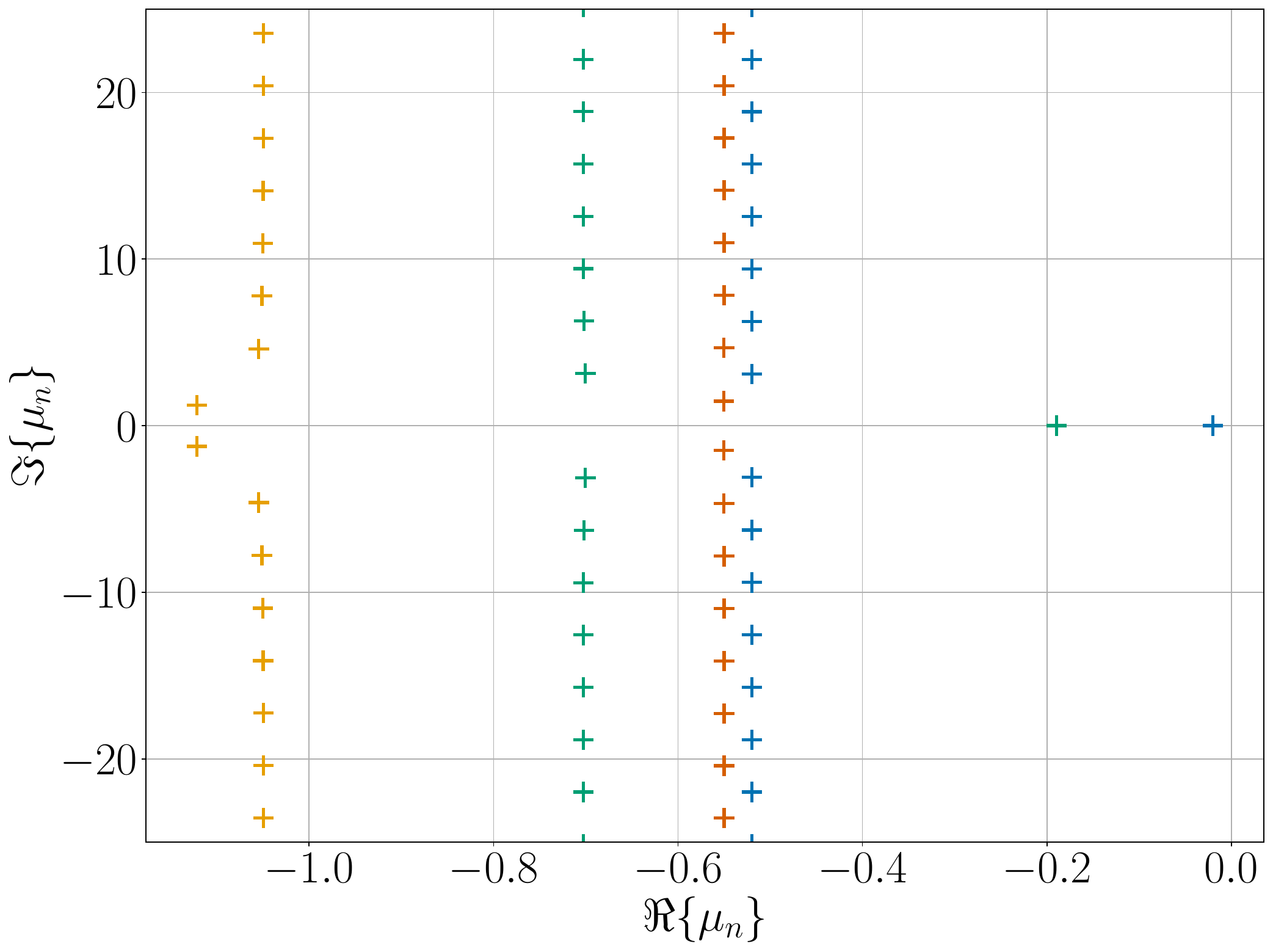} \centering
    \caption{Distribution of the spatial coefficients $ \qnu_n $ (\emph{left}) and eigenvalues $ \qmu_n $ (\emph{right}) on the complex plane under a fixed dimensionless relaxation time $ \htau = 1 $ for the Biot number $ \qBi = 0.02 $, $ 0.2 $, $ 2 $, and $ 20 $ (increase is visualized by the coloring from blue to red).}
    \label{fig:spect-Bi-test}
\end{figure}

Next, the system's dependence on the Biot number is investigated under a fixed dimensionless relaxation time of $ \htau = 1 $.
For analyzing
the effect of boundary cooling intensity, the Biot number is chosen to be $ \qBi = 0.02 $, $ 0.2 $, $ 2 $, and $ 20 $. The corresponding spectral evolution is graphically illustrated
in
the complex planes in Figure~\ref{fig:spect-Bi-test}, where the left panel displays the spatial coefficients $ \qnu_n $ and the right panel presents the eigenvalues $ \qmu_n $. 

At a low Biot number ($ \qBi = 0.02 $), the spatial coefficients lie tightly clustered near the real axis ($ \IM\{\qnu_n\} \approx 0 $). This behavior indicates that, under weak convective cooling, the system asymptotically approaches a symmetric
BC2--BC2 adiabatic--adiabatic
configuration, where the underlying differential operator is self-adjoint. As $ \qBi $ increases to $ \qBi = 0.2 $ and $ \qBi = 2 $, the eigenmodes
get vertically repelled
from the real axis. This growth of the imaginary parts reflects the intensifying non-selfadjoint character of the boundary coupling. However, upon further increasing the cooling intensity to $ \qBi = 20 $, a counterintuitive phenomenon emerges: the imaginary parts of the spatial arguments begin to contract, bending back toward the real axis. This spectral re-entry occurs because the extreme cooling limit ($ \qBi \to \infty $) formally recovers another self-adjoint constraint, namely a
BC1
condition
\1 1 {prescribed temperature}
at the right boundary. Consequently, the non-selfadjointness of the system does not monotonically scale with $ \qBi$ ; rather, it reaches a maximum in the intermediate convective regime and diminishes as either of the two self-adjoint operational limits is approached. This structural transition systematically governs the horizontal shifts along the real axis as well. In the low-Biot limit, the real parts of the roots are positioned near the integer multiples of $ \pi $, corresponding to the classical adiabatic eigensystem. As $ \qBi $ increases, the spectrum continuously shifts leftward, transitioning toward $ \left( n - \nicefrac{1}{2} \right) \pi $ with $ n \in \mathbb{Z} $. Notably, this parametric migration also reorganizes the low-index spectrum. While weak cooling preserves a purely real root at the lowest-indexed mode, increasing the Biot number drives this diffuse remnant toward the origin. At a critical Biot number value a qualitative mode transition occurs, the real part vanishes completely.

This non-monotonic transition is mirrored by the eigenvalues, too. Note that the choice of the Biot number does not modify the underlying oscillation frequencies ($ \IM \{ \qmu_n \} $). The characteristic shift 
observable on
the spatial coefficients $ \qnu_n $
is
clearly reflected in the discrete vertical positioning of the modes. For lower Biot numbers, the parabolic diffusion background is still represented by isolated real roots near the origin.

\begin{figure}[!ht]
	\includegraphics[width=0.45\textwidth]{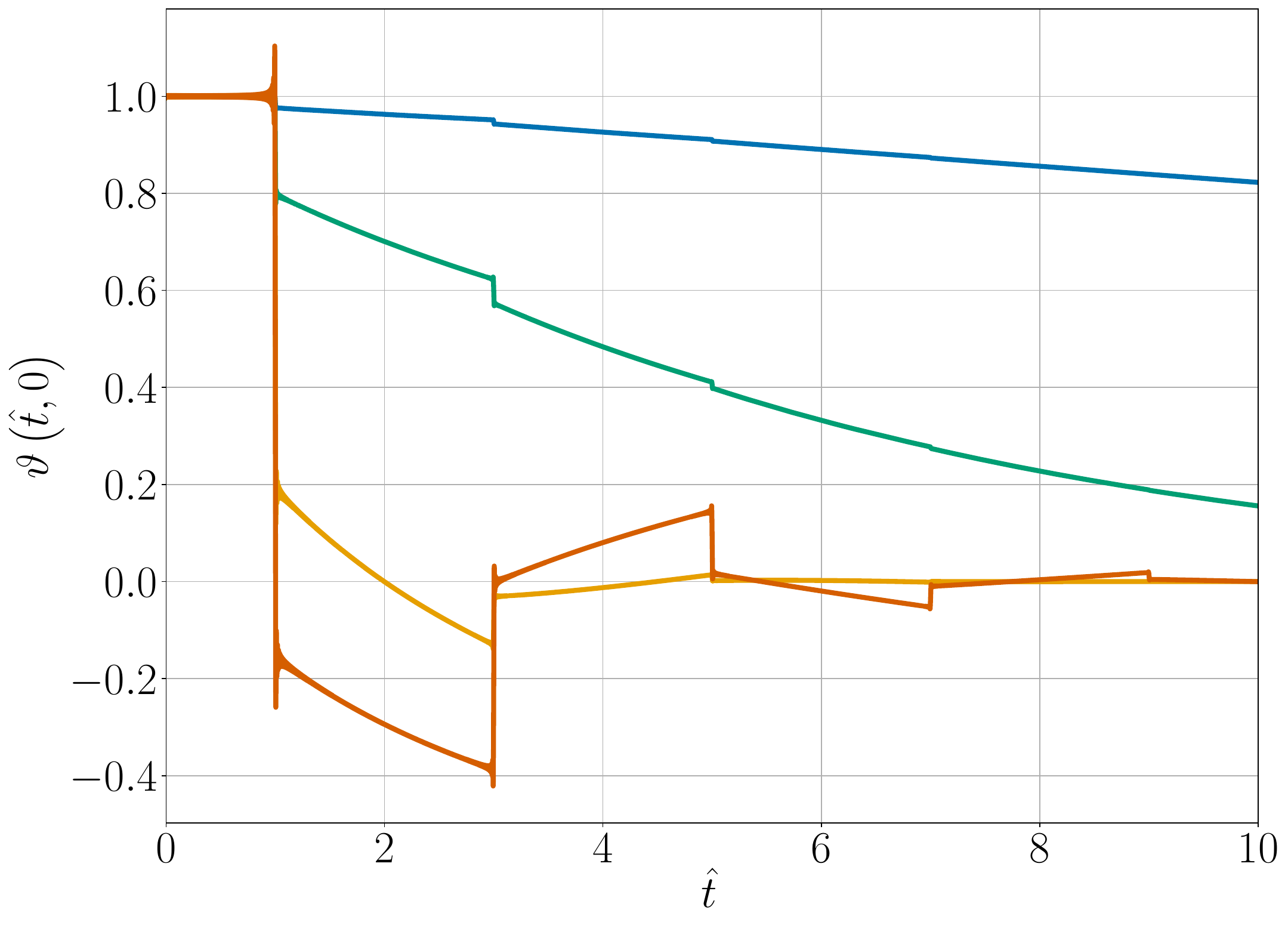} \centering
   \caption{Dimensionless transient temperature response $ \qtheta $ evaluated at the insulated boundary $ \hx = 0 $ under a fixed dimensionless relaxation time $ \htau = 1 $ for the Biot numbers $ \qBi = 0.02 $, $ 0.2 $, $ 2 $, and $ 20 $ (increase is visualized by the coloring from blue to red).}
    \label{fig:T_t_tau_1_Bi_test}
\end{figure}

The macroscopic temperature response at the insulated boundary ($\hx = 0$) is depicted in Figure~\ref{fig:T_t_tau_1_Bi_test}, perfectly mirroring the previously discussed spectral characteristics. All curves remain strictly flat until $ \hht=1 $, confirming a dispersion-free first wave arrival that is independent of the cooling intensity. Note that the negative dimensionless temperature represents transient cooling below the ambient temperature rather than an unphysical absolute temperature, although the actual physical realizability of such a ballistic sub-ambient state remains an open question.

A deeper analysis of the underlying transcendental equation reveals that the qualitative nature of the eigensystem is governed by a threshold. As presented in Appendix~\ref{sec:eigenvalues}, the boundary separating the two fundamentally different spectral regimes is defined by the dimensionless condition $ \htau \qBi^2 = 1 $ \1 2 {see
Eq.~\ree{eq:tr-eq-ome-param}{b}
}. This criterion establishes a direct link between the internal relaxation time and the external convective cooling intensity, dictating whether the lower-order transport modes remain diffusive or transform into purely wave-like configurations.

\begin{figure}[!ht]
	\includegraphics[height=0.35\textwidth]{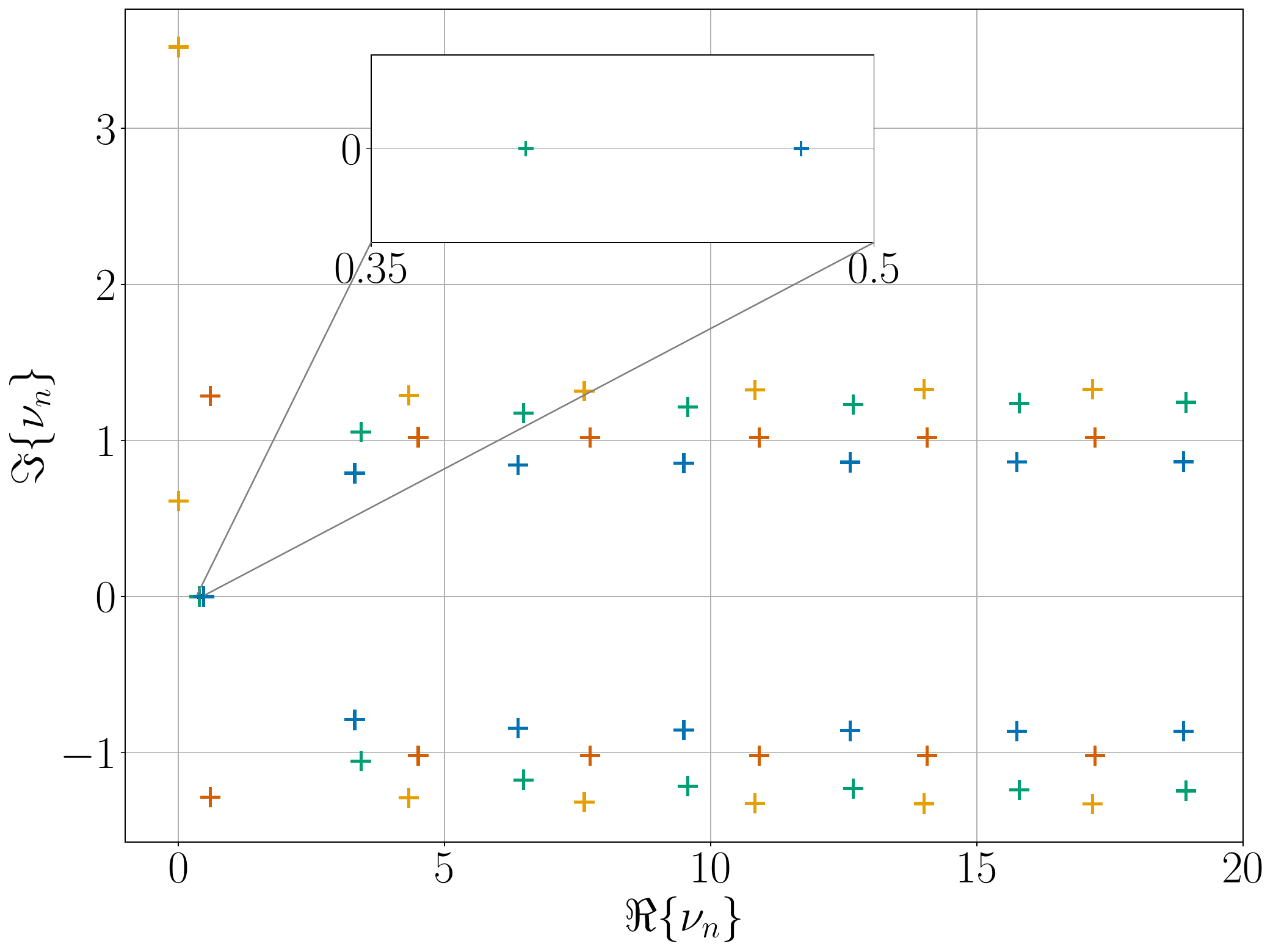} \centering
    \hfill
    \includegraphics[height=0.35\textwidth]{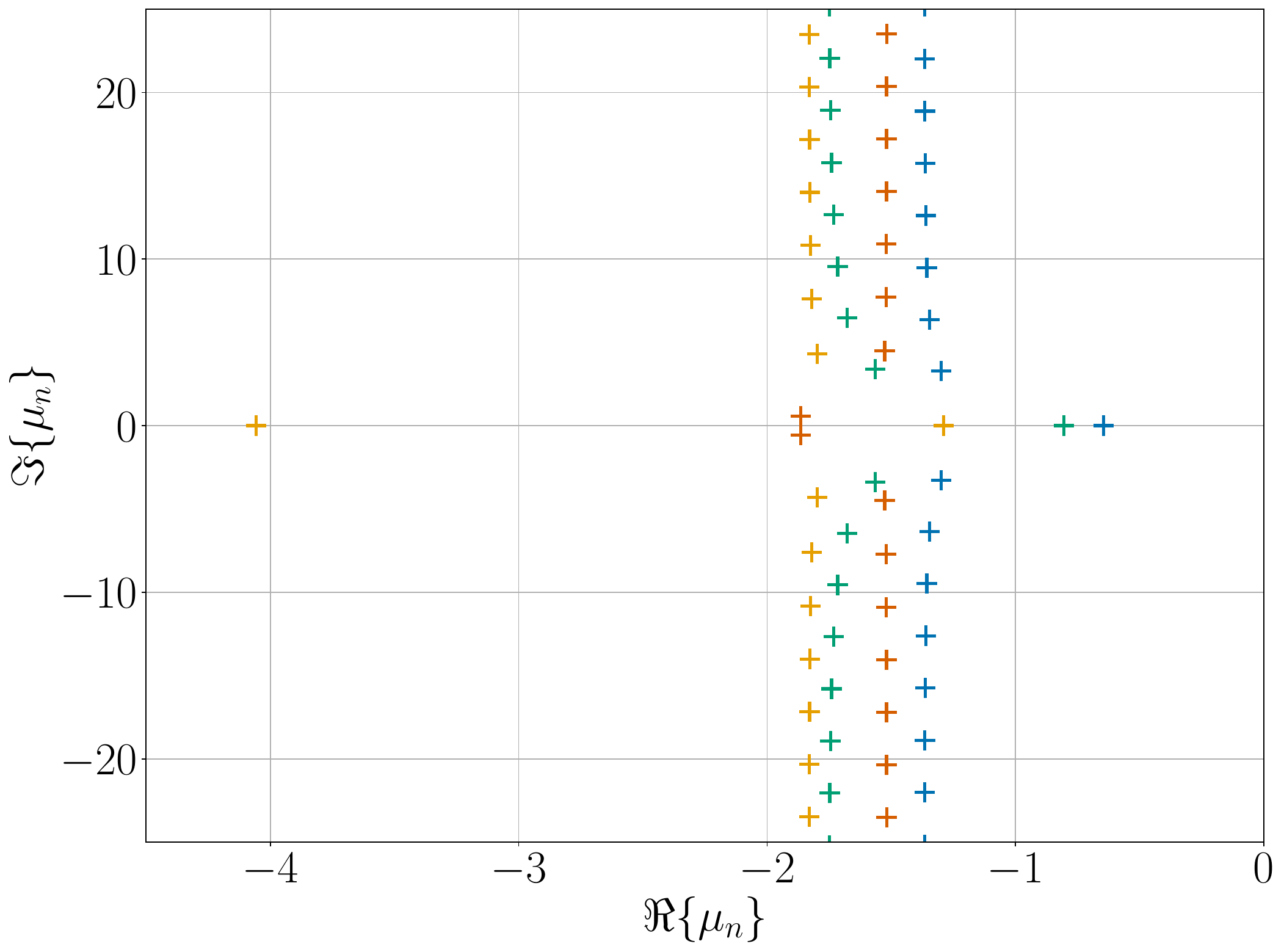} \centering
    \caption{Distribution of the spatial coefficients $ \qnu_n $ (\emph{left}) and eigenvalues $ \qmu_n $ (\emph{right}) on the complex plane under a fixed dimensionless relaxation time $ \htau = 1 $ for the Biot number $ \qBi = 0.7 $, $ 0.85 $, $ 1.15 $, and $ 1.3 $ (increase is visualized by the coloring from blue to red).}
    \label{fig:spect-Bi-zoomed_test}
\end{figure}

To graphically illustrate this transition, the spectral distribution of the spatial coefficients $ \qnu_n $ and the eigenvalues $ \qmu_n $, and the corresponding temporal evolution of the dimensionless temperature field is investigated under a fixed relaxation time of $ \htau = 1 $ with varying the Biot number near the threshold as $ \qBi = 0.7 $, $ 0.85 $, $ 1.15 $, and $ 1.3 $. The resulting spectra are shown in Figure~\ref{fig:spect-Bi-zoomed_test}. Since $ \htau = 1 $, the transition occurs at $ \qBi = 1 $. For sub-critical cooling intensities ($ \qBi < 1 $), the lowest-order mode manifests as a single real root. As the Biot number increases toward the threshold, this real root migrates toward the origin and upon entering the super-critical regime ($ \qBi > 1 $), the mode transitions onto the purely imaginary axis, splitting into either one or two purely imaginary roots. Finally, beyond a certain higher Biot number, these imaginary roots merge and break away from the imaginary axis, leaving a spectrum composed exclusively of complex conjugate pairs with non-zero real parts.

Correspondingly, this critical mode reorganization is distinctly reflected in the eigenvalues $\qmu_n$. When the system operates in the sub-critical cooling regime, the  real spatial coefficients yield purely real
eigenvalues. Crucially, these real modes are positioned to the right of the vertical complex asymptote, meaning they exhibit substantially weaker physical damping than the rest of the spectrum. When the cooling intensity is sufficiently high and the system operates deep within the super-critical regime (where all the spatial coefficients are general complex conjugate pairs), the eigenvalues corresponding to these lowest-order modes migrate significantly to the left of the main vertical asymptote in the spectrum. Consequently, these fundamental components undergo an acceleration in their energy decay rates, shifting from the least damped modes of the system to the most intensely damped ones.

\begin{figure}[!ht]
	\includegraphics[width=0.45\textwidth]{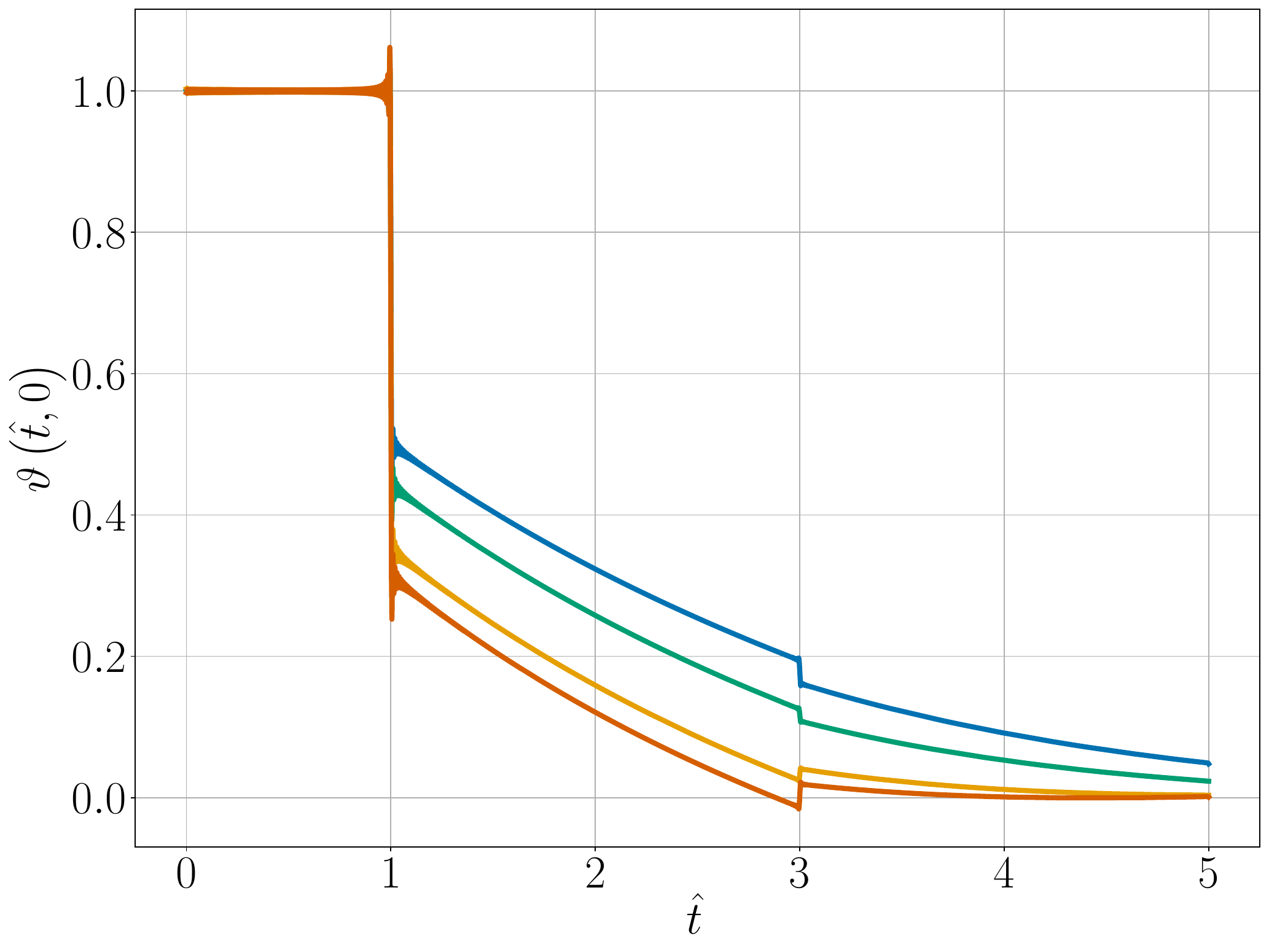} \centering
   \caption{Dimensionless transient temperature response $ \qtheta $ evaluated at the insulated boundary $ \hx = 0 $ under a fixed dimensionless relaxation time $ \htau = 1 $ for the Biot number $ \qBi = 0.7 $, $ 0.85 $, $ 1.15 $, and $ 1.3 $ (increase is visualized by the coloring from blue to red).}
    \label{fig:T_t_tau_1_Bi-zoomed_test}
\end{figure}

The physical manifestation of this transition at the insulated boundary ($ \hx = 0 $) is shown in Figure~\ref{fig:T_t_tau_1_Bi-zoomed_test}. All curves overlap during the initial transient until the sharp shock front arrives at the left boundary. Note that a qualitative inversion is observed at the second wave reflection (in this case, at $ \hht = 3 $): while sub-critical cooling ($ \qBi < 1 $) causes a further downward jump in temperature, super-critical intensities ($ \qBi > 1 $) trigger an opposite temperature change, manifesting as a distinct upward slope change upon reflection. This dynamic inversion is a direct physical consequence of the phase transition in the eigenfunctions, driven by the root migration between the $ \pi $-spaced and $ \pi/2 $-spaced spectral limits.

\subsubsection{Small-relaxation-time limit: near-Fourier behavior} \label{sec:MCV-vs-F}

When the relaxation time tends to zero, \ie $ \htau \to 0 $, the MCV constitutive equation \re{eq:q-MCV} gets reduced to Fourier's heat conduction law.
\1 1 {Notably, a \emp{hyperbolic} equation transitions thus into a \emp{parabolic} one.} Correspondingly, the transcendental equation \re{eq:MCV-eig-val-eq}
gets simplified
to
\begin{align}
    \label{eq:eig-eq-F}
    &&
    - \qnu_n \sin \qnu_n + \qBi \cos \qnu_n &= 0 & \Longleftrightarrow &&
    \frac{\qBi}{\tan \qnu_n} &= \qnu_n ,
    &&
\end{align}
which is actually the equation obtainable directly from the Fourier problem \1 1 {see, \eg \cite{carslaw1959conduction}, p121, Eq.~(6)}.
In parallel, \re{eq:nu-def}
gets simplified
to
\begin{align}
    \label{eq:eig-val-F-1}
    \qmu_n = - \qnu_n^2 \stackrel{\re{eq:eig-eq-F}}{=} - \qBi \frac{\qnu_n}{\tan \qnu_n} .
\end{align}
Consequently, \re{eq:MCV:eig-vals} remains valid. Crucially, in this asymptotic limit, the underlying differential operator of the investigated boundary value problem becomes self-adjoint. Consequently, the normalization factor \re{eq:MCV:norm-fac}
turns into
\begin{align}
    \qNsc_n & = - \frac{2 \qBi^2}{\tan^2 \qnu_n \left( 1 + \frac{\sin \left( 2 \qnu_n \right)}{2 \qnu_n} \right)} = - \frac{2 \qBi^2}{\tan^2 \qnu_n \left( 1 + \frac{\sin \qnu_n \cos \qnu_n}{\qnu_n} \right)} ,
\end{align}
which yields the expansion coefficients
\begin{align}
    \label{eq:c_n(0)-F}
    \qqc_n ( 0 ) = \qBi \frac{2 \cos \qnu_n}{\qnu_n + \sin \qnu_n \cos \qnu_n} \stackrel{\re{eq:eig-eq-F}}{=} \qnu_n \frac{2 \sin \qnu_n}{\qnu_n + \sin \qnu_n \cos \qnu_n} .
\end{align}
Finally, the dimensionless temperature field in the limit $ \htau \to 0 $ is recovered from \re{eq:MCV-dimless-sol} as
\begin{align}
    \qtheta \left( \hht , \hx \right) = \sum_{n = 1}^\infty \qqc_n ( 0 ) \ee^{\qmu_n \hht} \frac{\tan \qnu_n}{\qBi} \cos \left( \qnu_n \hx \right) \stackrel{\re{eq:eig-eq-F},\re{eq:eig-val-F-1},\re{eq:c_n(0)-F}}{=} \sum_{n = 1}^\infty \frac{2 \sin \qnu_n}{\qnu_n + \sin \qnu_n \cos \qnu_n} \ee^{- \qnu_n^2 \hht} \cos \left( \qnu_n \hx \right) .
\end{align}
This result is also identical to the
classic
analytical solution of Fourier's heat conduction problem, available in standard heat transfer textbooks
\1 1 {\eg \cite{carslaw1959conduction}, p122, Eq.~(12)}.  

\begin{figure}[!ht]
	\includegraphics[height=0.35\textwidth]{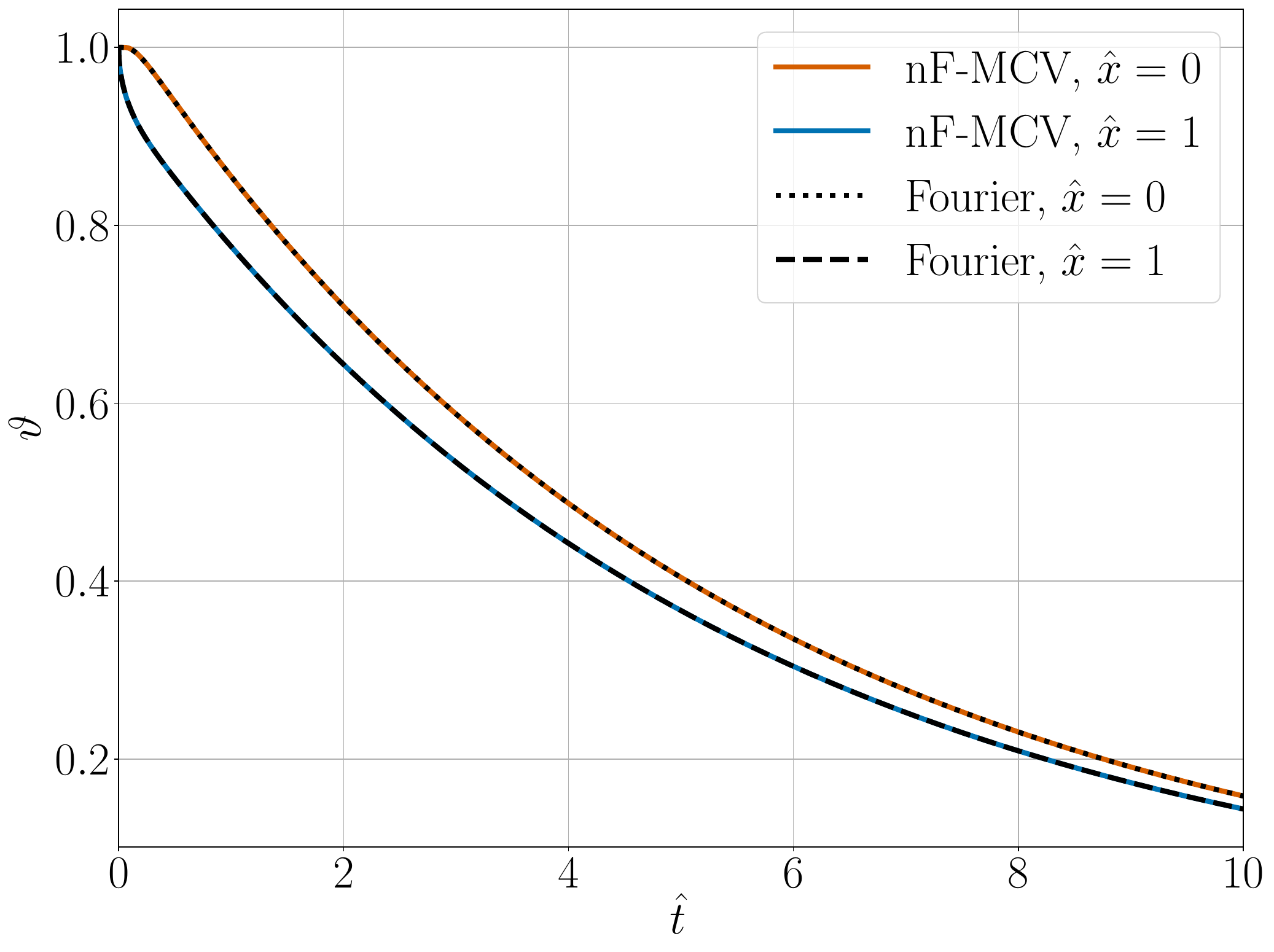} \centering
    \hfill
    \includegraphics[height=0.35\textwidth]{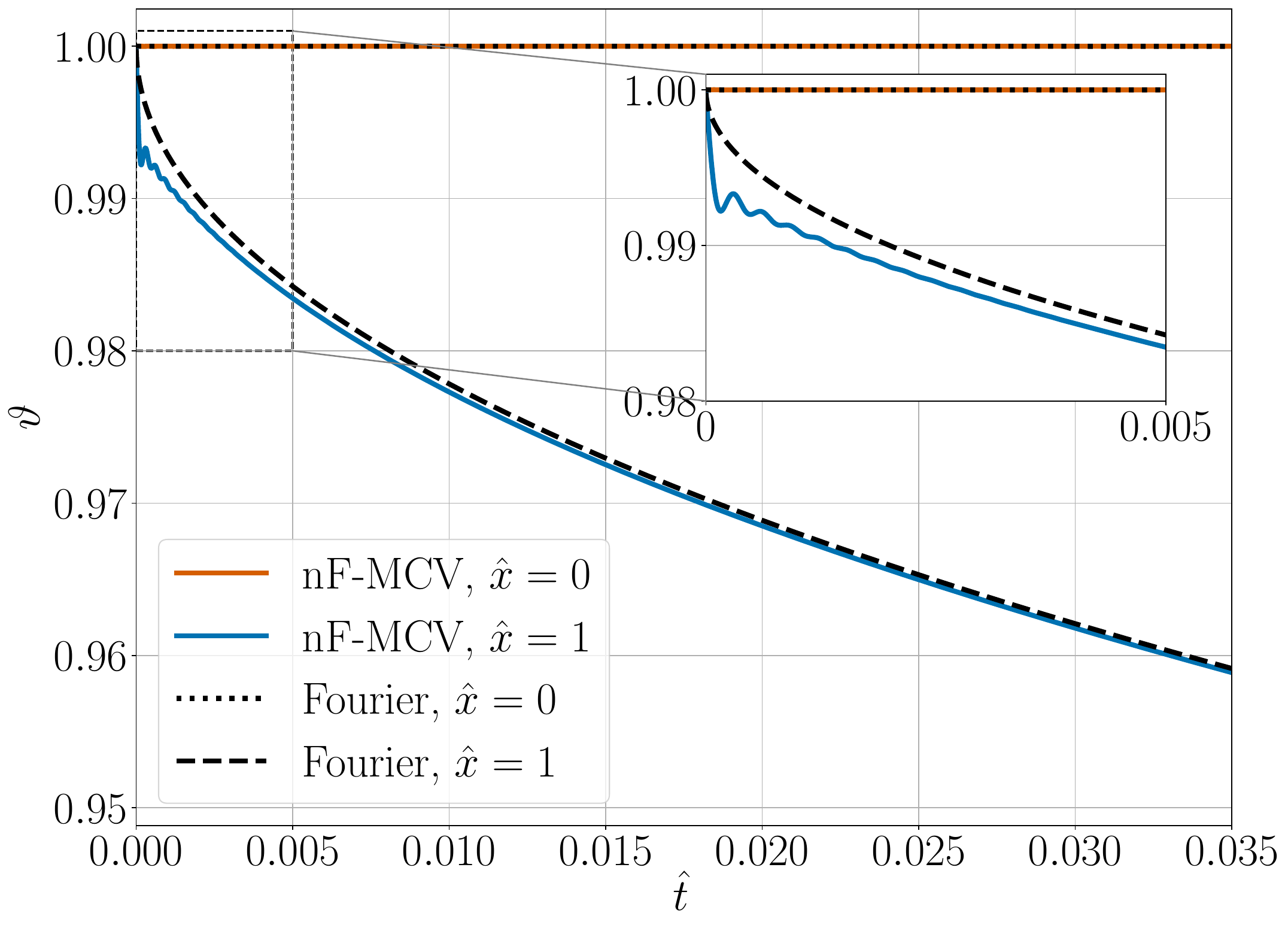} \centering
    \caption{Dimensionless temperature response in the small-relaxation-time limit evaluated at the boundaries with $ \qBi = 0.2 $ and $ \htau = 0.001 $. The solid lines correspond to the MCV model, while the dashed and dotted lines belongs to the classical Fourier solution. \emph{Left:} Long-time behavior. The perfect alignment of the profiles illustrates the smooth transition from hyperbolic wave propagation to parabolic diffusion. \emph{Right:} Short-time behavior, appearance of the hyperbolic characteristics.}
    \label{fig:T-t_MCV_vs_F}
\end{figure}

The asymptotic behavior in the small-relaxation-time limit is illustrated in Figure~\ref{fig:T-t_MCV_vs_F}, where the transient temperature response is evaluated for $ \qBi = 0.2 $ and a highly reduced dimensionless relaxation time of $ \htau = 0.001 $. The analytical solution of the MCV model is evaluated by considering the real
eigenvalues
along with $ 200 $ complex conjugate pairs, consequently, the same number of Fourier eigenmodes are utilized, together with the temporal and spatial resolution of $ \Delta \hht = 0.001 $ and $ \Delta \hx = 0.01 $, respectively. As observed, the MCV solution profiles computed via the biorthogonal eigenfunction expansion perfectly coincide with the classical Fourier solution throughout the entire transient process. Significant deviation can be observed only in the beginning of the process (calculated via $ \Delta \hht = 0.000005 $), where the hyperbolic characteristics of the near-Fourier MCV model emerges at the cooling boundary. Due to the extremely small relaxation time, the thermal propagation speed becomes nearly infinite, and the characteristic wave-like discontinuities and boundary reflections are suppressed almost instantaneously, therefore, wave propagation is totally unobservable at adiabatic boundary.

To quantitatively illustrate the hyperbolic-to-parabolic transition, Table~\ref{tab:F-vs-MCV} contrasts the first ten eigenmodes of the classical Fourier (F) framework with those of the near-Fourier behavior MCV model (nF-MCV) for $ \qBi = 0.2 $ and $ \htau = 0.001 $. For direct comparability, the indices of the classical Fourier eigenvalues are utilized here to label two corresponding MCV modes. In this small-relaxation-time regime, a clear splitting of the Fourier eigenmodes is observed within the MCV framework. Specifically, for lower-order modes ($ n \le 6 $) except for the first one, each single Fourier
eigenvalue root
splits into two distinct, purely real MCV roots. As the mode index increases, these real branches approach each other, leading to a localized amplification of the expansion coefficients near the transition boundary see Table~\ref{tab:F-vs-MCV} with $ n = 6 $). Beyond this critical index ($n \ge 7$), the roots branch into complex conjugate pairs. Structurally, this spectral transition fundamentally alters the spatial profiles of the eigenfunctions through the coefficients $ \qnu_n $. While the purely real roots yield standard, periodic trigonometric spatial distributions, the emergence of complex
eigenvalue roots
introduces hyperbolic components into the eigenfunctions.

Correspondingly, the eigenvalues $ \qmu_n $ governing the temporal behavior also separate into two distinct branches. One branch of these real eigenvalues asymptotically tracks the classical Fourier spectrum, representing the slow, physical diffusion process, whereas the other branch describes an extremely rapid damping. Beyond the previously introduced critical index, the
eigenvalues become complex, characterizing intensively relaxing oscillations that share nearly identical damping factors. 

\begin{figure}[!ht]
	\includegraphics[width=0.85\textwidth]{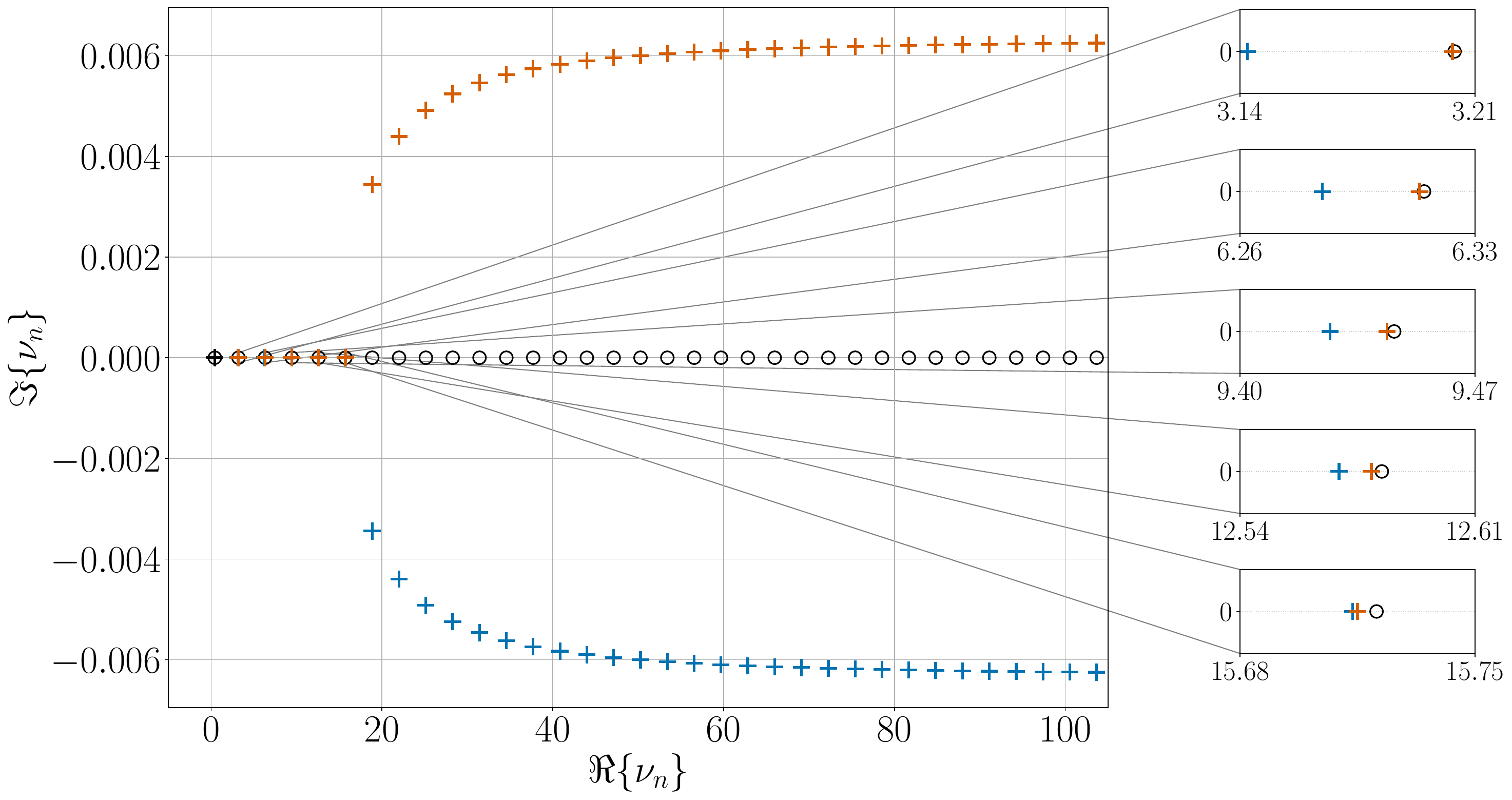}
    \caption{Distribution of the spatial coefficients $\qnu_n$ on the complex plane for $ \qBi = 0.2 $ and $ \htau = 0.001 $, demonstrating the near-Fourier asymptotic behavior. The classical Fourier roots and the near-Fourier MCV
    roots
    are denoted by black circles and by red and blue crosses for the two branches, respectively. The main plot illustrates how the lower-order roots lie strictly on the real axis before merging and transforming into complex conjugate pairs for higher values. The inset panels on the right provide detailed close-ups of the real axis, showcasing how the two distinct real MCV roots (red and blue crosses) approach each other, eventually merging at a critical point.}
    \label{fig:nu_MCV-vs-F}
\end{figure}

\begin{figure}[!ht]
	\includegraphics[width=0.65\textwidth]{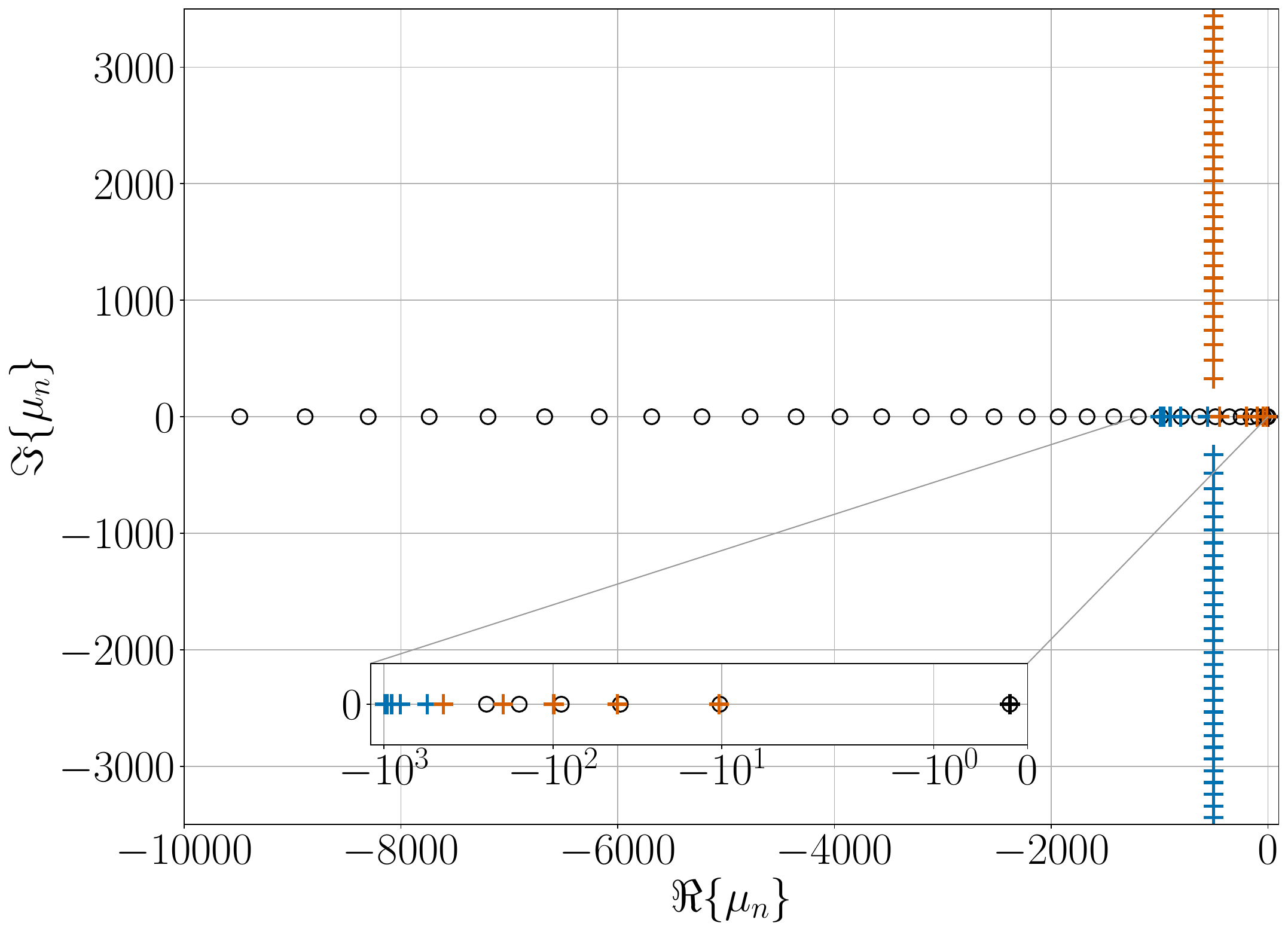}
    \caption{Distribution of the eigenvalues $ \qmu_n $ on the complex plane for $ \qBi = 0.2 $ and $ \htau = 0.001 $, demonstrating the near-Fourier asymptotic behavior. The classical Fourier
    eigenvalues
    and the near-Fourier MCV eigenvalues are denoted by black circles and by red and blue crosses for the two branches, respectively. The main plot visualizes the parabolic and hyperbolic behaviors of the respective models. The inset panel presents a detailed logarithmic view of the first six Fourier eigenvalues and their corresponding MCV counterparts along the real axis.}
    \label{fig:mu_MCV-vs-F}
\end{figure}

While the parabolic Fourier behavior manifests through real spatial coefficients
and
real
eigenvalues, the hyperbolic MCV characteristics express themselves through complex quantities. Visually, the MCV eigenvalues tend toward a vertical line in the complex plane, where the growing imaginary parts represent high-frequency, wave-like oscillations. This structural evolution is graphically illustrated on the complex planes in Figure~\ref{fig:nu_MCV-vs-F} for the spatial coefficients $ \qnu_n $, and in Figure~\ref{fig:mu_MCV-vs-F} for the
eigenvalues $ \qmu_n $, demonstrating how the
MCV beyond-Fourier
eigensystem encapsulates both the diffuse parabolic background and the high-frequency hyperbolic characteristics within a single mathematical framework.

\begin{table}[!ht]
    \centering
    \renewcommand{\arraystretch}{1.2}
    \centering
    \begin{tabular}{c||c|c||c|c||c|c}
    & \multicolumn{2}{c||}{$ \qnu_n$} & \multicolumn{2}{c||}{$ \qmu_n$} & \multicolumn{2}{c}{$ \qqc_n ( 0 ) $} \\
    $ n $ & F & nF-MCV & F & nF-MCV & F & nF-MCV \\
    \hline
    $ 1 $ & $ \hphantom{1}0.4328 $ & $ \hphantom{1}0.4328 $ & $ \hphantom{1}-0.1874 $ & $ -0.1874 $ & $ \hphantom{-}0.4463 $ & $ \hphantom{-}0.4463 $ \\
    \hline
    \multirow{2}{*}{$ 2 $} & \multirow{2}{*}{$ \hphantom{1}3.2039 $} & $ \hphantom{1}3.1422 $ & \multirow{2}{*}{$ \hphantom{1}-10.2652 $} & $ - 990.0274 $ & \multirow{2}{*}{$ - 0.1222 $} & $ \hphantom{-}0.0013 $ \\
    & & $ \hphantom{1}3.2033 $ & & $ - 10.3687 $ &  & $ - 0.1235 $ \\
    \hline
    \multirow{2}{*}{$ 3 $} & \multirow{2}{*}{$ \hphantom{1}6.3148 $} & $\hphantom{1}6.2845 $ & \multirow{2}{*}{$ \hphantom{1}-39.8773 $} & $ - 958.8086 $ & \multirow{2}{*}{$ \hphantom{-}0.0630 $} & $ - 0.0029 $ \\
    & & $ \hphantom{1}6.3135 $ & & $ - 41.5905 $ & & $ \hphantom{-}0.0659 $ \\
    \hline
    \multirow{2}{*}{$ 4 $} & \multirow{2}{*}{$ \hphantom{1}9.4459 $} & $ \hphantom{1}9.4269 $ & \multirow{2}{*}{$ \hphantom{1}-89.2259 $} & $ - 901.4161 $ & \multirow{2}{*}{$ - 0.0422 $} & $ \hphantom{-}0.0052 $ \\
    & & $ \hphantom{1}9.4439 $ & & $ - 98.9843 $ &  & $ - 0.0475 $ \\
    \hline
    \multirow{2}{*}{$ 5 $} & \multirow{2}{*}{$ 12.5823 $} & $ 12.5695 $ & \multirow{2}{*}{$ -158.3134 $} & $ - 803.3269 $& \multirow{2}{*}{$ \hphantom{-}0.0317 $} & $ - 0.0103 $ \\
    & & $ 12.5791 $ & & $ - 197.0720 $ &  & $ \hphantom{-}0.0421 $ \\
    \hline
    \multirow{2}{*}{$ 6 $}& \multirow{2}{*}{$ 15.7207 $} & $ 15.7136 $ & \multirow{2}{*}{$ -247.1399 $} & $ - 555.5160 $ & \multirow{2}{*}{$ - 0.0254 $} & $ \hphantom{-}0.1022 $ \\
    & & $ 15.7150 $ & & $ - 444.8828 $ & & $ - 0.1277 $ \\
    \hline
    \multirow{2}{*}{$ 7 $}& \multirow{2}{*}{$ 18.8602 $} & $ 18.8549 + 0.0034 \ii $ & \multirow{2}{*}{$ -355.7056 $} & $ - 500.1999 + 324.8164 \ii $ & \multirow{2}{*}{$ \hphantom{-}0.0212 $} & $ \hphantom{-}0.0106 - 0.0163 \ii $ \\
    & & $ 18.8549 - 0.0034 \ii $ & & $ - 500.1999 - 324.8164 \ii $ & & $ \hphantom{-}0.0106 + 0.0163 \ii $ \\
    \hline
    \multirow{2}{*}{$ 8 $}& \multirow{2}{*}{$ 22.0002 $} & $ 21.9957 + 0.0044 \ii $ & \multirow{2}{*}{$ -484.0105 $} & $ - 500.2000 + 483.5396 \ii $ & \multirow{2}{*}{$ - 0.0182 $} & $ - 0.0091 + 0.0094 \ii $ \\
    & & $ 21.9957 - 0.0044 \ii $ & & $ - 500.2000 - 483.5396 \ii $ & & $ - 0.0091 - 0.0094 \ii $ \\
    \hline
    \multirow{2}{*}{$ 9 $}& \multirow{2}{*}{$ 25.1407 $} & $ 25.1367 + 0.0049 \ii $ & \multirow{2}{*}{$ -632.0546 $} & $ - 500.2000 + 617.9439 \ii $ & \multirow{2}{*}{$ \hphantom{-}0.0159 $} & $ \hphantom{-}0.0080 - 0.0064 \ii $ \\
    & & $ 25.1367 - 0.0049 \ii $ & & $ - 500.2000 - 617.9439 \ii $ & & $ \hphantom{-}0.0080 + 0.0064 \ii $ \\
    \hline
    \multirow{2}{*}{$ 10 $}& \multirow{2}{*}{$ 28.2814 $} & $ 28.2779 + 0.0052 \ii $ & \multirow{2}{*}{$ -799.8379 $} & $ - 500.2000 + 741.3757 \ii $ & \multirow{2}{*}{$ - 0.0141 $} & $ - 0.0071 + 0.0048 \ii $ \\
    & & $ 28.2779 - 0.0052 \ii $ & & $ - 500.2000 - 741.3757 \ii $ & & $ - 0.0071 - 0.0048 \ii $
    \end{tabular}
    \caption{Comparison of the characteristics of the first ten eigenmodes between the classical Fourier (F) and the near-Fourier behavior MCV (nF-MCV) models for $ \qBi = 0.2 $ and $ \htau = 0.001 $.}
    \label{tab:F-vs-MCV}
\end{table}

\subsection{Modeling the flash method through a finite thermal penetration depth}  \label{flashinitcond}

The heat pulse experiment, also known as the flash method, proposed by Parker et al.~\cite{parker1961flash}, is the standard and most widely utilized experimental technique for measuring the thermal diffusivity of solid materials, particularly high-thermal-conductivity specimens such as metals, alloys, and advanced ceramics. The procedure starts from a homogeneous thermal equilibrium state, where the small, disk-shaped sample with a constant cross-section is characterized by a uniform initial temperature -- equal to the ambient temperature $ \qTinfty $ -- and zero initial heat current density. The measurement then involves subjecting the front face of the specimen to a short, high-intensity light pulse -- typically generated by a laser or a xenon flash lamp -- and recording the subsequent transient temperature response at the rear surface.

The idealized model of the flash experiment assumes an adiabatically insulated sample, except for the brief duration of the heat pulse. The finite energy transferred by the light pulse is classically modeled as an instantaneous and homogeneous excitation at the boundary interface, causing purely axial, one-dimensional heat conduction. In actual instruments, the pulse duration is finite, and the energy is absorbed within a thin but finite subsurface layer depending on the material's optical penetration depth, while convective boundary losses also appear
\cite{cape1963temperature,cowan1963pulse}. Now we analyze a model of the flash experiment in an MCV sample,\footnote{In a practically informative flash experiment for an MCV sample, the brief duration of the heat pulse should
also be smaller
than the MCV relaxation time \m { \qtau }. We assume this hereafter.} where the volumetric energy absorption is characterized by a finite thermal penetration depth resulting in a piecewise constant initial temperature. The convective heat loss is taken into account at the rear surface. The corresponding transient temperature response predicted by this benchmark model can be directly computed from our derived closed-form analytical
solution (as opposed to a numerical approach like in \cite{carr2025modelling}),
demonstrating its practical applicability.

The flash excitation is modeled by a heat pulse of finite energy $ \qQ $, applied instantaneously at the initial moment as $ \qQ \delta ( \qt ) $, where $ \delta ( \qt ) $ denotes the temporal Dirac delta distribution. The energy absorption occurring during the brief duration of the pulse -- treated here as instantaneous -- is accounted for by a homogeneous volumetric heat source density. This initial energy absorption is characterized by a finite thermal penetration depth \mm {  0 < \qxp < \qX \,,} within which
thermal 
energy is
assumed to be
uniformly distributed. Consequently, the total transferred energy of the heat pulse is expressed as
\begin{align}
    \qQ = \int\limits_{0^-}^\infty \qqA^{} \qA \delta ( \qt' ) \dd \qt' = \int\limits_{0^-}^\infty \left( \int\limits_0^{\qxp} \qqV^{} ( t' ) \qA \dd \qx \right) \dd \qt' = \int\limits_{0^-}^\infty \qqV^{} ( \qt' ) \qA \qxp \dd \qt' ,
\end{align}
where $ \qqA^{} $ denotes the homogeneous area-specific heat transferred through the front surface with the cross-section $ \qA $, and \m { \qqV^{} ( \qt' ) } abbreviates the homogeneous nonzero value of \mm {  \qqV^{} ( \qt' , \qx )} within \mm{ 0 \le \qx \le \qxp \,.}
Therefore, the spatio-temporal distribution of the volumetric heat source can be compactly expressed utilizing the Heaviside step function $ H $ as
\begin{align}
    \label{eq:q_V-flash}
    \qqV^{} ( \qt , \qx ) = \frac{\qqA^{}}{\qxp} \delta ( \qt ) \2 2 { 1 - H ( \qx - \qxp ) } .
\end{align}

Just after the heat pulse, the temperature and heat current density fields -- utilizing a fast--slow scale separation approximation \1 1 {see Appendix~\ref{FTinitial}} -- can be calculated from \re{eq:q_V-flash} and applied as an initial condition. In other words, the non-trivial flash excitation at the left boundary is formally reduced to an adiabatic one, transforming its energy impact directly into the bulk of the sample. The resulting initial conditions are
\begin{align}
    \qT ( 0 , \qx ) &= \qTinfty + \frac{\qqA^{}}{\qrho \qc \qxp} \2 2 { 1 - H ( \qx - \qxp ) } , \\
    \qq ( 0 , \qx ) &= 0 .
\end{align}
The detailed derivation of these initial conditions is given in Appendix~\ref{FTinitial}.

In order to introduce
the dimensionless variables, it is convenient to scale the problem with respect to the idealized, adiabatically insulated, configuration. If the sample remains completely insulated outside the brief duration of the heat pulse, the first law of thermodynamics yields the globally conserved energy balance as
\begin{align}
    \qrho \qc \qA \qX \left[ \qT ( \qt \to \infty ) - \qTinfty \right] = \qQ ,
\end{align}
where $ \qT ( \qt \to \infty ) $ represents the final homogeneous equilibrium temperature of the sample. Consequently, the uniform temperature rise achieved in this idealized setup provides a well-defined characteristic temperature scale, formulated as $ \qTheta_{\mathrm{u}} = \frac{\qqA^{}}{\qrho \qc \qX} $. Thereby, under this scaling, the dimensionless temperature field $ \qtheta \defeq \frac{\qT - \qTinfty}{\qTheta_{\rm u}} $ starts from a uniform initial value of zero and asymptotically tends toward a uniform value of unity in the ideal case. In our case, when convective heat loss is taken into account at the rear surface, the temperature field will eventually return to the uniform ambient baseline ($ \qtheta \to 0 $) in the
large-time
limit across the entire domain, after exhibiting distinct local maxima that vary from point to point. 

The corresponding dimensionless initial conditions are formulated as
\begin{align}
    \label{eq:IC-T-flash}
    \qtheta \left( 0 , \hx \right) &= \frac{1}{\hX_{\rm p}} \left[ 1 - H ( \hx - \hxp ) \right] , \\
    \label{eq:IC-q-flash}
    \qchi \left( 0 , \hx \right) &= 0 ,
\end{align}
where $ \hX_{\rm p} = \frac{\qxp}{\qX} $ denotes the dimensionless penetration depth. By applying \re{eq:c_n(0)}, the corresponding expansion coefficients can be explicitly given as
\begin{align}
    \label{eq:MCV-c_n(0)-flash}
    \qqc_n ( 0 ) = - \tcNsc_n \frac{\tan \tilde{\qnu}_n^*}{\qBi} \frac{\sin ( \hX_{\rm p} \tilde{\qnu}_n^* )}{\hX_{\rm p} \tilde{\qnu}_n^*}
    = - \tcNsc_n \frac{\tan {\qnu}_n}{\qBi} \frac{\sin ( \hX_{\rm p} {\qnu}_n )}{\hX_{\rm p} {\qnu}_n} .
\end{align}

An important remark must be highlighted regarding the convergence of the spectral sum for this specific benchmark configuration. Since the dimensionless penetration depth is localized near the left boundary (\ie $ \hX_{\rm p} \ll 1 $), the initial temperature field represents an exceptionally sharp, narrow spatial step. Reconstructing such a highly localized discontinuity requires a massive bandwidth of small-wavelength, consequently, high-frequency components. Specifically, the minimum spatial wavelength captured by the truncated spectrum must scale with the size of the heated zone; our experience indicates that the thermal profiles can be accurately resolved when the shortest wavelength in the summation is at most one-fifth of the dimensionless penetration depth $ \hX_{\rm p} $. Accordingly, a large number of eigenmodes must be retained in the analytical summation to achieve the spatial resolution necessary to accurately capture the initiation and subsequent sharp wave propagation of the thermal shock front without causing severe numerical truncation artifacts.

In the subsequent numerical calculations, the parametric study is conducted under a fixed dimensionless penetration depth of $ \hX_{\rm p} = 0.01 $ and a Biot number of $ \qBi = 0.2 $, while the relaxation time is varied as $ \htau = 0.01 $, $ 0.05 $, $ 0.1 $, and $ 0.5 $. To ensure the convergence near the initial sharp discontinuity, the analytical MCV solution is evaluated by considering the purely real
eigenvalues
along with $ 1000 $ complex conjugate pairs. The temporal resolutions for all illustrated fields are set to $ \Delta \hht = 0.001 $. For a direct comparison with the classical parabolic framework, the reference Fourier solution is generated utilizing exactly $ 1001 $ eigenmodes. This choice of truncated summation ensures a one-to-one compatibility with the $ \htau = 0.01 $ MCV case, which contains the maximum number of five purely real eigenvalues within the investigated parameter set.

\begin{figure}[!ht]
    \includegraphics[width=0.45\textwidth]{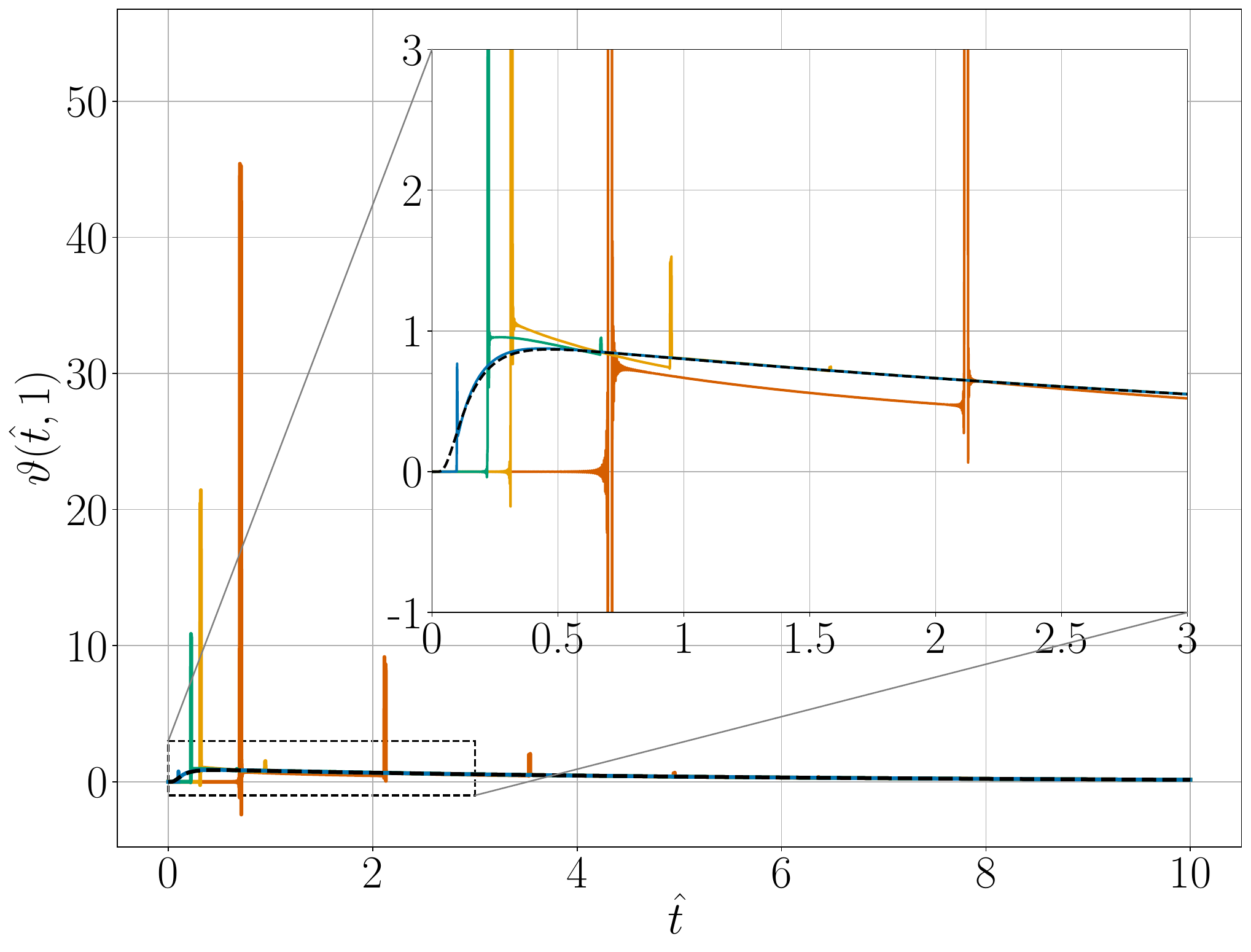} \centering
    \caption{Dimensionless rear side ($ \hx = 1 $) temperature response $ \qtheta $ for a fixed penetration depth of $ \hX_{\rm p} = 0.01 $ and a Biot number of $ \qBi = 0.2 $ under the dimensionless relaxation times $ \htau = 0.01 $, $ 0.05 $, $ 0.1 $, and $ 0.5 $ (increase is visualized by the coloring from blue to red).}
    \label{fig:T-t_MCV_vs_F_flash}
\end{figure}

The rear side ($ \hx = 1 $) temperature response is presented in Figure~\ref{fig:T-t_MCV_vs_F_flash}. The localized spatial discontinuity in the initial condition combined with the wave-like transport triggers intensive early-time temperature spikes, representing the physical arrival of the concentrated thermal shock at the rear boundary. As captured in the inset panel, the temperature remains strictly zero until the wave arrival time. Smaller relaxation times exhibit higher propagation velocities but undergo rapid thermal attenuation due to stronger damping scales. Conversely, larger relaxation times preserve the sharpness of the shock front, culminating in giant localized temperature peaks upon the first wave reflection. Most importantly, once these wave packets are dissipated by internal damping, all hyperbolic MCV profiles smoothly and perfectly converge onto the classical Fourier solution, confirming that the macroscopic diffusive behavior and the convective rear-surface cooling are accurately recovered in the long-time limit. Note that for larger relaxation times, the preservation of the steep shock gradients inherently triggers intense Gibbs phenomenon-driven oscillations near the peak values, which gradually decay as the front undergoes internal dissipation.

\begin{figure}[!ht]
    \includegraphics[width=0.45\textwidth]{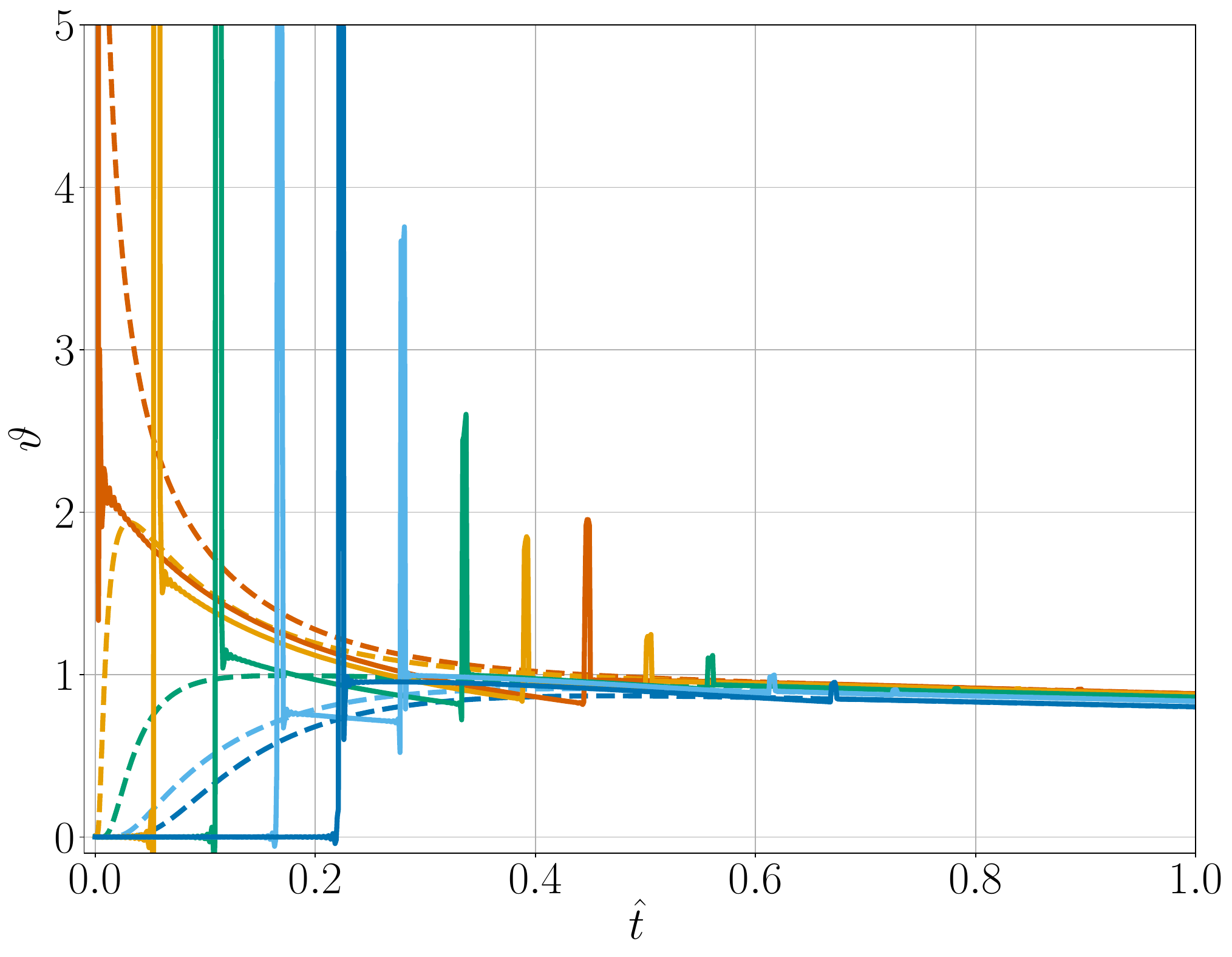} \centering
    \caption{Dimensionless transient temperature response $ \qtheta $ at the spatial coordinates $ \hx = 0 $, $ 0.25 $, $ 0.5 $, $ 0.75 $, and $ 1 $ (from red to blue, respectively) for $ \htau = 0.05 $ for a fixed relaxation time of $ \htau = 0.05 $, a penetration depth of $ \hX_{\rm p} = 0.01 $, and a Biot number of $ \qBi = 0.2 $. The solid lines represent the MCV analytical solution, while the dashed lines correspond to the Fourier solution.}
    \label{fig:T-t-xdisc_MCV_vs_F_flash}
\end{figure}

The spatial evolution of the hyperbolic transport mechanism is further illustrated in Figure~\ref{fig:T-t-xdisc_MCV_vs_F_flash}, which traces the transient temperature response at the discrete spatial coordinates $ \hx = 0 $, $ 0.25 $, $ 0.5 $, $ 0.75 $, and $ 1 $ for $ \htau = 0.05 $ against the classical Fourier solution. The profiles resolve the discrete spatial propagation of the localized thermal wave. The shock front arrives at each internal coordinate with a distinct time lag that scales linearly with distance and reaching the rear surface at $ \hht_{\rm arr} = \sqrt{\htau} \approx 0.22 $. As the disturbance travels through the sample, internal dissipation gradually attenuates the magnitude of the sharp temperature spikes. Upon reaching the convective rear boundary ($ \hx = 1 $), the wave reflects and propagates backward, triggering secondary, heavily damped temperature peaks at the internal points. After the wave propagation is suppressed, all hyperbolic fields perfectly collapse onto their respective parabolic Fourier counter-parts.

\begin{figure}[!ht]
    \includegraphics[width=0.45\textwidth]{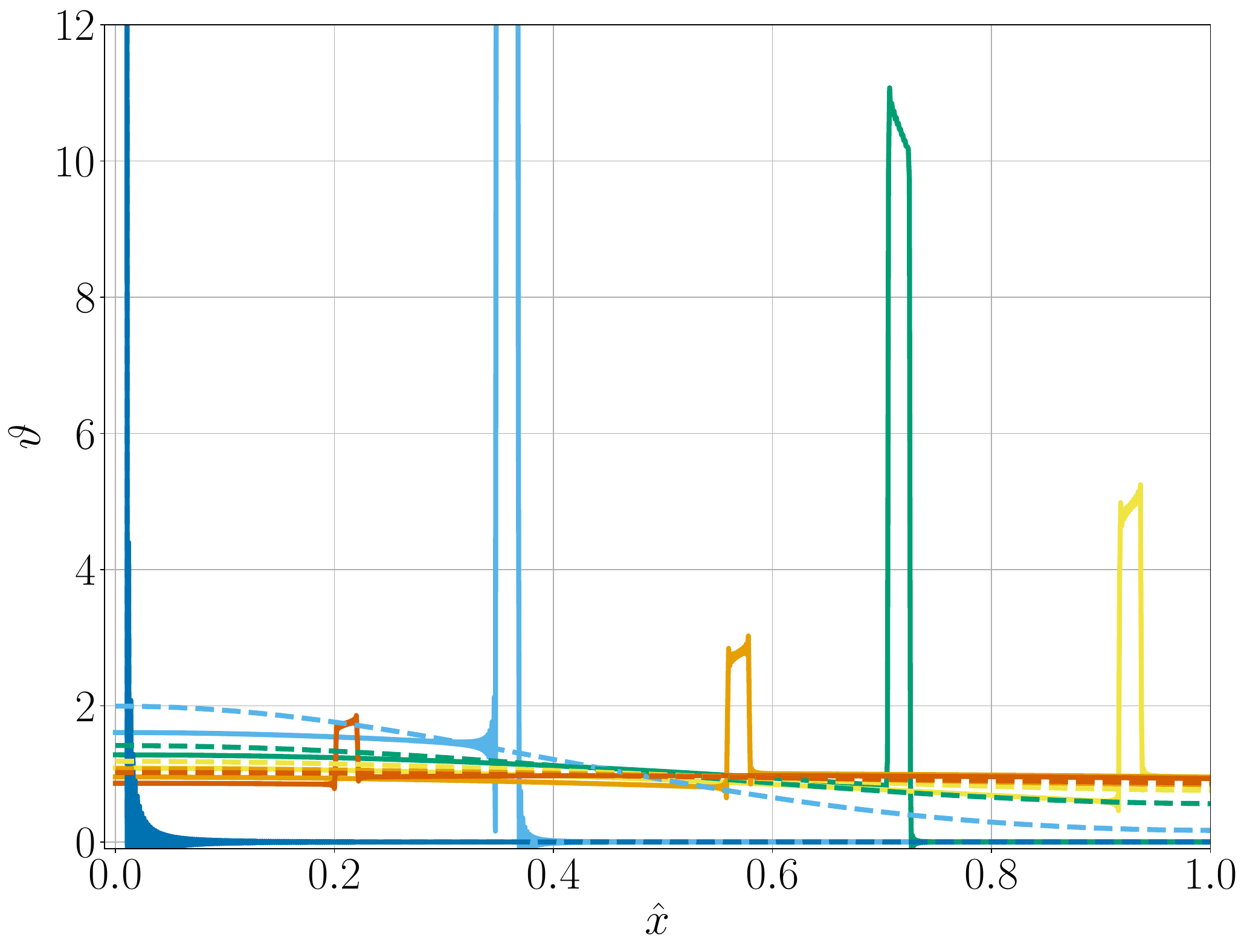} \centering
    \caption{Snapshots of the dimensionless temperature distribution $ \qtheta $ for a fixed penetration depth of $ \hX_{\rm p} = 0.01 $ and a Biot number of $ \qBi = 0.2 $ under the dimensionless relaxation times $ \htau = 0.01 $, $ 0.05 $, $ 0.1 $, and $ 0.5 $ (increase is visualized by the coloring from blue to red).}
    \label{fig:T-tx_MCV_vs_F_flash}
\end{figure}

The transport of thermal energy is demonstrated through the spatial temperature profiles presented in Figure~\ref{fig:T-tx_MCV_vs_F_flash}, captured at equal-time increments between $ \hht = 0 $ and $ \hht = 0.4 $ for $ \htau = 0.05 $. In stark contrast to the instantaneous, infinite-velocity smoothing predicted by the parabolic diffusion, the hyperbolic MCV wave propagation confines the initially deposited energy into a discrete, rectangular wave packet. As time progresses, this compact thermal shock wave propagates toward the right boundary at a constant finite velocity. The spatial width of this traveling energy packet remains invariant. As the wave front undergoes exponential attenuation due to internal relaxation mechanisms, it leaves behind a smooth thermal background that perfectly matches the parabolic diffusion profile.

\section{Conclusions}

The main results of
the present
this
work can be summarized as follows:
\begin{itemize}
    \item The one-dimensional MCV equation with an adiabatic boundary condition at one end and a heat-transfer boundary condition at the other one has been solved analytically.
    \item The heat-transfer boundary condition renders the underlying differential operator non-selfadjoint. Via a suitable weighted scalar product, the corresponding adjoint problem has been identified. The resulting left and right eigenfunctions were found to form a biorthogonal system, providing an analytical eigenfunction-series representation of both the temperature and heat current density fields.
    Completeness
    of the eigenfunction system
    has
    been
    demonstrated numerically.
    \item The parameter-dependent spectral structure has been investigated, including the occurrence of real, purely imaginary, and complex-conjugate eigenvalue roots, as well as the asymptotic distribution of the complex roots.
    \item The transition between near-parabolic (\ie near-Fourier) and highly hyperbolic regimes has been analyzed. The analytical solutions capture the transition from smooth diffusive relaxation to finite-speed thermal-wave propagation.
    \item The analytical framework has been applied to a model of the flash experiment incorporating rear-surface heat transfer and a finite optical penetration depth beneath the irradiated front surface. The pulse-induced initial temperature and heat-current-density fields have been derived analytically, and the resulting transient response has been evaluated using the biorthogonal eigenfunction expansion.
\end{itemize}

The
found
physical behavior of the system reflects finite-speed wave propagation and relaxation in the bulk, while the heat-transfer boundary condition introduces additional dissipation and modifies the reflection of the thermal waves. At a fixed Biot number, corresponding to a fixed intensity of surface heat transfer, the
temporal process
approaches the smooth diffusive behavior predicted by Fourier heat conduction in the near-parabolic regime, apart from rapidly decaying initial transients.
With larger and larger relaxation time parameter,
the
time evolution
is more and more
wave-like, thermal disturbances propagate at finite speed, undergo reflections at the boundaries, and gradually decay because of bulk relaxation and boundary heat loss.

The Biot number dependence and relaxation time dependence of the spectrum is even more apparent, and is not only quantitative but also qualitative.
Real eigenvalues represent the dominant diffusive
contribution, whereas complex-conjugate eigenvalue pairs describe damped
thermal-wave modes. The heat-transfer boundary condition can qualitatively
reorganize the lower-order part of the spectrum, resulting in the migration,
reduction in number, or complete disappearance of real eigenvalue roots. In the
latter case, purely imaginary eigenvalue roots can also emerge. 

The analysis of
the eigenvalues has revealed the
important
role of the parameter combination \m { \htau \qBi^2 } in the
parameter dependence of the structure of the spectrum
\1 2 {see \re{eq:mu-nu} and the detailed
Appendix~\ref{sec:eigenvalues}
}.
In terms of the dimensionful system parameters,
this combination reads
\begin{align}
    \htau \qBi^2 = \frac{\qtau}{\nicefrac{\qX^2}{\FTa}} \9 1 { \frac{\qX}{\nicefrac{\qlambda}{\FTalpha}} }^2 = \frac{\qtau}{ \frac{ \9 1 {\qlambda/\FTalpha}^2 }{a} }
\end{align}
\1 1 {see Section~\ref{probstatprep}}. The physical content is that the heat transfer boundary condition introduces the length scale \m { \qlambda/\FTalpha } and the corresponding time scale \m { \frac{ \9 1 {\qlambda/\FTalpha}^2 }{a} } \1 1 {which both are independent of sample size!}. Accordingly, the ratio of the relaxation time \m { \qtau } to the boundary time scale \m { \frac{ \9 1 {\qlambda/\FTalpha}^2 }{a} } is
just the important
parameter combination \m { \htau \qBi^2}. This observation leads to a rule-of-thumb message that apparent beyond-Fourier MCV effects can be expected when relaxation time is not much smaller than the heat-transfer related time scale.

The customarily considered simpler boundary conditions BC1 and BC2 \1 1 {\m { \qBi \to \infty } resp.\ \m { \qBi = 0 }} lead to self-adjoint problems. Our findings here indicate that BC3 with an appropriate intermediate finite \m { \qBi }, introducing considerable non-selfadjointness, may provide an efficient extra means to find and explore beyond-Fourier heat conduction.

In the present work, a single flash pulse was discussed. We expect that visibility of non-Fourier temporal behavior may be enhanced when one applies a -- for example periodic -- \emp{sequence} of pulses. Sensitivity of the corresponding response to the strength of the heat transfer parameter is a promising possibility that is worth elaborating in the future.

Regarding the mathematical aspects,
our findings and experience with the present non-selfadjoint problem
suggest
that it is
free of problematic aspects like noncompleteness of the eigenfunctions or challenged eigenfunction expansion convergence behavior.
We conjecture
that, for non-selfadjoint problems in general, noncompleteness of the eigenfunctions can only occur in the presence of
gain-type
boundary condition
\1 1 {in the optical language of \cite{kostenbauder1997eigenmode}}. In heat-conduction applications, BC3 is always loss-type \1 1 {dissipative, irreversible: ``heat flows from warmer to colder''}, otherwise the Second Law of thermodynamics were violated. Therefore, the physical perspective of the Second Law and the related properties of entropy may serve as a helpful tool for the mathematical analysis of non-selfadjoint problems in general.

The results presented here are expected to be successfully generalizable for other beyond-Fourier heat-conduction models as well, including the Guyer--Krumhansl structured equation \cite{guyer1966solution}, and the Jeffreys structured one \cite{jeffreys1917viscosity},
which two models are particularly relevant in the light of experiments, \eg regarding heterogeneous materials and structures \cite{mariano2022solutions,kovacs2024heat,feher2024dynamic,feher2024thermal},
including porous composites used in energetics-related applications such as gas storage \cite{gal2024thermal}.

\subsection*{Acknowledgment}

The authors thank David Krej{\v{c}}i{\v{r}}{\'{\i}}k for the valuable discussions and suggested references.

This research was supported by the Sustainable Development and Technologies National Programme of the Hungarian Academy of Sciences (FFT NP FTA)
and by the Hungarian Scientific Research Fund under grant agreements NKKP STARTING24 149487 and NKKP Advanced 150038.

\subsection*{Author contributions}

TF: Conceptual steps for the dual problem, completeness, the eigenvalue asymptotics, and the flash initial condition. MSz: Numerical implementations, parameter dependence investigations, completeness test, analytical and numerical eigenvalue investigations. The authors worked equally on the manuscript, and have read and agreed in its final form.

\appendix

\section{On the behavior of the eigenvalues} \label{sec:eigenvalues}

The transcendental equation \re{ftp} via trigonometric identities can be transformed into the form
\begin{align}
    \label{ftp:omega}
    \eif ( \qomega )
    \defeq
    \left( 1 - \qeps \cos \qomega \right) - \qBi \left( 1 + \qeps \right) \frac{\sin \qomega}{\qomega}
    \stackrel{!}{=} 0
\end{align}
with
\begin{align}
 &&&&
    \label{eq:tr-eq-ome-param}
    \qomega_n & \defeq 2 \qnu_n ,
 &
    \qeps & \defeq \frac{1 - \htau \qBi^2}{1 + \htau \qBi^2} .
 &&&&
\end{align}
Based on its definition, \m { -1 < \qeps < 1 }.
For simplicity, the solutions \m { \qomega_n } will also be called eigenvalue roots.

\null\par
\begin{figure}[!ht]
	\includegraphics[width=0.32\textwidth]{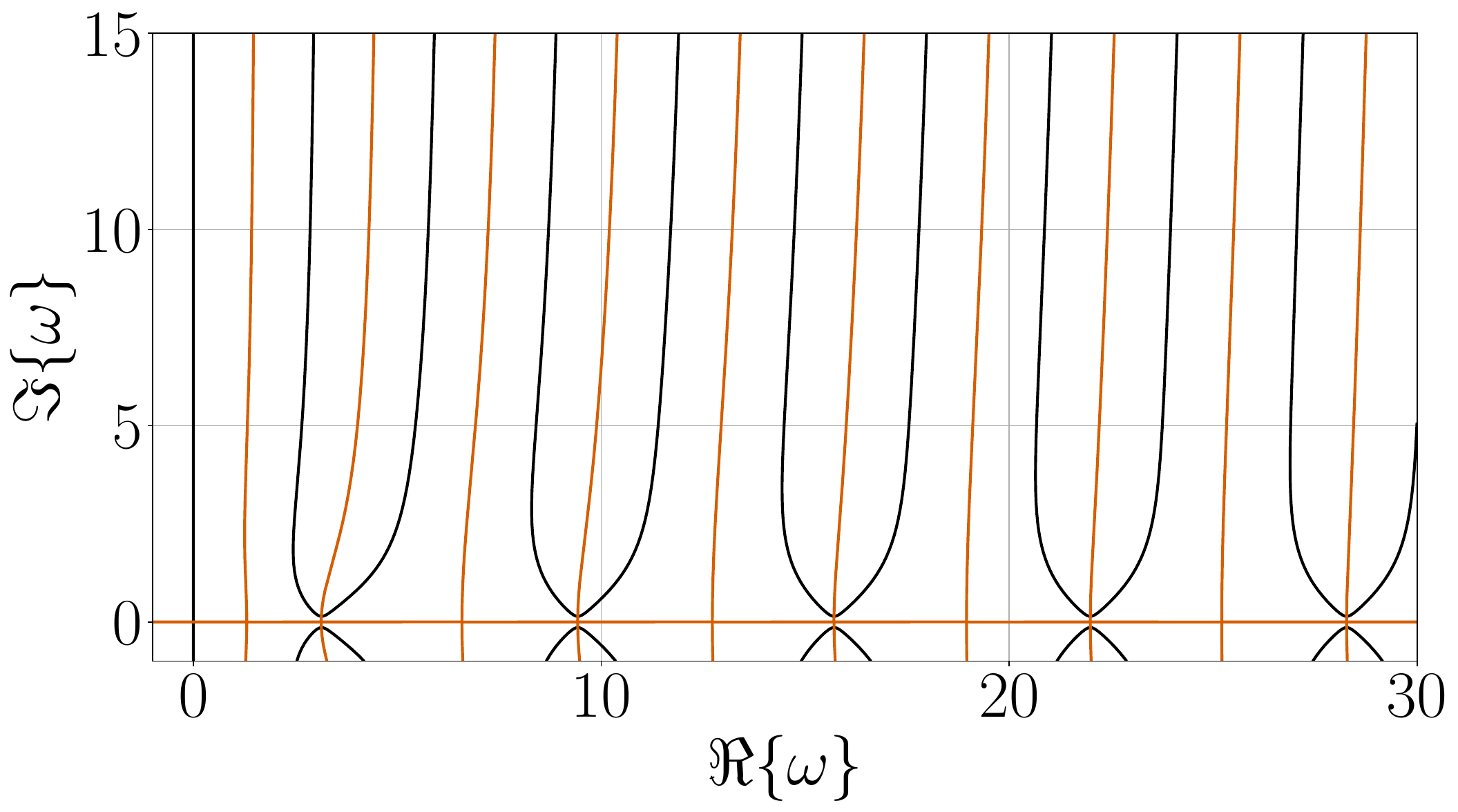} 
    \hfill
    \includegraphics[width=0.32\textwidth]{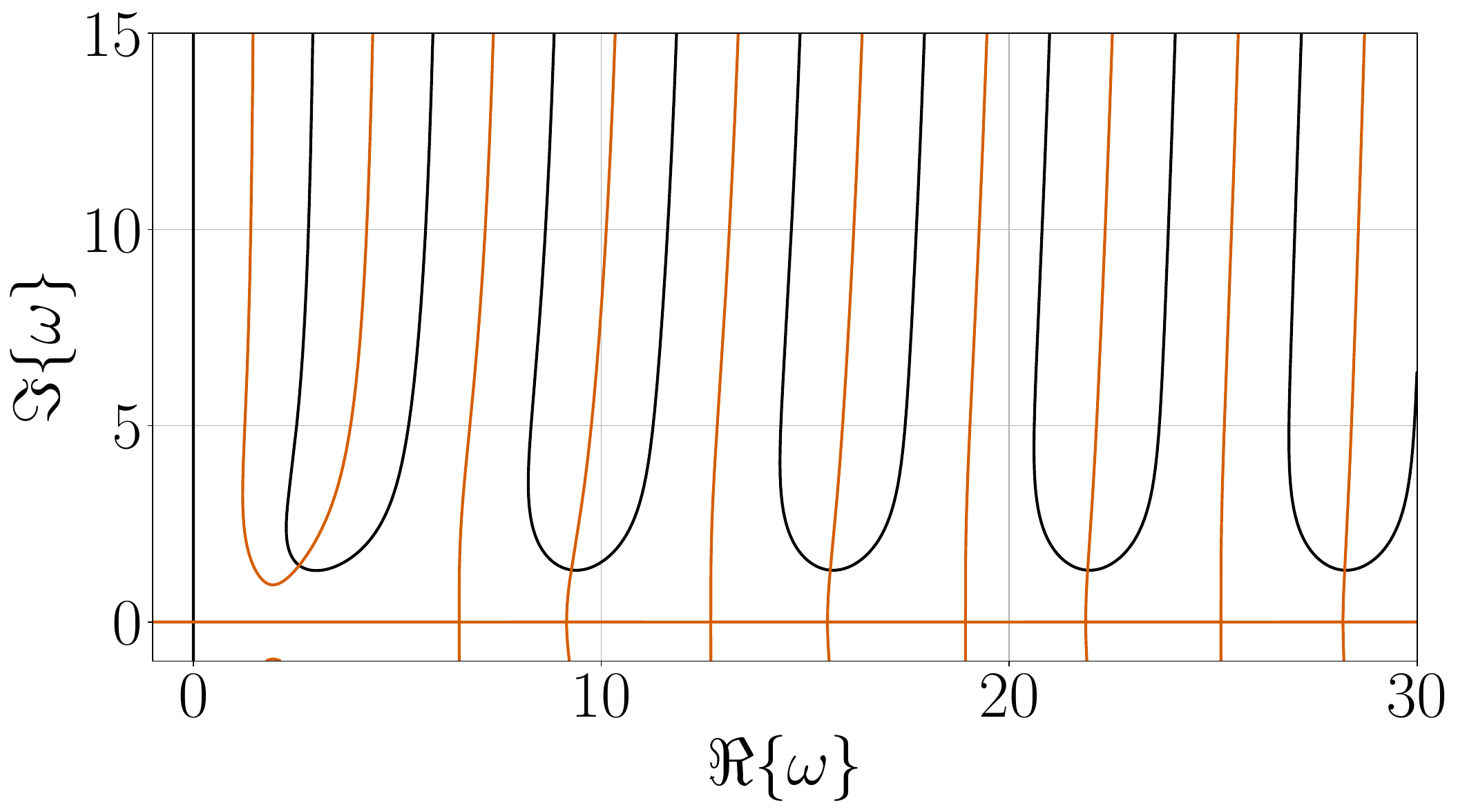} 
    \hfill
    \includegraphics[width=0.32\textwidth]{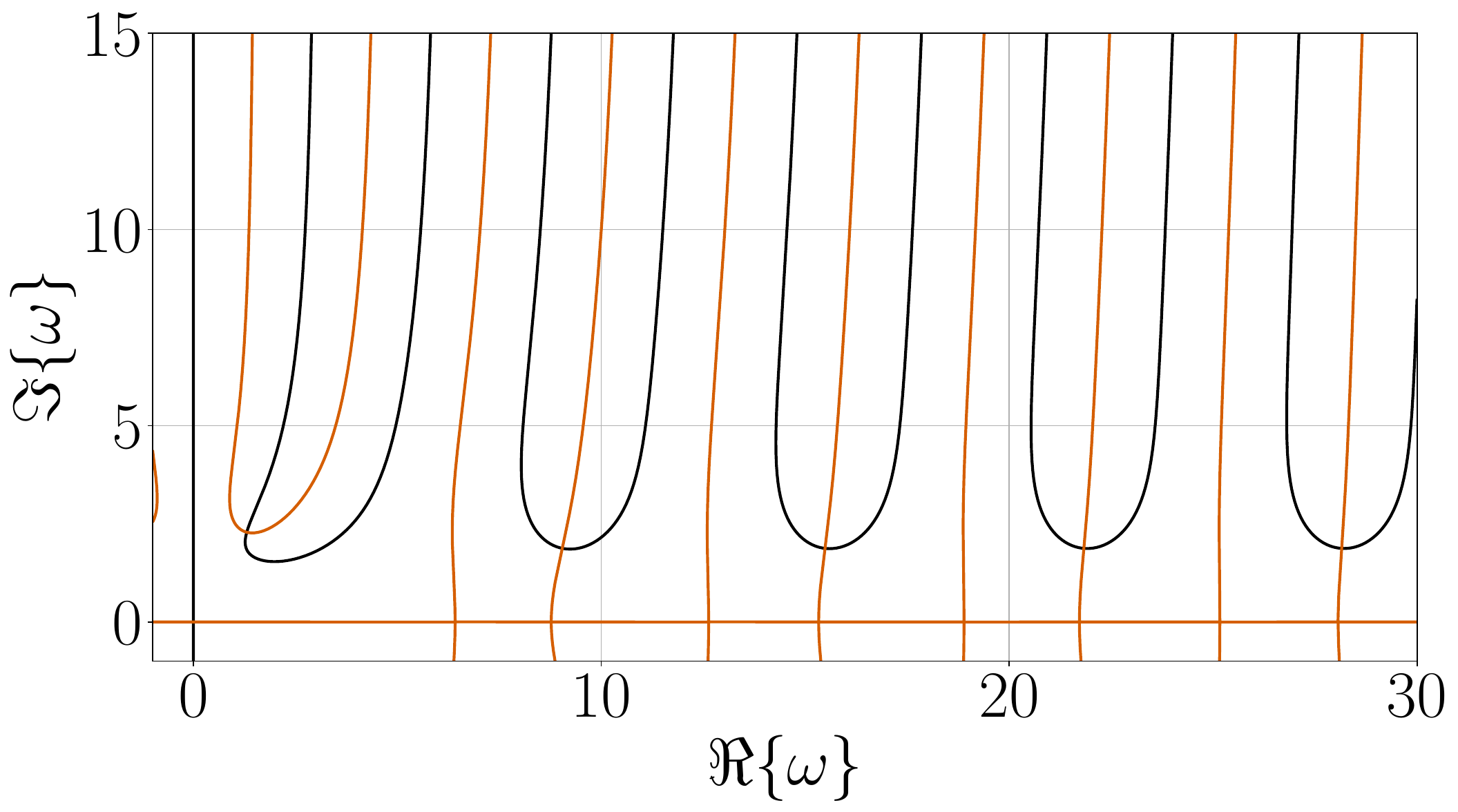} \\
    \includegraphics[width=0.32\textwidth]{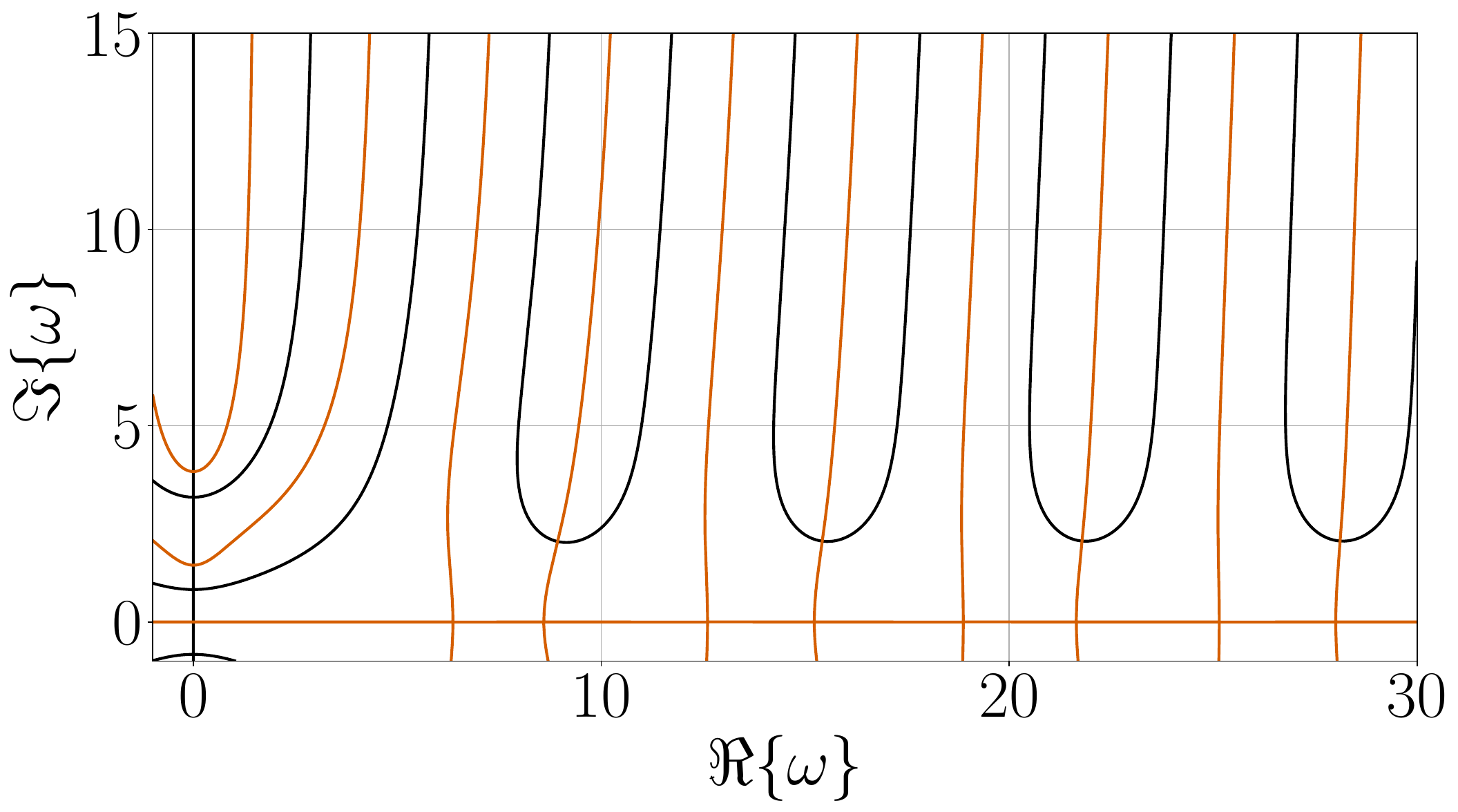} 
    \hfill
    \includegraphics[width=0.32\textwidth]{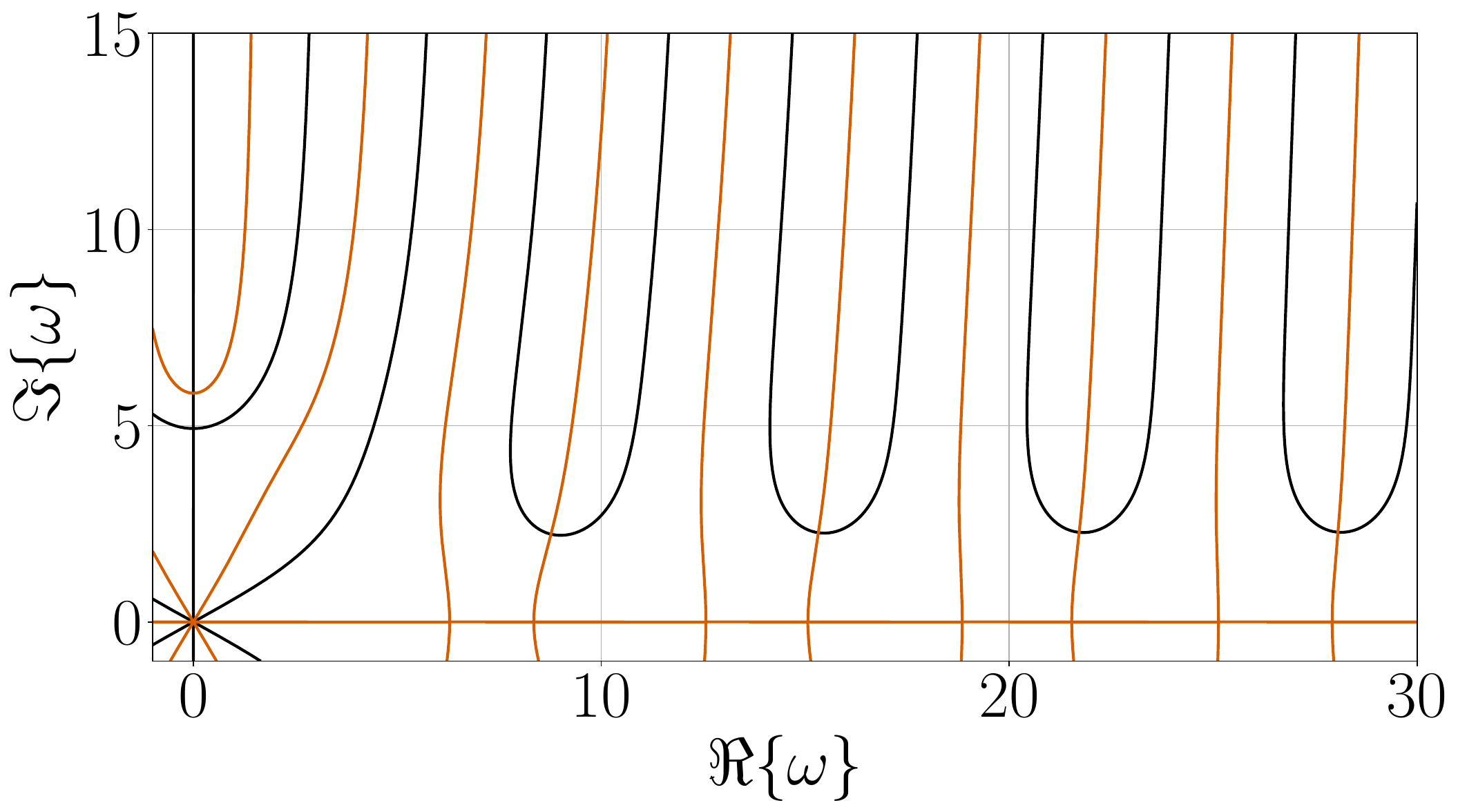}
    \hfill
    \includegraphics[width=0.32\textwidth]{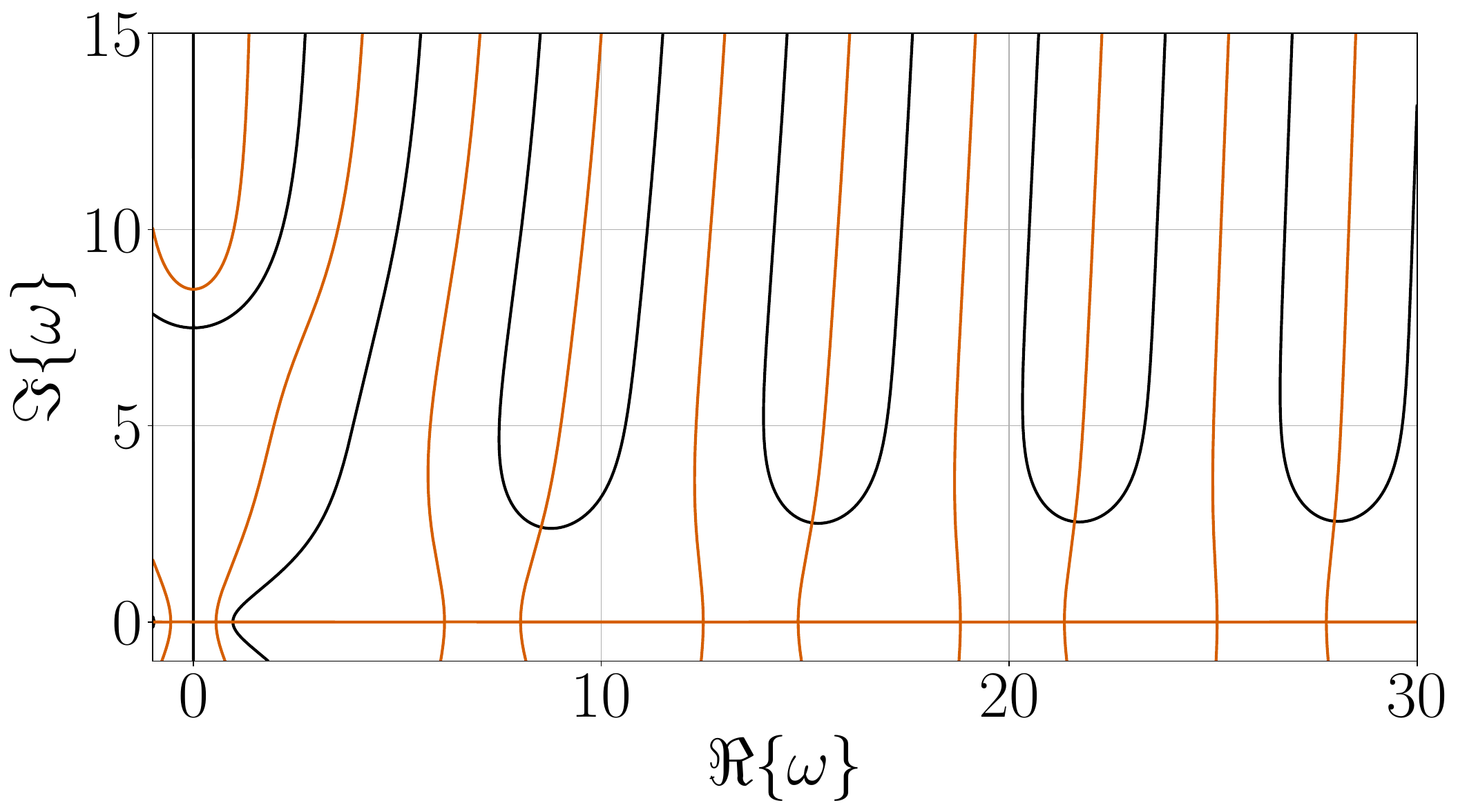} \\
    \includegraphics[width=0.32\textwidth]{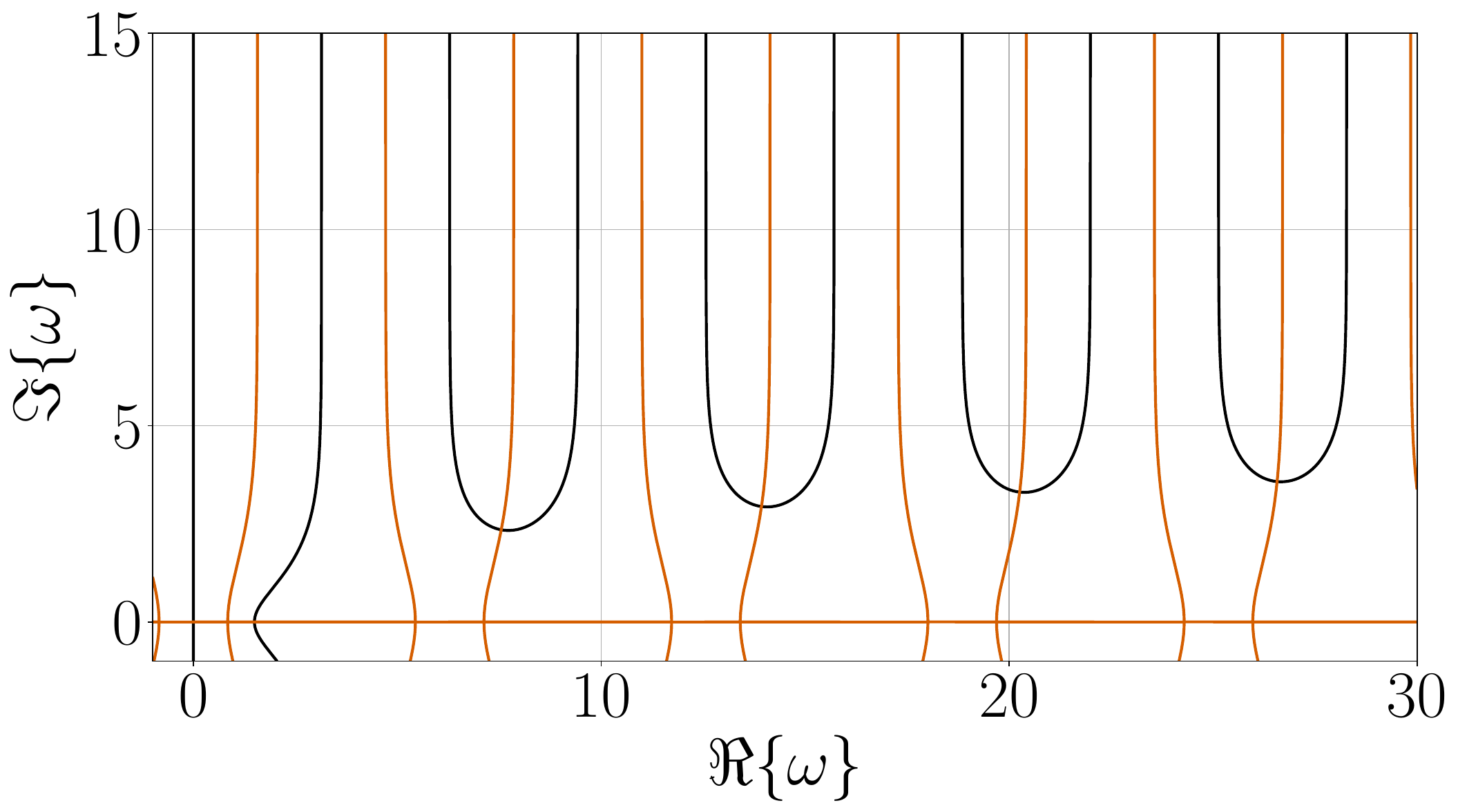}
    \hfill
    \includegraphics[width=0.32\textwidth]{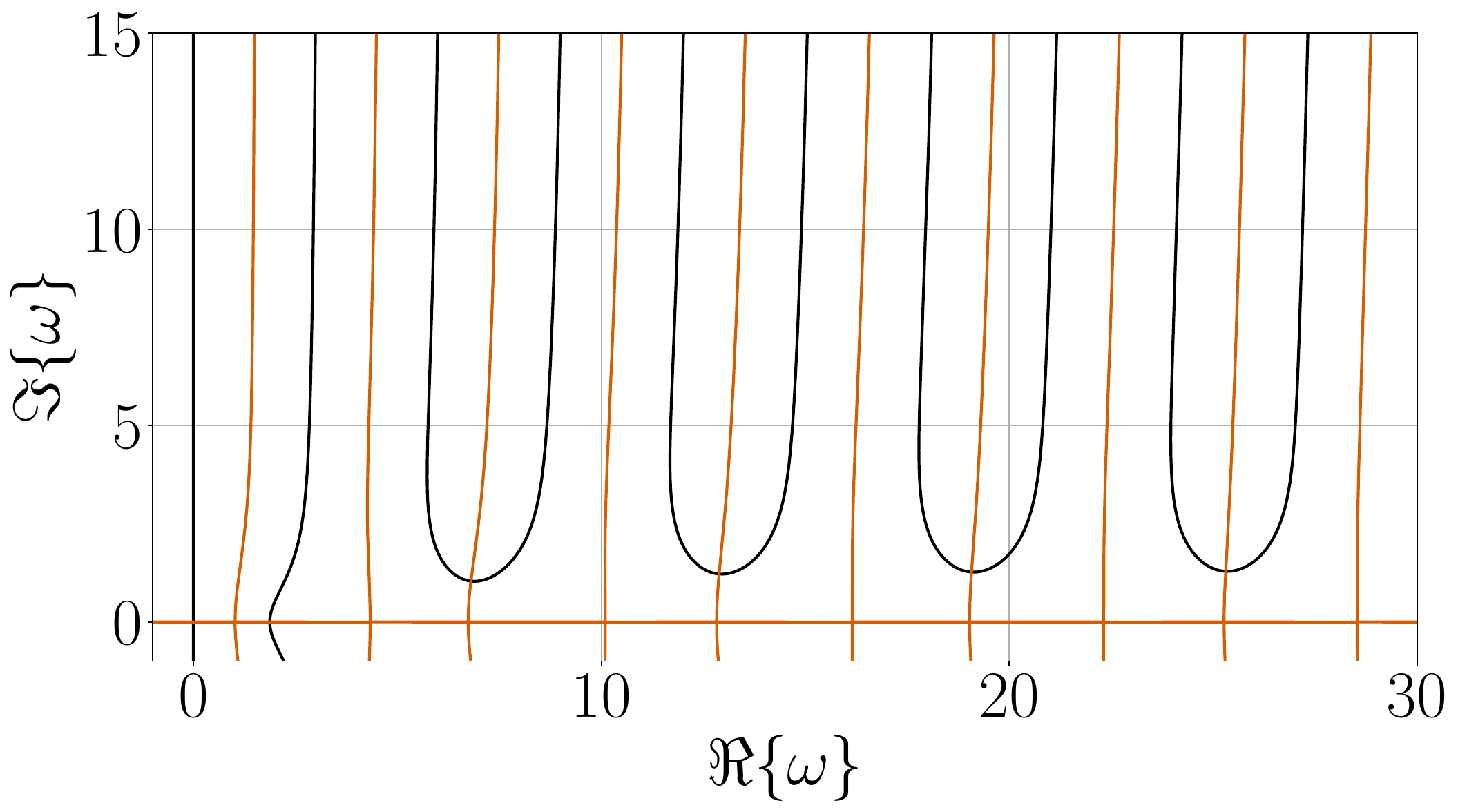} 
    \hfill
    \includegraphics[width=0.32\textwidth]{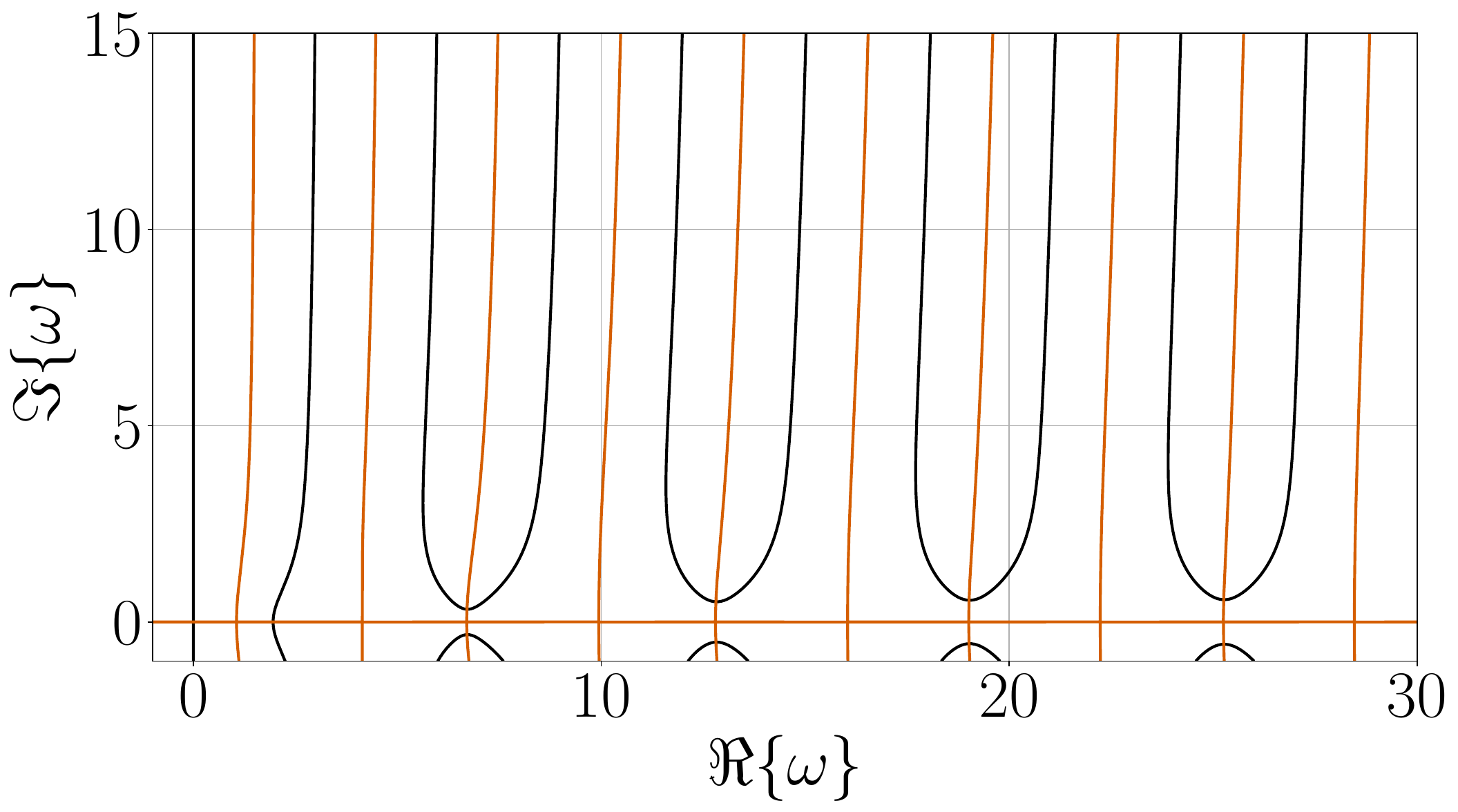} \\
    \includegraphics[width=0.32\textwidth]{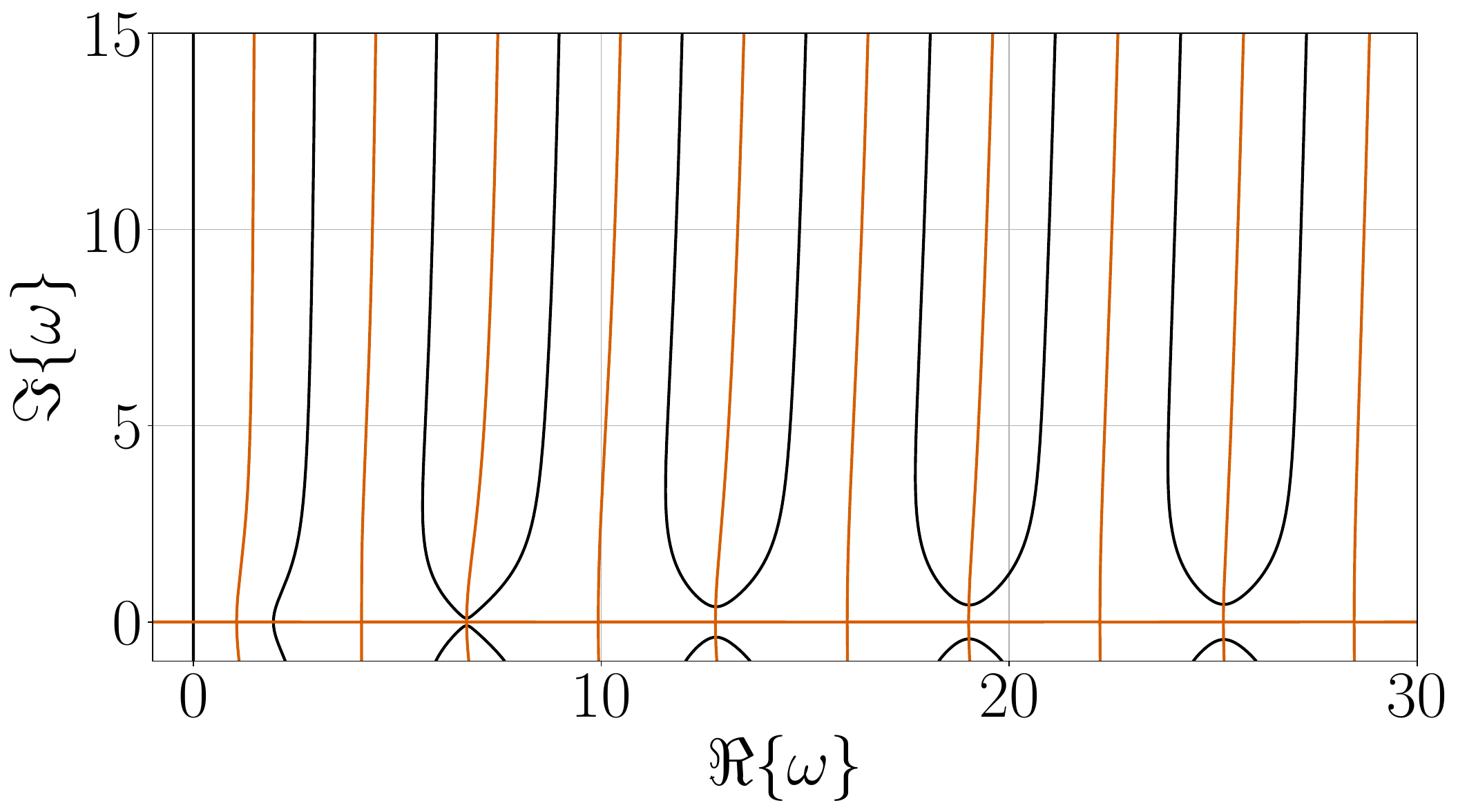} 
    \hfill
    \includegraphics[width=0.32\textwidth]{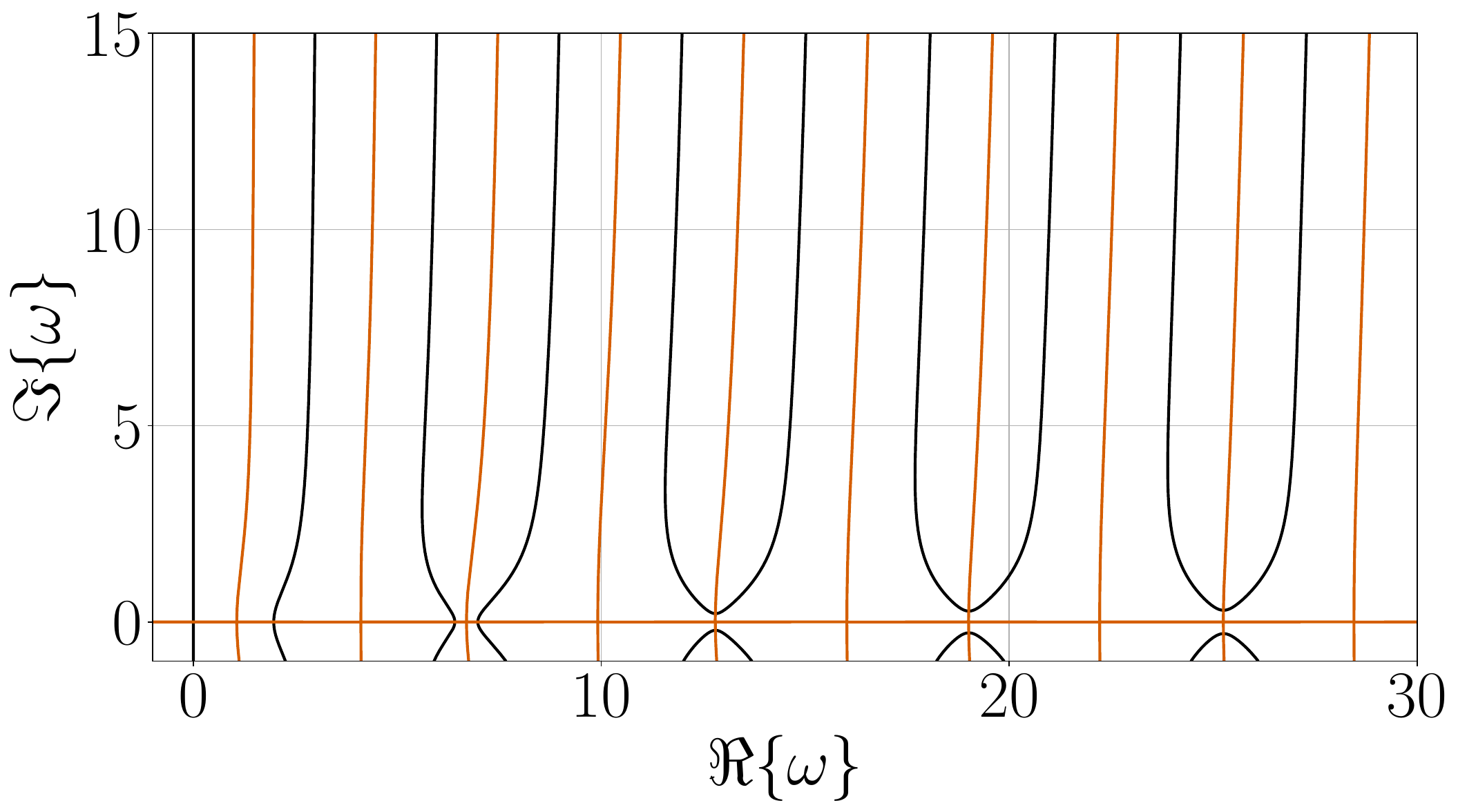} 
    \hfill
    \includegraphics[width=0.32\textwidth]{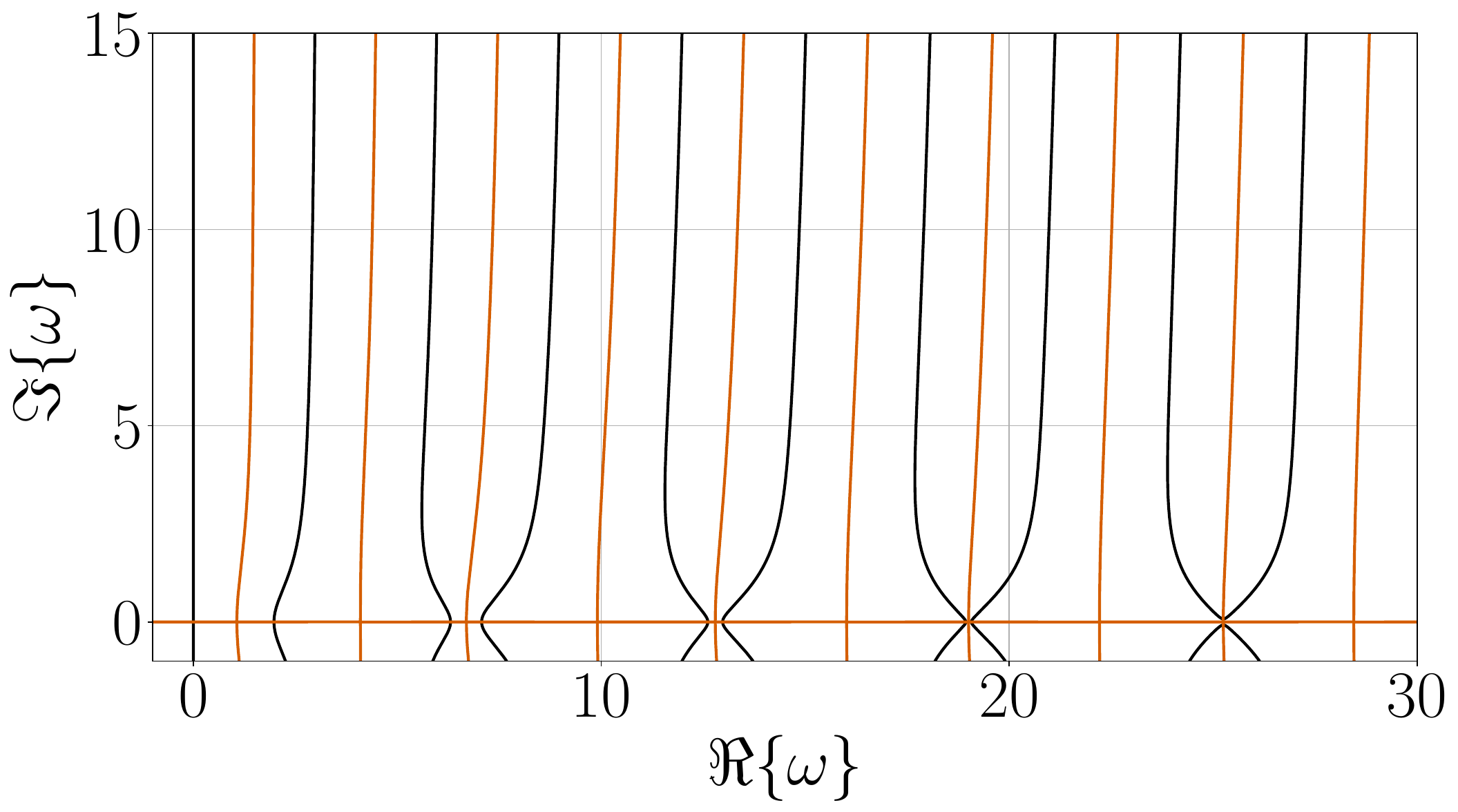}
    \caption{The zero-level curves of the real (black line) and imaginary parts (red line) of \re{ftp:omega} in the complex plane with $ \qBi = 1.5 $ and different sets of $ \qeps $. \emph{%
     From left to right, and broken to multiple rows in downward sequence%
    }: $ \qeps = - 0.99, \ - 0.5, \ - 0.3, \ - 0.25, \ - 0.2, \ - 0.15, \ 0, \ 0.5, \ 0.85, \ 0.9, \ 0.95, \ 0.99 $.}
    \label{fig:ImRe-plot}
\end{figure}

As presented in Section~\ref{sec:adjoint}, the operator associated with the investigated problem is neither self-adjoint nor anti-self-adjoint. Consequently, its spectrum is in general complex, although real or purely imaginary
eigenvalues
may appear for particular combinations of the parameters $ \qeps $ and $ \qBi $. Moreover, the eigenvalues are sensitive to these parameters, so even small variations in $ \qeps $ or $ \qBi $ may lead to significant spectral shifts. This behavior is illustrated in Figure~\ref{fig:ImRe-plot}, where the zero-level curves of the real and imaginary parts of the transcendental equation \re{ftp:omega} in the complex plane are shown for different parameter sets. The intersection points of these curves correspond to the eigenvalue locations.

Our procedure is as follows:
 \begin{enumerate}
 \item
First, we identify the purely imaginary
eigenvalue roots
(if they exist; at most two such
eigenvalue roots
may occur, see below). 
 \item
Then we
determine
the
possible
real
eigenvalue roots
we have not found any condition limiting their possible number).
 \item
Finally, among the
\1 0 {infinitely many}
complex
eigenvalue roots,
we determine a finite subset using the Newton method.
 \end{enumerate}

\subsection
{Imaginary eigenvalues}

If purely imaginary eigenvalues exist then,
by
via
the substitution $ \qomega = \ii \eibeta $, the transcendental equation \re{ftp:omega} can be transformed into the equivalent form
\begin{align}
    \label{ftp:beta}
    \eig ( \eibeta )
    \defeq
    \IM \{ \eif \0 1 { \ii \eibeta } \} = \left( 1 - \qeps \cosh \eibeta \right) - \qBi \left( 1 + \qeps \right) \frac{\sinh \eibeta}{\eibeta}
    \stackrel{!}{=} 0
    ,
\end{align}
and we seek for real solutions $ \eibeta $  of \re{ftp:beta}.

First, let us investigate the limit of validity of the parameter $ \qeps $, \ie $ \qeps = \pm 1 $. If $ \qeps = 1 $, then $ \eig ( 0 ) = - 2 \qBi < 0 $ and $ \eig $ has a negative slope for all positive $ \eibeta $. If $ \qeps = -1 $, then $ \eig(0) = 2 > 0 $ and $ \eig $ has a positive slope for all positive $ \eibeta $. Consequently, if $ \qeps = \pm 1 $, then no real roots of \re{ftp:beta} exist. If $ \qeps = 0 $, then $ \eig ( 0 ) = 1 - \qBi $ and $ \eig $ has a negative slope for all positive $ \eibeta $.

In general, the function $ \eig $ starts from the value 
\begin{align}
    \eig ( 0 ) = 1 - \qBi - \qeps \left( 1 + \qBi \right) ,
\end{align}
whose sign, as well as the slope of $ \eig $, depends on the values of $ \qeps $ and $ \qBi $. The initial value of $ \eig $ is positive if $ \qeps \le \frac{1 - \qBi}{1 + \qBi} $. The slope of the function is deduced from its Taylor expansion around zero, which, using $ \cosh \eibeta = \sum_{k=0}^\infty \frac{1}{(2k)!} \eibeta^{2k} $ and $ \sinh \eibeta = \sum_{k=0}^\infty \frac{1}{(2k + 1)!} \eibeta^{2k + 1} $, can be given as
\begin{align}
    \label{ftp:beta-taylor}
    \eig = \eig ( 0 ) - \sum_{k=1}^\infty \frac{1}{ (2k + 1)!} \left[ \left( 1 + 2k + \qBi \right) \qeps + \qBi \right] \eibeta^{2k} .
\end{align}
If $ 0 \le \qeps $, then the slope of $ \eig $ is negative for all positive $ \eibeta $; in the opposite case, positive Taylor coefficient may exist. A lower bound on $ \qeps $ for which the slope of $ \eig $ may remain negative for all positive $ \eibeta $ is obtained from
\begin{align}
    \left( 1 + 2k + \qBi \right) \qeps + \qBi > 0 , \qquad \qquad 
    \text{\ie} \qquad \qquad
    0 > \qeps > - \frac{\qBi}{1 + 2k + \qBi} > - \frac{\qBi}{3 + \qBi} .
\end{align}

\begin{figure}[!ht]
	\includegraphics[width=0.32\textwidth]{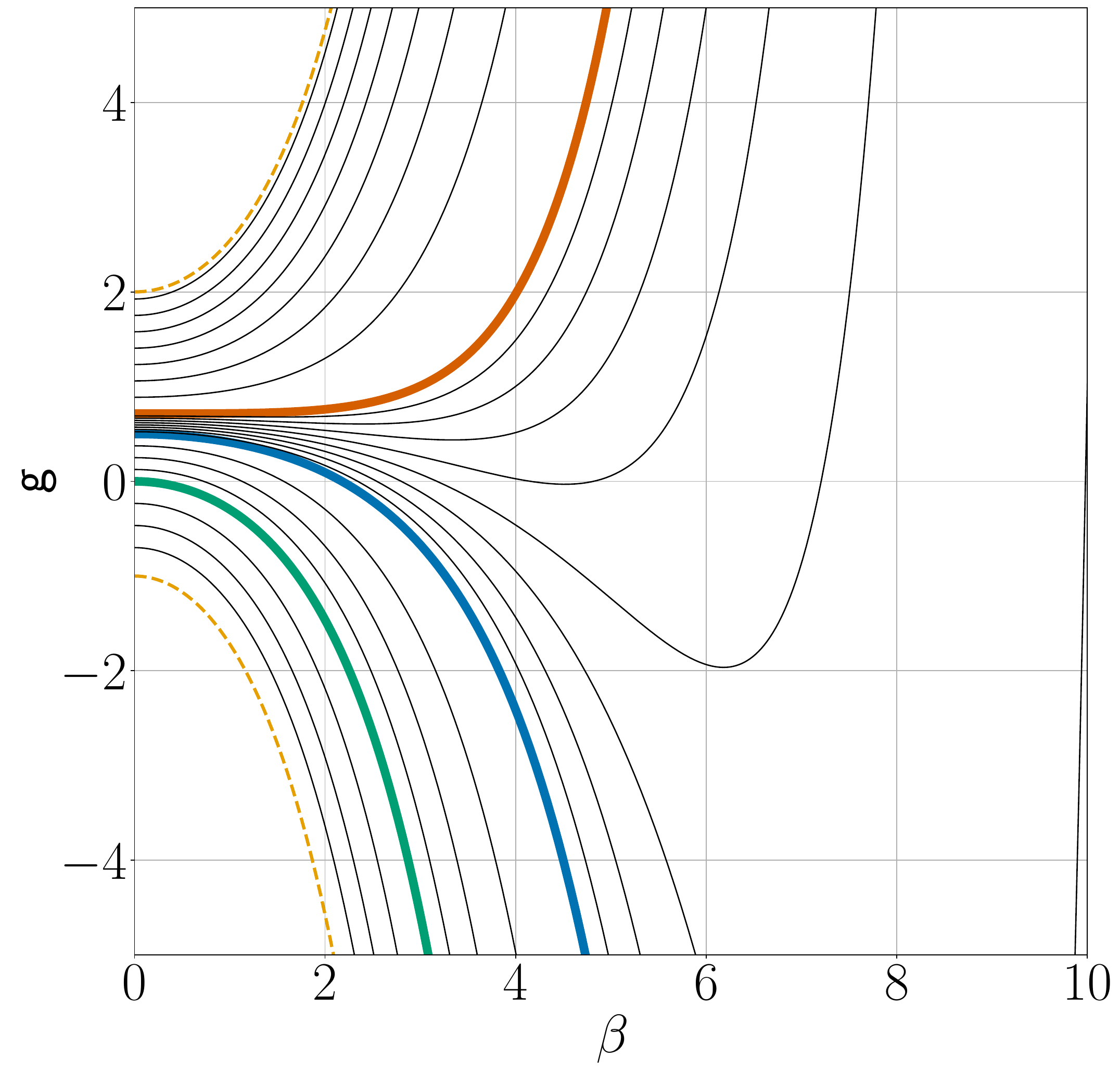} 
    \hfill
    \includegraphics[width=0.32\textwidth]{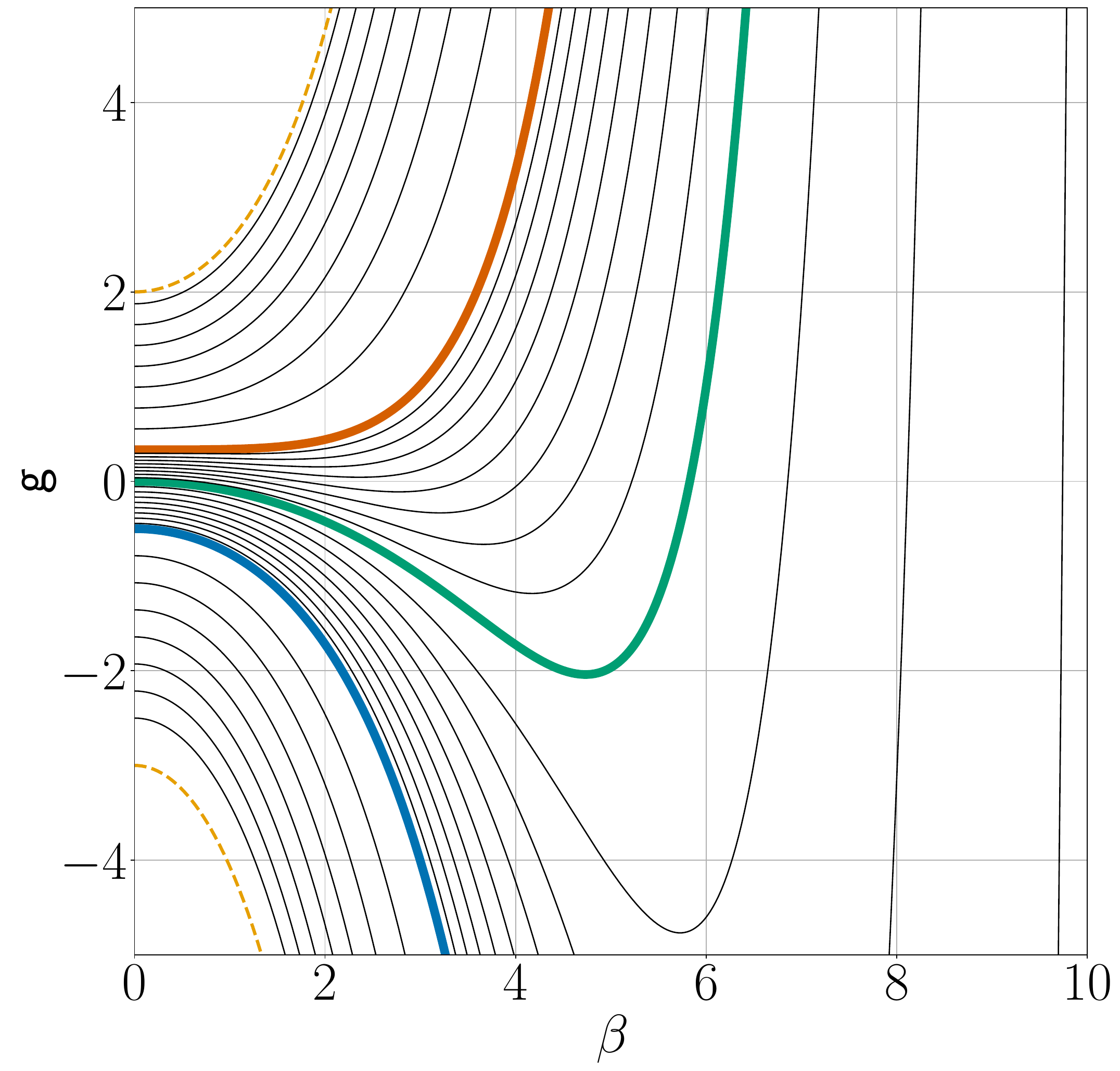} 
    \hfill
    \includegraphics[width=0.32\textwidth]{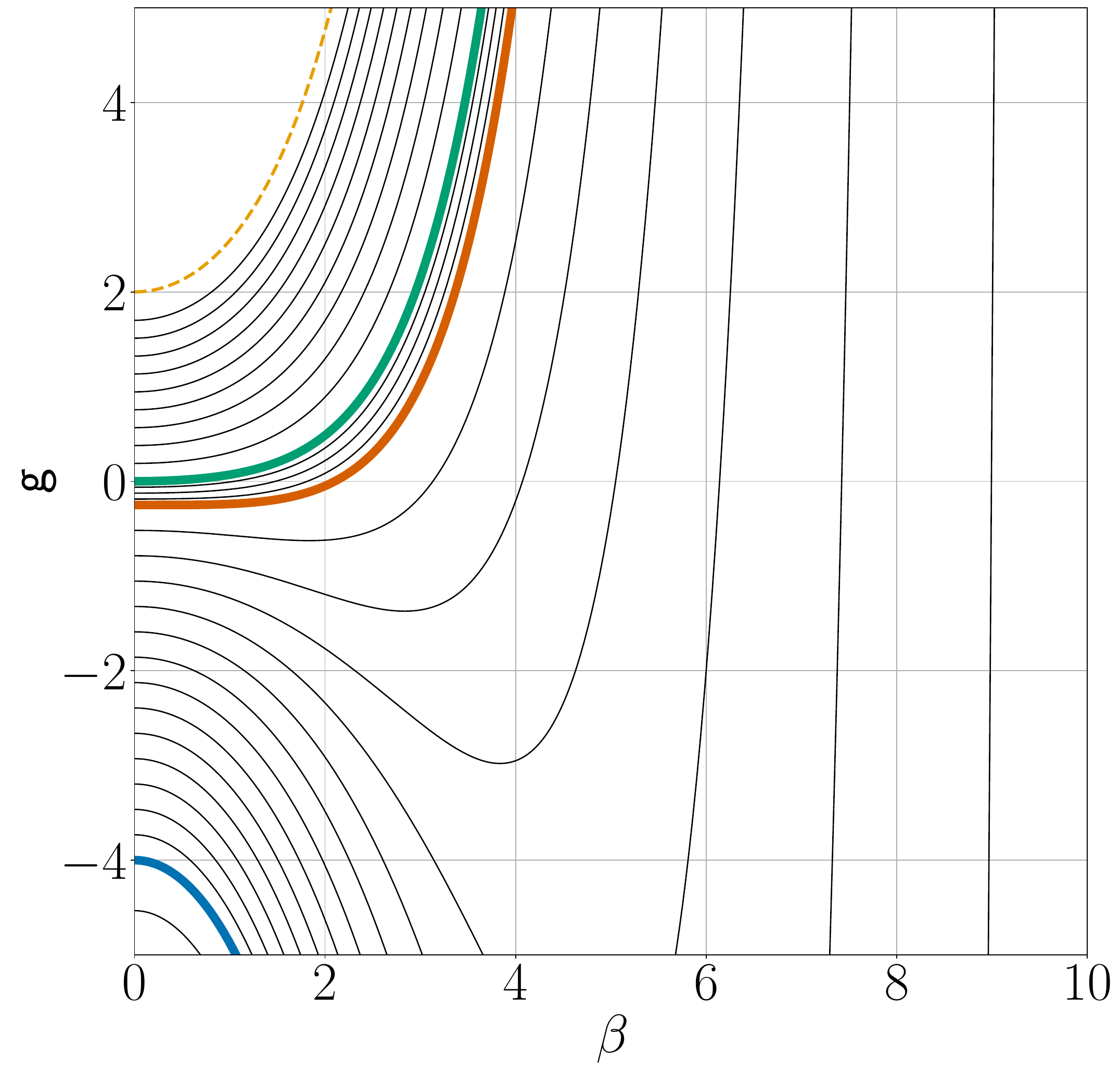} 
    \caption{The separating curves $ \eig \left( \eibeta ; \qeps_1 \right) $ (blue line), $ \eig \left( \eibeta ; \qeps_2 \right) $ (red line) and $ \eig \left( \eibeta ; \qeps_3 \right) $ (green line) together with some $ \eig \left( \eibeta ; \qeps \right) $ for arbitrary values of $ \qeps $ within the interval $ ( -1 , 1 ) $. The dashed orange lines denotes $ \eig \left( \eibeta ; \pm 1 \right) $. \emph{From left to right:} $ \qBi = 0.5 $, $ \qBi = 1.5 $ and $ \qBi = 5 $. As a visual guide, the initial value of the curves increases as the parameter $ \qeps $ decreases.}
    \label{fig:Im-eigval}
\end{figure}

Consequently, three important separating curves can be distinguished through the parameters
\begin{align}
    \qeps_1 = 0 , \qquad \qquad  
    \qeps_2 = - \frac{\qBi}{3 + \qBi} , \qquad \qquad  
    \qeps_3 = \frac{1 - \qBi}{1 + \qBi} ,
\end{align}
which read
\begin{align}
    \eig \left( \eibeta ; \qeps_1 \right) &= 1 - \qBi \frac{\sinh \eibeta}{\eibeta} , \\
    \eig \left( \eibeta ; \qeps_2 \right) &= 1 + \frac{\qBi}{3 + \qBi} \cosh \eibeta - \frac{3 \qBi }{3 + \qBi} \frac{\sinh \eibeta}{\eibeta} , \\
    \eig \left( \eibeta ; \qeps_3 \right) &= 1 - \frac{1 - \qBi}{1 + \qBi}  \cosh \eibeta - \frac{2\qBi}{1 + \qBi} \frac{\sinh \eibeta}{\eibeta} .
\end{align}
The possible existence of the roots and their number are determined by the relative position of these three curves. First let us observe that
\begin{align}
    &&
    \eig \left( 0 ; \qeps_1 \right) &= 1 - \qBi , &
    \text{which is positive, if} && \qBi &< 1 ,
    &&
    \\
    &&
    \eig \left( 0 ; \qeps_2 \right) &= 1 - \frac{2 \qBi }{3 + \qBi} \in ( - 1 , 1 ) , &
    \text{which is positive, if} && \qBi &< 3 ,
    &&
\end{align}
while $ \eig \left( 0 ; \qeps_3 \right) \equiv 0 $ (since the corresponding $ \qeps_3 $ parameter is defined through this relationship). The slope of $ \eig \left( \eibeta ; \qeps_1 \right) $ is negative and the slope of $ \eig \left( \eibeta ; \qeps_2 \right) $ is positive (which
can be easily proven through the Taylor expansion \re{ftp:beta-taylor}, since the Taylor coefficients evaluated via $ \qeps_2 $ are $ \frac{2 \qBi (k - 1)}{3 + \qBi} $ with $ k = 1 , 2 , \dots $) for all positive $ \eibeta $. In general, the slope of $ \eig \left( \eibeta ; \qeps_2 \right) $ is indefinite, however, for $ \qBi \le 1 $ it is strictly negative and for $ \qBi \ge 3 $ it is strictly positive. Let us furthermore note that if $ \qBi = 1 $, then $ \qeps_1 = \qeps_3 $ and $ \eig \left( \eibeta ; \qeps_1 \right) = \eig \left( \eibeta ; \qeps_3 \right) $, and if $ \qBi = 3 $, then $ \qeps_2 = \qeps_3 $ and $ \eig \left( \eibeta ; \qeps_2 \right) = \eig \left( \eibeta ; \qeps_3 \right) $. These are illustrated in Figure~\ref{fig:Im-eigval}.

Finally, one can draw the following conclusions:
\begin{itemize}
    \item If $ 0 < \qBi < 1 $ then $ \qeps_2 < \qeps_1 = 0 < \qeps_3 $; correspondingly,
    \begin{itemize}
        \item if $ \qeps_1 = 0 < \qeps < \qeps_3 $ then
        one root exists,
        \item if $ \qeps_2 < \qeps < \qeps_1 = 0 $
        then there is one root for a certain \m { \tilde\qeps }, and zero or two roots for \m { \qeps > \tilde\qeps } resp.\ \m { \qeps < \tilde\qeps };
        \item for all other $ \qeps $ values there
        are no roots.
    \end{itemize}
    \item If $ \qBi = 1 $ then $ \qeps_2 < \qeps_3 = \qeps_1 = 0 $; correspondingly $ \eig \left( \eibeta ; \qeps_1 \right) = \eig \left( \eibeta ; \qeps_3 \right) $ and 
    \begin{itemize}
        \item if $ \qeps_2 < \qeps < \qeps_3 = \qeps_1 = 0 $
        then there is one root for a certain \m { \tilde\qeps }, and zero or two roots for \m { \qeps > \tilde\qeps } or \m { \qeps < \tilde\qeps };
        \item for all other $ \qeps $ values
        there are no roots.
    \end{itemize}
    \item If $ 1 < \qBi < 3 $ then $ \qeps_2 < \qeps_3 < \qeps_1 = 0 $; correspondingly,
    \begin{itemize}
        \item if $ \qeps_3 < \qeps < \qeps_1 = 0 $ then one root exists,
        \item if $ \qeps_2 < \qeps < \qeps_3 $
        then there is one root for a certain \m { \tilde\qeps }, and zero or two roots for \m { \qeps > \tilde\qeps } or \m { \qeps < \tilde\qeps };
        \item for all other $ \qeps $ values there are no roots.
    \end{itemize}
    \item If $ \qBi = 3 $ then $ \qeps_2 = \qeps_3 < \qeps_1 = 0 $; correspondingly $ \eig \left( \eibeta ; \qeps_2 \right) = \eig \left( \eibeta ; \qeps_3 \right) $ and
    \begin{itemize}
        \item if $ \qeps_2 = \qeps_3 < \qeps < \qeps_1 = 0 $ then one root exists,
        \item for all other $ \qeps $ values there are no roots.
    \end{itemize}
    \item If $ \qBi > 3 $ then $ \qeps_3 < \qeps_2 < \qeps_1 = 0 $; correspondingly
    \begin{itemize}
        \item if $ \qeps_3 < \qeps < \qeps_1 = 0 $ then there is one root,
        \item for all other $ \qeps $ values no roots exist.
    \end{itemize}

\end{itemize}

In light of our previous statements, the roots can be determined using any standard numerical root-finding algorithm.

\subsection{Real eigenvalues}

\begin{figure}[!ht]
	\includegraphics[width=0.45\textwidth]{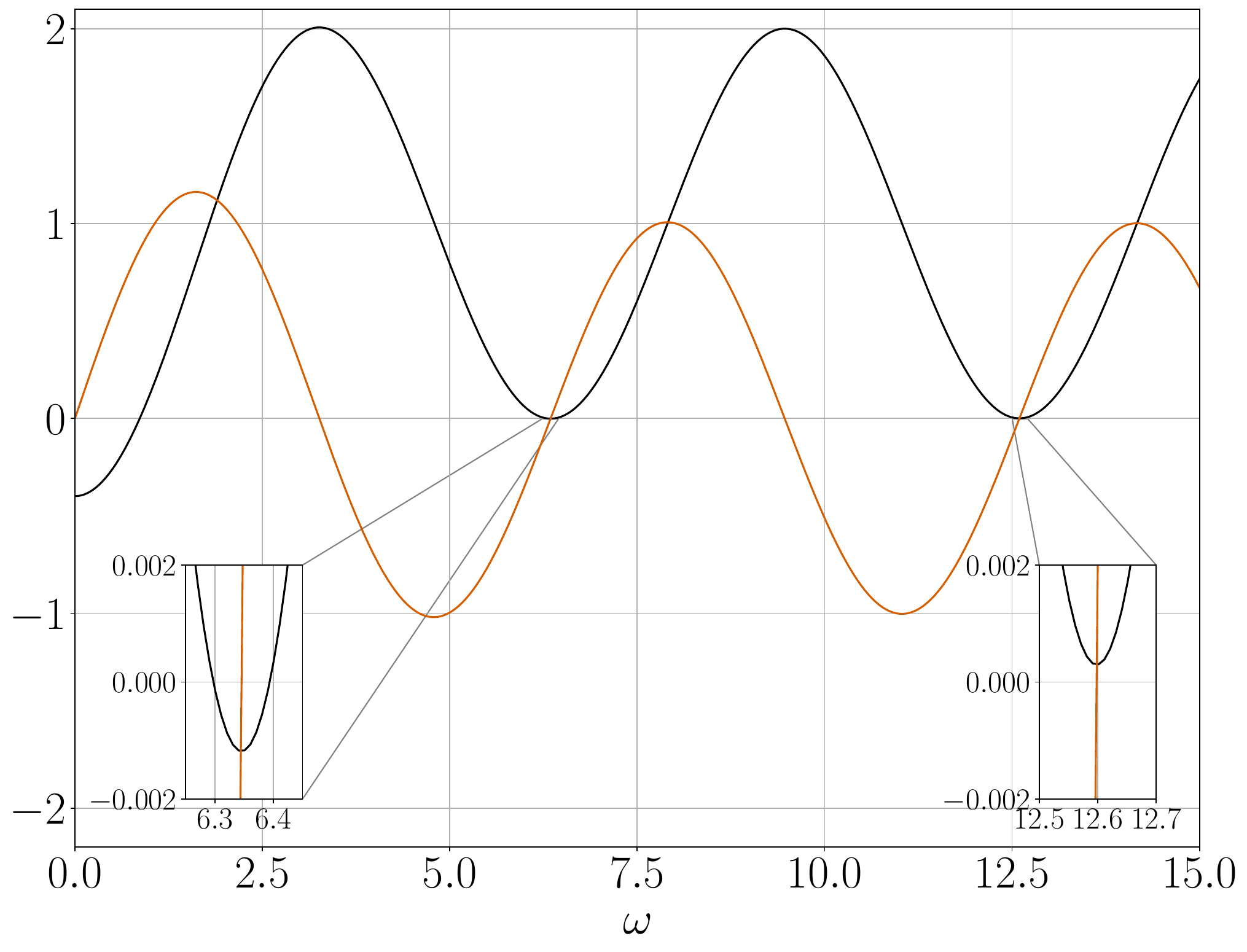} 
    \hfill
    \includegraphics[width=0.45\textwidth]{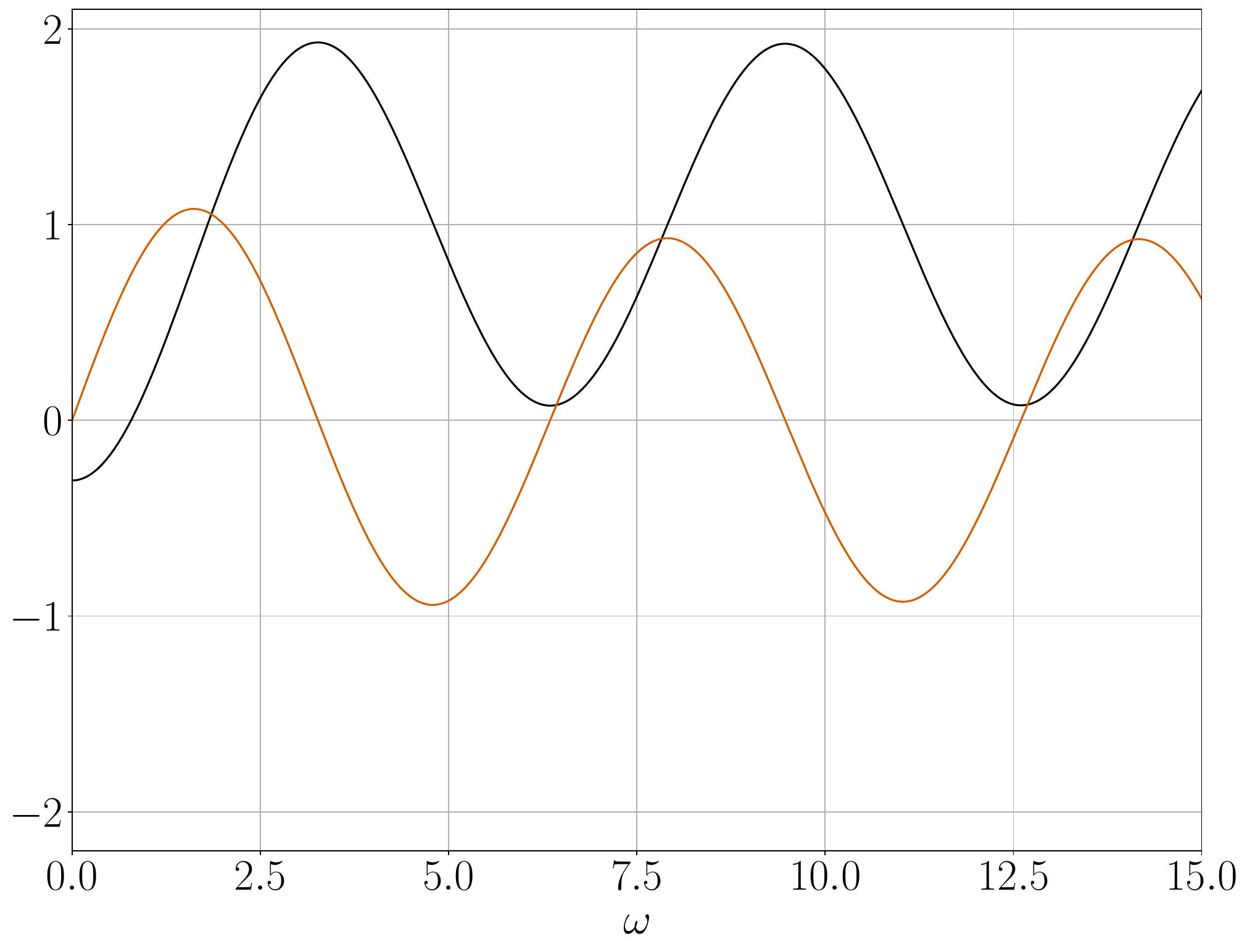} 
    \caption{The graph of $ \eif $ (black line) and $ \frac{\dd \eif}{\dd \omega} $ (red line) for the real variable $ \omega $. \emph{Left:} $ \qBi = 0.2 $, $ \qeps = 0.9992 $. \emph{Right:} $ \qBi = 0.2 $, $ \qeps = 0.9231 $.}
    \label{fig:Re-eigval}
\end{figure}

Since neither the existence nor the number of real
eigenvalue roots
is known, the transcendental equation \re{ftp:omega} and its derivative \wrt $ \omega $ are investigated simultaneously. Let us choose a ``small'' step $ \Delta \omega $ and evaluate $ \eif ( 0^{+} + k \Delta \omega ) $ and $ \frac{\dd \eif}{\dd \omega} ( 0^{+} + k \Delta \omega ) $ for $ k = 1, 2, \dots $ (here $ 0^{+} $ denotes a ``small'' positive number to avoid division by zero, which we omit in the following for the sake of clarity). If the derivative changes its sign between the steps $ k $ and $ k + 1 $, then $ \eif $ has a local extremum between $ k \Delta \omega $ and $ ( k + 1 ) \Delta \omega $. The location of this local extremum, denoted by $ \omega_{\mathrm{loc}}^{l} $ (with $ l = 1, 2, \dots $), is determined numerically via the bisection method within the interval $ [\, k \Delta \omega , \, (k+1) \Delta \omega \,]$, after which the function value $ \eif ( \omega_{\mathrm{loc}}^{l} ) $ is computed. Roots of $ \eif ( \omega ) $ may exist between two consecutive extrema if $ \operatorname{sgn} \eif ( \omega_{\mathrm{loc}}^{l} ) \neq \operatorname{sgn}\eif( \omega_{\mathrm{loc}}^{l+1} ) $. If this condition holds, then a root of $ \eif $ can be located using the bisection method within the interval $[\, \omega_{\mathrm{loc}}^{l} , \, \omega_{\mathrm{loc}}^{l+1} \,]$. The procedure is continued until the function values at two consecutive extrema have the same sign. These are illustrated in Figure~\ref{fig:Re-eigval}.

\subsection{Asymptotic behavior of the eigenvalues}

Let us now investigate the asymptotic behavior of \re{ftp:omega}. First, \re{ftp:omega} is multiplied by $ \omega $, then by the substitution $ \qomega = \eialpha + \ii \eibeta $ with $ \eialpha , \eibeta \in \mathbb{R} $ the real and imaginary parts are separated, finally, the obtained equation is divided by $ \eialpha $, which results 
\begin{align}
    \nonumber
    & \left( 1 - \left( \qeps \cos \eialpha + \frac{\qBi ( 1 + \qeps )}{\eialpha} \sin \eialpha \right) \cosh \eibeta - \frac{\eibeta}{\eialpha} \qeps \sin \eialpha \sinh \eibeta \right) \\
    \label{eq:tr-eq-asym}
    & \hskip 3ex + \ii \left( \frac{\eibeta}{\eialpha} + \left( \qeps \sin \eialpha - \frac{\qBi ( 1 + \qeps )}{\eialpha} \cos \eialpha \right) \sinh \eibeta - \frac{\eibeta}{\eialpha} \qeps \cos \eialpha \cosh \eibeta \right) = 0 .
\end{align}
When $ \eialpha \rightarrow \infty $ then \re{eq:tr-eq-asym} tends to
\begin{align}
    \left( 1 - \qeps \cos \eialpha \cosh \eibeta \right) + \ii \qeps \sin \eialpha \sinh \eibeta = 0 ,
\end{align}
which has to be fulfilled separately for the real and imaginary parts, \ie
\begin{align}
    \label{eq:tr-eq-asym-re}
    & \RE: \qquad \qquad
    1 - \qeps \cos \eialpha \cosh \eibeta = 0 , \\
    \label{eq:tr-eq-asym-im}
    & \IM: \qquad \qquad
    \qeps \sin \eialpha \sinh \eibeta = 0 .
\end{align}
Equation \re{eq:tr-eq-asym-im} is fulfilled if $ \eialpha_1 = 2 k \pi $ or $ \eialpha_2 = \pi + 2 k \pi $ with $ k = 0 , 1 , 2 , \dots $. Since $ \eibeta = \operatorname{arcosh} \frac{1}{\qeps \cos \eialpha} = \operatorname{arcosh} \frac{1}{\pm \qeps} \in \mathbb{R} $, consequently, if $ \qeps > 0 $ then $ \eialpha_1 $, in the opposite case $ \eialpha_2 $ have to be chosen. Summarizing, the asymptotic behavior of the eigenvalues reads
\begin{align}
    \label{eq:eig-asym}
    \omega_k^\infty =
    \begin{cases}
        2 k \pi \pm \ii \operatorname{arcosh} \frac{1}{\qeps} , & \text{if } \qeps > 0 , \\
        \left( \pi + 2 k \pi \right) \pm \ii \operatorname{arcosh} \frac{1}{- \qeps} , & \text{if } \qeps < 0 
    \end{cases}
\end{align}
with $ k \in \mathbb{Z}^+ $. Table~\ref{tab:eig-val} numerically
illustrates
these results. 
\1 2 {N%
aturally, much higher floating-point accuracy is advisable \1 1 {and has been used throughout this paper} -- the numbers in Table~\ref{tab:eig-val} have been truncated to already show the tendencies but to still fit into a compact and
easily comprehensible
Table.}

\begin{table}[!ht]
    \centering
    \renewcommand{\arraystretch}{1.2}
    \centering
    \begin{tabular}{c||c|c}
    & $ \qeps = -0.5 $ & $ \qeps = 0.5 $ \\
    \hline
    $ \omega_0 $ & & $ 1.8716 $ \\
    $ \omega_1 $ & $ \hphantom{1}2.5946 \pm 1.4410 \ii $ & $ \hphantom{1}6.8038 \pm 1.0411 \ii $ \\
    $ \omega_2 $ & $ \hphantom{1}9.2633 \pm 1.3250 \ii $ & $ 12.8879 \pm 1.2222 \ii $\\
    $ \omega_3 $ & $ 15.6120 \pm 1.3198 \ii $ & $ 19.0762 \pm 1.2711 \ii $ \\
    $ \omega_4 $ & $ 21.9228 \pm 1.3184 \ii $ & $ 25.3065 \pm 1.2903 \ii $ \\
    $ \omega_5 $ & $ 28.2212 \pm 1.3178 \ii $ & $ 31.5564 \pm 1.2996 \ii $
    \end{tabular}
    \hspace{2cm}
    \centering
    \begin{tabular}{c||c|c}
    & $ \qeps = -0.5 $ & $ \qeps = 0.5 $ \\
    \hline
    $ \omega_6 $ & $ 34.5141 \pm 1.3175 \ii $ & $ 37.8169 \pm 1.3048 \ii $ \\
    $ \omega_7 $ & $ 40.8039 \pm 1.3174 \ii $ & $ 44.0836 \pm 1.3080 \ii $ \\
    $ \omega_8 $ & $ 47.0920 \pm 1.3173 \ii $ & $ 50.3543 \pm 1.3101 \ii $ \\
    $ \omega_9 $ & $ 53.3790 \pm 1.3172 \ii $ & $ 56.6278 \pm 1.3115 \ii $ \\
    $ \omega_{10} $ & $ 59.6651 \pm 1.3171 \ii $ & $ 62.9031 \pm 1.3125 \ii $ \\
    $ \omega_{11} $ & $ 65.9507 \pm 1.3171 \ii $ & $ 69.1799 \pm 1.3133 \ii $
    \end{tabular}
    \caption{The first 11 complex
    eigenvalue root
    pairs of the transcendental equation \re{ftp:omega},
    obtained for $ \qeps = -0.5 $ and $ \qeps = 0.5 $, using $ \qBi = 1.5 $ in both cases. The imaginary part of the asymptotic 
    eigenvalue roots
    are $ \operatorname{arcosh} \frac{1}{0.5} = 1.3170 $. For $ \qeps = 0.5 $, one additional real
    eigenvalue root
    is present, which is denoted by $ \omega_0 $.}
    \label{tab:eig-val}
\end{table}

Practically, after determination of the purely imaginary and real 
eigenvalue roots,
the complex
eigenvalue roots
are sought for via the Newton iteration method initialized by the asymptotic values given in \re{eq:eig-asym}.

\section{Initial conditions generated by a pulse}  \label{FTinitial}

The question we address here is as follows: What is the temperature field \m { \qT } and the heat current density field \m { \qq } after a volumetric heat-up \1 1 {a radiative pulse penetrating into the sample, like in the flash experiment} in an MCV
modellable sample if initially temperature gradient and heat current density were zero, and during the pulse there was no considerable time for any locally added energy to diffuse?

The practical benefit of knowing these is that then an analytical or numerical calculation can be started from these fields as initial conditions for the rest of the process, while the boundary conditions and volumetric heat source are already \emp{much simpler} \1 1 {time-independent and zero, respectively}. In other words, we can transform the time-dependent external \1 1 {boundary + volumetric} effect of the pulse into initial conditions for the rest of the process. Naturally, this simplification would be available only under the assumptions above.

In formulation, the question is as follows. 
Corresponding to the sample size \m { \qxchar } as charcateristic length, there is a characteristic time scale, the diffusive
time scale,
 \begin{align}  \label{}
\qtdiff \defeq \f{\qxchar^2}{\FTa}
 \end{align}
\m { \2 1 {r\text{ecall } \FTa = \f{\qlambda}{\qrho \qc} }}.
The duration of the pulse \m { \qtp } is assumed to fulfil
 \begin{align}  \label{pulseshorterthanFourier}
\qtp \ll \qtdiff \,,  \qquad \text{in other form,} \qquad  \f{\qtp}{\qtdiff} \ll 1 \,.
 \end{align}
\1 1 {No condition will be needed for the MCV parameter  \m { \qtau }.}
Under this assumption, we are interested in
\mm { \qq \1 1 {\qtp, \qx}
}
and \mm { \qT \1 1 {\qtp, \qx}
}
if, for \mm { \qt \le 0 \,, }
 \begin{align}  \label{FTTjEnull}
 &&
\qT \1 1 {\qt, \qx} & = \qTnull \,,  &  \qq \1 1 {\qt, \qx} & = 0 \,.
 &&
 \end{align}
As a consequence, for \mm { \qt < 0 } \1 1 {and in the one-sided limit \mm { \qt = -0 \,,} in other customary notation,
\mm { \qt \mathrel{\vcenter{\hbox{$\scriptstyle \nearrow$}}} 0 }},
 \begin{align}  \label{FTpdpdtTjEnull}
 &&
\FTpdpdt{\qT} \1 1 {\qt, \qx} & = 0 \,,  &  \FTpdpdt{\qq} \1 1 {\qt, \qx} & = 0 \,.
 &&
 \end{align}

The local/differential energy balance considered here reads
 \begin{align}  \label{FTenergybalance}
\qrho \qc \FTpdpdt{\qT} = - \FTpdpdx{\qq} + \qqV \,,
 \end{align}
where the volumetric heating rate density \1 1 {volumetric heat source} distribution \m { \qqV } is assumed to be nonzero only for \mm { 0 \le \qt \le \qtp \,, } during which one
possible
example is penetration decreasing exponentially spatially -- typical for gradually absorbed irradiation that propagates very fast \1 1 {at the order of the speed of light} within
a homogeneous
sample --, \ie  
\mm {
\qqV \1 1 {\qt, \qx} =
\qqV \1 1 {\qt, 0}
\cdot
\exp{ \1 1 {{- \qx}/{\qxpexp}}}
 }
with a pulse penetration characteristic length \m { \qxpexp }. The considerations below won't assume
any
specific
form, nevertheless;
in particular, the
spatial pattern taken
in Section~\ref{flashinitcond} is
one
legitimate example covered here.

Before starting, let us make two preparatory steps that will prove beneficial in what follows. The first is that, instead of temperature \m { \qT }, one can also use the temperature increase \mm { \qTheta \defeq \qT - \qTnull } induced by the pulse. Since this is a shift by a constant, we have
 \begin{align}  \label{FTTTet}
 &&
\FTpdpdt{\qT} & = \FTpdpdt{\qTheta} \,,  &  \FTpdpdx{\qT} & = \FTpdpdx{\qTheta} \,.
 &&
 \end{align}
The second is that it will be advantageous to switch to the following dimensionless time and
space variables, respectively:
 \begin{align}  \label{dimlesstx}
 &&
\ct & \defeq \f{\qt}{\qtp} \,,  &   \hx &
= \f{\qx}{\qxchar} \,;
& \Longrightarrow &&
\FTpdpdt{} & = \f{1}{\qtp} \FTpdpdtdimless{} \,, & \FTpdpdx{} & = \f{1}{\qxchar} \FTpdpdxdimless{} \,.
 &&
\end{align}
With these, the energy balance \eqref{FTenergybalance} \1 1 {as an important example} becomes
 \begin{align}  \label{FTenergybalancedimless}
 &&
 &&
\qrho \qc \f{1}{\qtp} \FTpdpdtdimless{\qT} & = - \f{1}{\qxchar} \FTpdpdxdimless{\qq} + \qqV \,,  &
\FTpdpdtdimless{\qT} & = - \f{\qtp}{\qxchar} \f{1}{\qrho \qc} \FTpdpdxdimless{\qq} + \f{\qtp}{\qrho \qc} \qqV \,.
 &&
 &&
 \end{align}

\subsection{The Fourier case}  \label{FTinitialFourier}

Let us start with Fourier's model for heat conduction,
 \begin{align}  \label{FTFouriereqn}
&&
\qq & = - \qlambda \FTpdpdx{\qT}
 & \text{%
or,
in dimensionless space variable,}
 \qquad
\qq & = - \f{\qlambda}{\qxchar} \FTpdpdxdimless{\qT} \,.
&&
 \end{align}
This latter formula \ree{FTFouriereqn}{b} can be directly substituted into the energy balance \ree{FTenergybalancedimless}{b}, with the result -- written in terms of \m { \qTheta } --
 \begin{align}  \label{FTThetabalanceFourier}
\FTpdpdtdimless{\qTheta} = \f{\qtp}{\qxchar} \f{1}{\qrho \qc} \f{\qlambda}{\qxchar} \FTpdpdxxdimless{\qTheta} + \f{\qtp}{\qrho \qc} \qqV
= \f{\FTa \qtp}{\qxchar^2} \FTpdpdxxdimless{\qTheta} + \f{\qtp}{\qrho \qc} \qqV
 &  \nonumber \\
=
\underbrace{
\f{\qtp}{\qtdiff
\vphantom{|_|}  
}
}_{%
\makebox[2.6 em]
{\m{\scriptstyle \ll 1 \text{ \1 1 {cf.~\ree{pulseshorterthanFourier}{b}}}}}%
}
\FTpdpdxxdimless{\qTheta} + \f{\qtp}{\qrho \qc} \qqV
 \nonumber
 \\
\approx \f{\qtp}{\qrho \qc} \qqV & \,,
 \end{align}
where in the last step we neglected the term with coefficient \mm { \ll 1 }. This is the place where we impose our assumption that, during the pulse duration, there is negligible diffusion of internal energy via heat conduction. In closer detail: we have made our variables dimensionless with respect to the time and space scales \emp{of interest for us}. Then the corresponding dimensionless time and space derivatives, \m { \FTpdpdtdimless{\qTheta} } and \m { \FTpdpdxxdimless{\qTheta} }, are expected to be of the same order. A multiplier \m { \qtp/\qtdiff \ll 1 } then makes the \m { \FTpdpdxxdimless{\qTheta} }-containing term negligible with respect to the \m { \FTpdpdtdimless{\qTheta} } term.
In yet another presentation, one may assume the solution to be a power series in \m { \f{\qtp}{\qtdiff} },
 \begin{align}  \label{FTTetpower}  \textstyle
\qTheta \0 1 { \ct, \hx } =
\qTheta_0 \0 1 { \ct, \hx } +
\9 1 { \f{\qtp}{\qtdiff} } \qTheta_1 \0 1 { \ct, \hx } +
\9 1 { \f{\qtp}{\qtdiff} }^2  \qTheta_2 \0 1 { \ct, \hx } + \cdots \,.
 \end{align}
Substituting this ansatz into the middle row of \eqref{FTThetabalanceFourier} and imposing that the \lhss power series has to be equal to the \rhss power series term-by-term, we find
 \begin{align}  \label{FTpowzero}
\FTpdpdtdimless{\qTheta_0} = \f{\qtp}{\qrho \qc} \qqV
 \end{align}
as the leading-order approximation for \mm { \FTpdpdtdimless{\qTheta} \,.}

Integrating in the dimensionless time variable yields then
 \begin{align}  \label{FTThetaF}
\qTheta \0 1 { \ct, \hx } & \approx \f{\qtp}{\qrho \qc}
\int_0^{\ct} \dd \ct' \qqV  \0 1 { \ct', \hx } \,.
 \end{align}

The corresponding approximate time evolution of \m { \qq } can be obtained applying \ree{FTFouriereqn}{b} on this \1 1 {recall \mm { \FTa = \f{\qlambda}{\qrho \qc} }}:
 \begin{align}  \label{FTjEF}
\qq \0 1 { \ct, \hx } \approx
- \f{\FTa \qtp}{\qxchar} \int_0^{\ct} \dd \ct' \FTpdpdxdimless{\qqV} \0 1 { \ct', \hx } \,.
 \end{align}

Finally, taking \mm { \ct = 1 } in \eqref{FTThetaF}--\eqref{FTjEF} here provides \m { \qTheta } and \m { \qq }, respectively, at the end of the pulse, as the initial status for the rest of the process.

Rewriting \re{FTThetaF}--\re{FTjEF} to the original dimensional variables \m { \qt, \qx } is straightforward, via \mm { \qtp \dd \ct' = \dd \qt' \,, } \mm { \f{1}{\qxchar} \FTpdpdxdimless{} = \FTpdpdx{} \,. }

The interpretation of the results is that temperature increases locally due to the volumetric heat source only \1 1 {essentially}, there is no \1 1 {observable} diffusion \1 1 {spatial spreading} of energy. In parallel, heat current density merely obeys the Fourier-law prediction corresponding to the resulting temperature profile \1 1 {\mm { \qq = - \qlambda \FTpdpdx{\qTheta} }}.

A side remark here is that, in applications -- including flash experiment situations -- where the strength of the pulse is not known but only its spatial distribution is available, the multiplier standing before the integrand sign in \eqref{FTThetaF} \1 2 {or the corresponding one in \eqref{FTjEF}} is not important -- only the spatial distribution of \m { \qTheta } \1 1 {and of \m { \qq }} is important.

\subsection{The MCV case}  \label{FTinitialMCV}

In the MCV case, we have
 \begin{align}  \label{FTMCVeqn}
\qq + \qtau \FTpdpdt{\qq} = - \qlambda \FTpdpdx{\qT}
 \end{align}
with \mm { \qtau > 0 \,.}
\1 1 {As already mentioned, the
relationship of \m { \qtau } to \m { \qtp }, and to \m { \qtdiff }, can be arbitrary.}
Because of the presence of a time derivative on the \lhs, adapting the approach shown for the Fourier case would lead to a more complicated treatment. For a more step-by-step presentation, let us now start with what happens to \m { \qq }, instead.

The key observation will be that the \lhss here can be rewritten as
 \begin{align}  \label{FTobservation}
\qq + \qtau \FTpdpdt{\qq} = \qtau \ee^{- \f{\qt}{\qtau}} \FTpdpdt{} \9 1 { \ee^{\f{\qt}{\qtau}} \qq } \,.
 \end{align}
In the dimensionless time and space variables
\eqref{dimlesstx}
, \eqref{FTMCVeqn} is transformed to
 \begin{align}  \label{FTMCVeqndimless}
\qq + \f{\qtau}{\qtp} \FTpdpdtdimless{\qq} = - \qlambda \f{1}{\qxchar} \FTpdpdxdimless{\qT} \,.
 \end{align}
Differentiating the latter with respect to \m { \ct } enables us to eliminate \m { \qT }:
 \begin{align}  \label{FTMCVeqndimlessd}
\FTpdpdtdimless{} \9 1 { \qq + \f{\qtau}{\qtp} \FTpdpdtdimless{\qq} }
&
\stackrel{\hphantom{\text{\ree{FTenergybalancedimless}{b}}}}{=}
- \f{\qlambda}{\qxchar} \f{\partial^2 \qT}{\partial \ct \partial \hx} =
- \f{\qlambda}{\qxchar} \f{\partial^2 \qT}{\partial \hx \partial \ct}
 \nonumber
 \\
& \stackrel{\text{\ree{FTenergybalancedimless}{b}}}{=}
- \f{\qlambda}{\qxchar} \FTpdpdxdimless{} \9 1 { - \f{\qtp}{\qxchar} \f{1}{\qrho \qc} \FTpdpdxdimless{\qq} + \f{\qtp}{\qrho \qc} \qqV }
 \nonumber
 \\
&
\stackrel{\hphantom{\text{\ree{FTenergybalancedimless}{b}}}}{=}
\bitttttt  
\f{\FTa \qtp}{\qxchar^2}
\bitttttt  
\FTpdpdxxdimless{\qq}
- \f{\FTa \qtp}{\qxchar} \FTpdpdxdimless{\qqV}
 \nonumber
 \\
&
\stackrel{\hphantom{\text{\ree{FTenergybalancedimless}{b}}}}{=}
\underbrace{
\f{\qtp}{\qtdiff
\vphantom{|_|}  
}
}_{\ll 1
}
\FTpdpdxxdimless{\qq}
- \f{\FTa \qtp}{\qxchar} \FTpdpdxdimless{\qqV}
 \nonumber
 \\
&
\x 5.5 em  
\approx - \f{\FTa \qtp}{\qxchar} \FTpdpdxdimless{\qqV} \,,
 \end{align}
where in the last step we neglected the term with coefficient \mm { \ll 1 \,, } as before in the Fourier case.

Integrating the approximate equality
obtained in \eqref{FTMCVeqndimlessd} in the integration variable \m { \ct' } between \m { 0 } and \m { \ct } \1 1 {in short, \mm { \int_0^{\ct} \dd \ct' }} yields, with \mm { \ctau = \f{\qtau}{\qtp} },
 \begin{align}  \label{FTjEintegrA}
\0 1 { \qq + \ctau \FTpdpdtdimless{\qq} } \0 1 { \ct, \hx } -
\underbrace{ \0 1 { \qq + \ctau \FTpdpdtdimless{\qq} } \0 1 { 0, \hx } }_{ 0
\text{ \1 1 {cf.~\ree{FTTjEnull}{b}--\ree{FTpdpdtTjEnull}{b}}} } \approx
- \f{\FTa \qtp}{\qxchar} \int_0^{\ct} \dd \ct' \FTpdpdxdimless{\qqV}  \0 1 { \ct', \hx } \,.
 \end{align}
Applying on the \lhss the dimensionless version of \eqref{FTobservation}, we have
 \begin{align}  \label{FTobservationd}
\ctau \ee^{- \f{\ct}{\ctau}} \FTpdpdtdimless{} \9 1 { \ee^{\f{\ct}{\ctau}} \qq \0 1 { \ct, \hx } } \approx
- \f{\FTa \qtp}{\qxchar} \int_0^{\ct} \dd \ct' \FTpdpdxdimless{\qqV}  \0 1 { \ct', \hx } \,,
 \end{align}
or, after straightforward multiplication,
 \begin{align}  \label{FTobservationdd}
\FTpdpdtdimless{} \9 1 { \ee^{\f{\ct}{\ctau}} \qq \0 1 { \ct, \hx } } \approx
- \f{\FTa \qtp}{\ctau \qxchar}  \ee^{\f{\ct}{\ctau}} \int_0^{\ct} \dd \ct' \FTpdpdxdimless{\qqV}  \0 1 { \ct', \hx } \,.
 \end{align}
Integrating now in dimensionless time as \mm { \int_0^{\ct} \dd \ct'' } yields
 \begin{align}  \label{FTjEintegrB}
\ee^{\f{\ct}{\ctau}} \qq \0 1 { \ct, \hx } - \underbrace{ \qq \0 1 { 0, \hx } }_{ 0 \text{ \1 1 {cf.~\ree{FTTjEnull}{b}} } } \approx
- \f{\FTa \qtp}{\ctau \qxchar} \int_0^{\ct} \dd \ct''
\ee^{\f{\ct''}{\ctau}} \int_0^{\ct''} \dd \ct' \FTpdpdxdimless{\qqV}  \0 1 { \ct', \hx } \,,
 \end{align}
from which we find
 \begin{align}  \label{FTjEintegrBB}
\qq \0 1 { \ct, \hx } \approx
- \f{\FTa \qtp}{\ctau \qxchar} \int_0^{\ct} \dd \ct'' \bit
\ee^{\f{\ct'' - \ct}{\ctau}} \int_0^{\ct''} \dd \ct' \FTpdpdxdimless{\qqV} \0 1 { \ct', \hx } \,.
 \end{align}
This double integral is of the form \1 1 {focusing on the time dependence only; \m { \dot\FTintbypartsf } representing the factor \m { \ee^{\cdots} } and \m { \FTintbypartsg } representing \m { \int_0^{\ct''} \dd \ct' \cdots }}
 \begin{align}  \label{FTintbyparts}
\int_0^{\ct} \dd \ct''
\bit \dot{\FTintbypartsf} \0 1 { \ct'' } \bit \FTintbypartsg \0 1 { \ct'' }
& = 
\int_0^{\ct} \dd \ct'' \0 2 {
\9 1 {\FTintbypartsf \FTintbypartsg} \dot{} - \FTintbypartsf \dot{\FTintbypartsg}
\vphantom{\big]}  
} \0 1 { \ct'' }
=
\2 2 { \FTintbypartsf \FTintbypartsg
}_{0}^{\ct}
- \int_0^{\ct} \dd \ct''
\bit \FTintbypartsf \0 1 { \ct'' } \bit \dot{\FTintbypartsg} \0 1 { \ct'' }
 \qquad \nonumber
 \\
& =
\FTintbypartsf \0 1 { \ct } \FTintbypartsg \0 1 { \ct } -
\FTintbypartsf \0 1 { 0 } \FTintbypartsg \0 1 { 0 }
- \int_0^{\ct} \dd \ct'
\bit \FTintbypartsf \0 1 { \ct' } \bit \dot{\FTintbypartsg} \0 1 { \ct' } \,,
 \end{align}
where we have applied integration by parts, and overdot abbreviates time derivative. Thanks to this, \eqref{FTjEintegrBB} turns out to be rewritable as the single integral
 \begin{align}  \label{FTjEintegrBBB}
\qq \0 1 { \ct, \hx } \approx
\f{\FTa \qtp}{\qxchar} \int_0^{\ct} \dd \ct' \bit
\9 1 { \ee^{\f{\ct' - \ct}{\ctau}} - 1 } \FTpdpdxdimless{\qqV} \0 1 { \ct', \hx } \,.
 \end{align}
It is worth noting that combining this with \eqref{FTjEintegrA} provides \mm { \FTpdpdtdimless{\qq} \0 1 { \ct, \hx } } as well:
 \begin{align}  \label{FTjEintegrAB}
\FTpdpdtdimless{\qq} \0 1 { \ct, \hx } \approx
- \f{\FTa \qtp}{\ctau \qxchar} \int_0^{\ct} \dd \ct' \bit
\ee^{\f{\ct' - \ct}{\ctau}} \FTpdpdxdimless{\qqV} \0 1 { \ct', \hx } \,.
 \end{align}
This, naturally, coincides with the outcome of directly differentiating \eqref{FTjEintegrBBB} with respect to \m { \ct }. \1 1 {Note that the upper limit is also \m { \ct } dependent. However, its contribution happens to be zero.}

Finally, taking \mm { \ct = 1 } in \eqref{FTjEintegrBBB}--\eqref{FTjEintegrAB} provides \m { \qq } and \m { \FTpdpdtdimless{\qq} }, respectively, at the end of the pulse.

Turning towards the temperature increase \mm { \qTheta = \qT - \qTnull } induced by the pulse, \ree{FTenergybalancedimless}{b} remains valid in the substitution \mm { \qT \rightarrow \qTheta }, which then can be integrated in \m {\ct} to reach \m { \qTheta }. The details are as follows.

With \mm { \qT \rightarrow \qTheta } in \ree{FTenergybalancedimless}{b},
and substituting \eqref{FTjEintegrBBB}, we have
 \begin{align}  \label{FTThetabalance}
\FTpdpdtdimless{\qTheta} \0 1 { \ct, \hx } & \approx - \f{\qtp}{\qxchar} \f{1}{\qrho \qc}
\f{\FTa \qtp}{\qxchar} \int_0^{\ct} \dd \ct' \bit
\9 1 { \ee^{\f{\ct' - \ct}{\ctau}} - 1 }
\FTpdpdxxdimless{\qqV}
\0 1 { \ct', \hx }
+ \f{\qtp}{\qrho \qc} \qqV \0 1 { \ct, \hx }
 \nonumber
 \\
&
\quad
= \f{\qtp}{\qrho \qc} \9 2 { -
\underbrace{
\f{\qtp}{\qtdiff
}
}_{\ll 1}
\int_0^{\ct} \dd \ct' \bit \9 1 { \ee^{\f{\ct' - \ct}{\ctau}} - 1 }
\FTpdpdxxdimless{\qqV}
\0 1 { \ct', \hx }
+ \qqV \0 1 { \ct, \hx } }
\bitttttt  
 \nonumber
 \\
& \quad \approx \f{\qtp}{\qrho \qc} \qqV \0 1 { \ct, \hx }
\,.
 \end{align}
Here, again, the diffusion-related term drops out in the \mm { {\qtp}/{\qtdiff} \ll 1 } approximation.
Integrating in the dimensionless time variable yields then
 \begin{align}  \label{FTTheta}
\qTheta \0 1 { \ct, \hx } & \approx \f{\qtp}{\qrho \qc}
\int_0^{\ct} \dd \ct' \qqV  \0 1 { \ct', \hx } \,.
 \end{align}
Thes outcomes \eqref{FTThetabalance}--\eqref{FTTheta} are the same as \eqref{FTTheta} in the Fourier case. For \m { \qq }, on the other side, there is significant difference between the two cases.

Again, taking \mm { \ct = 1 } tells \m { \qTheta } and \m { \FTpdpdtdimless{\qTheta} }, respectively, at the end of the pulse.

Rewriting the results to the dimensional variables \m { \qt, \qx } is straightforward as before, via \mm { \qtp \dd \ct' = \dd \qt' \,, } \mm { \f{1}{\qxchar} \FTpdpdxdimless{} = \FTpdpdx{} \,. }

The interpretation of the MCV results is that temperature increases only locally again, and heat current density is the solution of \re{FTMCVeqn} as an ordinary differential equation in time, with the temperature profile prescribed by the local increase induced by the volumetric heat source.

Now, returning to the dimensional variables and considering the specific volumetric heat source introduced in \re{eq:q_V-flash}, the temperature and heat current density fields just after the heat pulse can be calculated from \re{FTThetabalance} and \re{FTjEintegrBBB}:
\begin{align}  \label{HeaC}
    \Theta ( \qtp , \qx ) & = \frac{1}{\qrho \qc} \int\limits_{0^-}^{\qtp} \frac{\qqA^{}}{\qxp} \delta ( \qt' ) \2 2 { 1 - H ( \qx - \qxp ) } \dd \qt' = \frac{\qqA^{}}{\qrho \qc \qxp} \2 2 { 1 - H ( \qx - \qxp ) } ,
 \\  \label{HeaD}
    \qq ( \qtp , \qx ) & = \FTa \int\limits_{0^-}^{\qtp} \left( \ee^{\frac{\qt' - \qtp}{\qtau}} - 1 \right) \frac{\qqA^{}}{\qxp} \delta ( \qt' ) \frac{\pd}{\pd \qx} \2 2 { 1 - H ( \qx - \qxp ) } \dd \qt' = 
    \FTa \frac{\qqA^{}}{\qxp}
    \left( 1 - \ee^{- \frac{\qtp}{\qtau}} \right) \frac{\pd}{\pd \qx} H ( \qx - \qxp ) .
\end{align}
\newline
In the limit of a temporal Dirac delta excitation, \ie $ \qtp \to 0 $, we have \mm {  1 - \ee^{- \frac{\qtp}{\qtau}} \to 0}, and
the initial conditions
\begin{align}  \label{HeaA}
    \qTheta ( 0^+ , \qx ) &= \frac{\qqA^{}}{\qrho \qc \qxp} \2 2 { 1 - H ( \qx - \qxp ) } , \\  \label{HeaB}
    \qq ( 0^+ , \qx ) &= 0 
\end{align}
are obtained.

The outcome on \m { \qq } is, actually, independent of the spatial penetration profile of the pulse, as long as the pulse is Dirac delta-type in time.

\section{On the finite-difference numerical solution} \label{sec:num-meth}

The numerical solutions presented in Section~\ref{sec:num-ex} to `verify' the analytical results are computed using an implicit numerical integration scheme on a spatially staggered grid. Let $ \Delta \hht $ and $ \Delta \hx $ denote the non-dimensional time and space steps, respectively \1 2 {along \re{nondim} again}. Consequently, the discrete time and space coordinates are defined as $ \hht^j = j \Delta \hht , \ j = 0 , \dots , J $ and $ \hx_n = n \Delta \hx , \ n = 0 , \dots , N $ with $ J , N \in \mathbb{N} $. Spatial discretization is implemented on a staggered grid; thus, the temperature and heat current density values are shifted by a half space step relative to each other, yielding the notation
\begin{align}
    &&
    \qtheta \9 1 { \hht^j_{} , \hx_{n + \nicefrac{1}{2}}^{} } &
    \eqdef
    \qtheta^j_{n + \nicefrac{1}{2}} , &
    \qchi \9 1 { \hht^j_{} , \hx_n^{} } &
    \eqdef
    \qchi^j_n .
    &&
\end{align}
The partial time derivatives are approximated using a weighted one-step scheme, while a second-order accurate central difference method is applied to the spatial derivatives. Consequently, the discretized forms of the energy balance \re{eq:MCV-nondim-1} and the constitutive equation \re{eq:MCV-nondim-2} read as
\begin{align}
    \label{eq:MCV-disc-1}
    \frac{\qtheta_{n + \nicefrac{1}{2}}^{j+1} - \qtheta_{n + \nicefrac{1}{2}}^{j}}{\Delta \hht} &= \left( 1 - \varphi \right) \frac{\qchi_{n}^j - \qchi_{n + 1}^j }{\Delta \hx} + \varphi \frac{\qchi_{n}^{j+1} - \qchi_{n + 1}^{j+1} }{\Delta \hx} , \\
    \label{eq:MCV-disc-2}
    \htau \frac{\qchi_n^{j+1} - \qchi_n^j}{\Delta \hht} + \left[ \left( 1 - \varphi \right) \qchi_n^j + \varphi \qchi_n^{j+1} \right] &= \left( 1 - \varphi \right) \frac{\qtheta_{n - \nicefrac{1}{2}}^j - \qtheta_{n + \nicefrac{1}{2}}^j }{\Delta \hx} + \varphi \frac{\qtheta_{n-\nicefrac{1}{2}}^{j+1} - \qtheta_{n + \nicefrac{1}{2}}^{j+1} }{\Delta \hx} ,
\end{align}
where the parameter $ 0 \le \varphi \le 1 $ controls the degree of impliciteness.

The choice of the spatial grid layout is dictated by the boundary conditions applied at the domain boundaries. At the left boundary (\ie $ \hx = 0 $), a
zero
BC2 \1 1 {prescribed zero heat current density}
is
imposed
according to \re{ftj}. This allows the corresponding heat current node to be fixed to zero (and thus eliminated from the degrees of freedom). Conversely, the right boundary (\ie $ \hx = 1 $) is subjected to a
BC3 \1 1 {boundary heat current density is proportional to boundary temperature excess \wrt ambient temperature},
for which we introduce a ghost cell in a way that we
maintain second-order spatial accuracy. To implement this condition, the physical boundary at $ \hx = 1 $ is aligned with the last temperature node $ \qtheta_{N-\nicefrac{1}{2}} $
(consequently, the spatial domain is divided into a non-integer number ($ N-\nicefrac{1}{2} $) of grid cells).
Discretizing the boundary condition \re{ftk} at this position yields
\begin{align}
    \label{eq:BC-x=1-disc}
    \qchi_{N - \nicefrac{1}{2}}^j = \qBi \qtheta_{N - \nicefrac{1}{2}}^j .
\end{align}
However, due to the half-step shift inherent to the staggered grid layout, no explicit heat current node exists exactly at this boundary.
In order to
resolve this, the boundary heat
current density
is approximated as the average of the
closest
internal heat current node and an external ghost node value, defined as
\begin{align}
    \label{eq:BC-x=1-ave}
    \qchi_{N - \nicefrac{1}{2}}^j = \frac{\qchi_{N - 1}^j + \qchi_{N}^j}{2} .
\end{align}
By combining \re{eq:BC-x=1-disc} and \re{eq:BC-x=1-ave}, the fictitious heat current density value at the ghost node $ \qchi_{N}^j $ can be expressed as
\begin{align}
    \qchi_{N}^j = 2 \qBi \qtheta^j_{N - \nicefrac{1}{2}} - \qchi_{N - 1}^j .
\end{align}
Finally, substituting this ghost value into the discretized energy balance equation \re{eq:MCV-disc-1} eliminates the fictitious node,
ensuring a closed, second-order accurate system of equations at the right boundary,
\begin{align}
    \frac{\qtheta_{N - \nicefrac{1}{2}}^{j+1} - \qtheta_{N - \nicefrac{1}{2}}^{j}}{\Delta \hht} &= \left( 1 - \varphi \right) \frac{2 \left( \qchi_{N - 1}^j - \qBi \qtheta^j_{N - \nicefrac{1}{2}} \right)}{\Delta \hx} + \varphi \frac{2 \left( \qchi_{N - 1}^{j+1} - \qBi \qtheta^{j+1}_{N - \nicefrac{1}{2}} \right)}{\Delta \hx} .
\end{align}

Introducing the vector $ \tensorr{Y} $ and the matrix $ \tensorr{A} $ as
\begin{align}
    \tensorr{Y}^j &= \begin{pmatrix} \qtheta_{\nicefrac{1}{2}}^{j} \\ \qchi_{1}^{j} \\ \qtheta_{\nicefrac{3}{2}}^{j} \\ \qchi_{2}^{j} \\ \vdots \\ \qchi_{N - 2}^{j} \\ \qtheta_{N - \nicefrac{3}{2}}^{j} \\ \qchi_{N - 1}^{j} \\ \qtheta_{N - \nicefrac{1}{2}}^{j} \end{pmatrix} , &
    \tensorr{A} &=
    \begin{pmatrix}
        0 & - \frac{\Delta \hht}{\Delta \hx} \\
        \frac{\Delta \hht}{\ctau \Delta \hx} & - \frac{\Delta \hht}{\ctau} & - \frac{\Delta \hht}{\ctau \Delta \hx} \\
        & \frac{\Delta \hht}{\Delta \hx} & 0 & - \frac{\Delta \hht}{\Delta \hx} \\
        & & \frac{\Delta \hht}{\ctau \Delta \hx} & - \frac{\Delta \hht}{\ctau} & - \frac{\Delta \hht}{\ctau \Delta \hx} \\
        & & & \ddots & \ddots & \ddots \\
        & & & & \frac{\Delta \hht}{\ctau \Delta \hx} & - \frac{\Delta \hht}{\ctau} & - \frac{\Delta \hht}{\ctau \Delta \hx} \\
        & & & & &  \frac{\Delta \hht}{\Delta \hx} & 0 & - \frac{\Delta \hht}{\Delta \hx} \\
        & & & & & & \frac{\Delta \hht}{\ctau \Delta \hx} & - \frac{\Delta \hht}{\ctau} & - \frac{\Delta \hht}{\ctau \Delta \hx} \\
        & & & & & & & 2 \frac{\Delta \hht}{\Delta \hx} & - 2 \qBi \frac{\Delta \hht}{\Delta \hx}
    \end{pmatrix}
\end{align}
the system of difference equations is given in the form
\begin{align}
    \tensorr{Y}^{j+1} = \tensorr{Y}^{j} + \tensorr{A} \left[ ( 1 - \varphi) \tensorr{Y}^{j} + \varphi \tensorr{Y}^{j+1} \right] ,
\end{align}
from which the values of the field variables in the $ j + 1 $-th time step can be calculated as
\begin{align}
    \tensorr{Y}^{j+1} = \left( \tensorr{1} - \varphi \tensorr{A} \right)^{-1} \left[ \tensorr{1} + ( 1 - \varphi)\tensorr{A} \right] \tensorr{Y}^{j} ,
\end{align}
where $ \tensorr{1} $ denotes the $ 2 N \times 2 N $ identity matrix.

Selecting $ \varphi = \frac{1}{2} $ corresponds to the Crank--Nicolson method, which provides a second-order accurate approximation in time. In principle, this choice would be preferred. However, our numerical experience indicated that while the implicit Euler scheme ($ \varphi = 1 $) introduces only a slight smoothing of the wave front due to numerical dissipation, the Crank--Nicolson method suffers from artificial oscillations triggered by numerical dispersion. Consequently, 
in order to suppress these dispersion errors and preserve the physical fidelity of the wave-like profiles, the numerical computations presented throughout this work were performed using the implicit Euler method.

\bibliography{mcv-dual}

@article{auriault2016cattaneo,
  title={Cattaneo--{V}ernotte equation versus {F}ourier thermoelastic hyperbolic heat equation},
  author={Auriault, J.-L.},
  journal={International Journal of Engineering Science},
  volume={101},
  pages={45--49},
  year={2016},
  doi={10.1016/j.ijengsci.2015.12.002},
  publisher={Elsevier}
}

@article{badeau2025spectral,
    author  = {Badeau, Roland},
    title   = {On the spectral decomposition of the complex {R}obin {L}aplacian},
    journal = {The Journal of the Acoustical Society of America},
    volume  = {158},
    number  = {1},
    pages   = {838-848},
    year    = {2025},
    month   = {07},
    issn    = {0001-4966},
    doi     = {10.1121/10.0037233},
    note    = {Post-publication corrections in
https://perso.telecom-paristech.fr/rbadeau/Badeau-JASA-2025-preprint3.pdf}
}

@article{bogli2022eigenvalues,
author    = {B{\"{o}}gli, Sabine and Kennedy, James B. and Lang, Robin},
title     = {On the eigenvalues of the {R}obin {L}aplacian with a complex parameter},
journal   = {Analysis and Mathematical Physics},
issn      = {1664-235X},
volume    = {12},
year      = {2022},
issue     = {1},
pages     = {39},
numpages  = {60},
doi       = {10.1007/s13324-022-00646-0},
url       = {https://link.aps.org/doi/10.1103/PhysRev.148.778}
}

@article{cape1963temperature,
  title={Temperature and finite pulse-time effects in the flash method for measuring thermal diffusivity},
  author={Cape, J. A. and Lehman, G. W.},
  journal={Journal of Applied Physics},
  volume={34},
  number={7},
  pages={1909--1913},
  year={1963},
  doi={10.1063/1.1729711},
  publisher={American Institute of Physics}
}

@article{carr2025modelling,
  title={Modelling and analysis of laser flash experiments using the {C}attaneo heat equation},
  author={Carr, Elliot J.},
  journal={Applied Mathematical Modelling},
  volume={155},
  pages={116727},
  year={2025},
  doi={10.1016/j.apm.2025.116727},
  publisher={Elsevier}
}

@book{carslaw1959conduction,
    title     = {Conduction of heat in solids},
    author    = {Carslaw, H. S. and Jaeger, J. C.},
    year      = {1959},
    publisher = "Oxford University Press",
    address   = "London",
    edition   = {2}
}

@article{cattaneo1948sulla,
    author = {Cattaneo, Carlo},
    title = {Sulla conduzione del calore},
    journal = {Atti del Seminario Matematico e Fisico dell'Universit{\`{a}} di Modena},
    volume = {3},
    pages = {83--101},
    year = {1948},
    doi = {10.1007/978-3-642-11051-1_5}
}

@article{chandrasekharaiah1998hyperbolic,
  title     = {Hyperbolic thermoelasticity: A review of recent literature},
  author    = {Chandrasekharaiah, D. S.},
  journal   = {Applied Mechanics Reviews},
  volume    = {51},
  number    = {12},
  pages     = {705--729},
  year      = {1998},
  month     = {12},
  doi       = {10.1115/1.3098984},
  publisher = {Elsevier}
}

@article{cowan1963pulse,
  title={Pulse method of measuring thermal diffusivity at high temperatures},
  author={Cowan, Robert D.},
  journal={Journal of Applied Physics},
  volume={34},
  number={4},
  pages={926--927},
  year={1963},
  doi={10.1063/1.1729564},
  publisher={American Institute of Physics}
}

@article{feher2024dynamic,
    author   = {Feh{\'{e}}r, Anna and Kov{\'{a}}cs, R{\'{o}}bert},
    title    = {On the dynamic thermal conductivity and diffusivity observed in heat pulse experiments},
    journal  = {Journal of Non-Equilibrium Thermodynamics},
    volume   = {49},
    pages    = {161--170},
    year     = {2024},
    issn     = {0340-0204},
    doi      = {10.1515/jnet-2023-0119}
}

@article{feher2024thermal,
    author   = {Feh{\'{e}}r, A. and Mar{\'{o}}ti, J. E. and Tak{\'{a}}cs, D. M. and Orbulov, I. N. and Kov{\'{a}}cs, R.},
    title    = {Thermal and mechanical properties of {A}l{S}i7{M}g matrix syntactic foams reinforced by {A}l$_2${O}$_3$ or {S}i{C} particles in matrix},
    journal  = {International Journal of Heat and Mass Transfer},
    volume   = {226},
    pages    = {125446},
    year     = {2024},
    issn     = {0017-9310},
	eissn    = {1879-2189},
    doi      = {10.1016/j.ijheatmasstransfer.2024.125446}
}

@book{fourier1822theorie,
title={Th{\'{e}}orie analytique de la chaleur},
author={Fourier, J.},
year={1822},
publisher={Chez Firmin Didot, P{\`e}re et Fils},
series={Libraires pour les Math{\'{e}}matiques, l'Artchitecture Hydraulique et la Marine, Rue Jacob, No. 24}
}

@article{fulop2018emergence,
author   = {F{\"u}l{\"o}p, Tam{\'a}s and Kov{\'a}cs, R{\'o}bert and Lovas, {\'A}d{\'a}m and Rieth, {\'A}gnes and Fodor, Tam{\'a}s and Sz{\"u}cs, M{\'a}ty{\'a}s and V{\'a}n, P{\'e}ter and Gr{\'o}f, Gyula},
title    = {Emergence of Non-{F}ourier Hierarchies},
journal  = {Entropy},
year     = {2018},
volume   = {20},
number   = {11},
pages    = {832},
numpages = {13},
doi      = {10.3390/e20110832}
}

@article{gal2024thermal,
  title={Thermal diffusity in copper benzene-1,3,5-tricarboxylate--reduced graphite oxide mechanical composites},
  author={G{\'{a}}l, M{\'{a}}rton and Samaniego Andrade, Samantha K. and Feh{\'{e}}r, Anna and Farkas, Attila and Madar{\'{a}}sz, J{\'{a}}nos and Horv{\'{a}}th, Lili and Gordon, P{\'{e}}ter and Kov{\'{a}}cs, R{\'{o}}bert and L{\'{a}}szl{\'{o}}, Krisztina},
  journal={Journal of Thermal Analysis and Calorimetry},
  volume={149},
  number={12},
  pages={5971--5983},
  year={2024},
  doi={10.1007/s10973-024-13021-x},
  publisher={Springer}
}

@inbook{grad1958principles,
author    = {Grad, Harold},
title     = {Principles of the Kinetic Theory of Gases},
pages     = {205--294},
editor    = {Fl{\"{u}}gge, S.},
booktitle = {Encyclopedia of Physics},
volume    = {3},
year      = {1958},
publisher = {Springer-Verlag},
address   = {Berlin G{\"{o}}ttingen Heidelberg},
isbn      = {978-3-642-45894-1},
doi       = {10.1007/978-3-642-45892-7}
}

@article{guyer1966solution,
author    = {Guyer, R. A. and Krumhansl, J. A.},
title     = {Solution of the linearized phonon {B}oltzmann equation},
journal   = {Physical Review},
volume    = {148},
year      = {1966},
issue     = {2},
pages     = {766--778},
doi       = {10.1103/PhysRev.148.766},
url       = {https://link.aps.org/doi/10.1103/PhysRev.148.766},
publisher = {American Physical Society}
}

@article{jeffreys1917viscosity,
    author  = {Jeffreys, Harold},
    title   = {The Viscosity of the {E}arth. ({T}hird {P}aper.)},
    journal = {Monthly Notices of the Royal Astronomical Society},
    volume  = {77},
    number  = {5},
    pages   = {449--456},
    year    = {1917},
    month   = {03},
    issn    = {0035-8711},
    doi     = {10.1093/mnras/77.5.449},
    eprint  = {https://academic.oup.com/mnras/article-pdf/77/5/449/4098268/mnras77-0449.pdf}
}

@article{jou1988extended,
doi = {10.1088/0034-4885/51/8/002},
url = {https://doi.org/10.1088/0034-4885/51/8/002},
year = {1988},
month = {aug},
volume = {51},
number = {8},
pages = {1105--1179},
author = {Jou, D. and Casas-V{\'{a}}zquez, J. and Lebon, G.},
title = {Extended irreversible thermodynamics},
journal = {Reports on Progress in Physics}
}

@article{kapitza1941heat,
  title={Heat transfer and superfluidity of helium {II}},
  author={Kapitza, P. L.},
  journal={Physical Review},
  volume={60},
  number={4},
  pages={354--355},
  year={1941},
  doi={10.1103/PhysRev.60.354},
  publisher={APS}
}

@article{kostenbauder1997eigenmode,
    author    = {Kostenbauder, Adnah and Sun, Yan and Siegman, A. E.},
    title     = {Eigenmode expansions using biorthogonal functions: complex-valued {H}ermite--{G}aussians},
    journal   = {Journal of the Optical Society of America A},
    publisher = {Optica Publishing Group},
    volume    = {14},
    number    = {8},
    pages     = {1780--1790},
    year      = {1997},
    month     = {Aug},
    url       = {https://opg.optica.org/josaa/abstract.cfm?URI=josaa-14-8-1780},
    doi       = {10.1364/JOSAA.14.001780}
}

@article{kovacs2024heat,
author   = {Kov{\'{a}}cs, R{\'{o}}bert},
title    = {Heat equations beyond {F}ourier: {F}rom heat waves to thermal metamaterials},
journal  = {Physics Reports},
volume   = {1048},
year     = {2024},
pages    = {1--75},
doi      = {10.1016/j.physrep.2023.11.001}
}

@article{krejcirik2006closed,
  author  = {Krej{\v{c}}i{\v{r}}{\'{\i}}k, D. and B{\'{\i}}la, H. and Znojil, M.},
  journal = {Journal of Physics A: Mathematical and General},
  number  = {32},
  pages   = {10143},
  title   = {Closed formula for the metric in the {H}ilbert space of a {PT}-symmetric model},
  volume  = {39},
  month   = {jul},
  year    = {2006},
  doi     = {10.1088/0305-4470/39/32/S15}
}

@article{krejcirik2014similarity,
  author  = {Krej{\v{c}}i{\v{r}}{\'{\i}}k, David and Siegl, Petr and {\v{Z}}elezn{\'{y}}, Jakub},
  title   = {On the Similarity of {S}turm--{L}iouville Operators with Non-{H}ermitian Boundary Conditions to Self-Adjoint and Normal Operators},
  journal = {Complex Analysis and Operator Theory},
  pages   = {255--281},
  volume  = {8},
  year    = {2014},
  doi     = {10.1007/s11785-013-0301-y}
}

@article{landau1941theory,
  title={Theory of the superfluidity of helium {II}},
  author={Landau, L.},
  journal={Physical Review},
  volume={60},
  number={4},
  pages={356--358},
  year={1941},
  doi={10.1103/PhysRev.60.356},
  publisher={APS}
}

@article{mariano2022solutions,
    author   = {Mariano, Paolo Maria and Polikarpus, Julia and Spadini, Marco},
    title    = {Solutions of linear and nonlinear schemes for non-{F}ourier heat
conduction},
    journal  = {International Journal of Heat and Mass Transfer},
    volume   = {183},
    pages    = {122193},
    numpages = {11},
    year     = {2022},
    doi      = {10.1016/j.ijheatmasstransfer.2021.122193}
}

@article{maxwell1867dynamical,
    author   = {Maxwell, James Clerk},
    title    = {{IV}. {O}n the dynamical theory of gases},
    journal  = {Philosophical Transactions of the Royal Society of London},
    issue    = {157},
    pages    = {49--88},
    year     = {1867},
    month    = {Dec},
    issn     = {0261-0523},
    doi      = {10.1098/rstl.1867.0004},
    url      = {https://doi.org/10.1098/rstl.1867.0004},
    eprint   = {https://royalsocietypublishing.org/rstl/article-pdf/doi/10.1098/rstl.1867.0004/1455921/rstl.1867.0004.pdf},
    note     = "Eq. (143)"
}

@article{nascimento2024exact,
  title={Exact solution of {M}axwell--{C}attaneo--{V}ernotte model: {D}iffusion versus second sound},
  author={Nascimento, J.A.R. and Ramos, Anderson de Jesus Ara{\'u}jo and Campelo, Anderson D.S. and Freitas, M.M.},
  journal={Applied {M}athematics {L}etters},
  volume={156},
  pages={109135},
  year={2024},
  doi={10.1016/j.aml.2024.109135},
  publisher={Elsevier}
}

@article{parker1961flash,
    author = {Parker, W. J. and Jenkins, R. J. and Butler, C. P. and Abbott, G. L.},
    title = {Flash method of determining thermal diffusivity, heat capacity, and thermal conductivity},
    journal = {Journal of Applied Physics},
    volume = {32},
    number = {9},
    pages = {1679--1684},
    year = {1961},
    publisher = {AIP},
    doi = {10.1063/1.1728417}
}

@book{ruggeri2015rational,
    title     = {Rational Extended Thermodynamics beyond the monatomic gas},
    author    = {Ruggeri, Tommaso and Sugiyama, Masaru},
    year      = {2015},
    publisher = "Springer International Publishing",
    address   = "{H}eidelberg, {N}ew {Y}ork, {D}ordrecht, {L}ondon"
}

@article{seshadri2006eigenmode,
author = {Seshadri, S. R.},
journal = {Journal of the Optical Society of America A},
number = {6},
pages = {1523--1527},
publisher = {Optica Publishing Group},
title = {Eigenmode expansions using biorthogonal functions: complex-valued {H}ermite--{G}aussians: comment},
volume = {23},
month = {Jun},
year = {2006},
url = {https://opg.optica.org/josaa/abstract.cfm?URI=josaa-23-6-1523},
doi = {10.1364/JOSAA.23.001523}
}

@article{van2012universality,
  author    = {V{\'{a}}n, Peter and F{\"{u}}l{\"{o}}p, Tamas},
  title     = "Universality in Heat Conduction Theory: Weakly Nonlocal
Thermodynamics",
  journal   = {Annalen der Physik},
  year      = {2012},
  volume    = {524},
  number    = {8},
  pages     = {470--478},
  doi       = {10.1002/andp.201200042}
}

@article{vernotte1958paradoxes,
title={Les paradoxes de la th{\'{e}}orie continue de l{\'{e}}quation de la chaleur},
author={Vernotte, Pierre},
journal={Comptes Rendus Hebdomadaires des S{\'{e}}ances de l'Acad{\'{e}}mie des Sciences},
volume={246},
pages={3154--3155},
year={1958}
}

@article{zhang2025analytical,
author   = {Zhang, Huimin and Zhang, Zhenhui and Zhou, Shaoling},
title    = {An Analytical Solution for the Uncertain Damped Wave Equation},
journal  = {Symmetry},
issn     = {2073-8994},
volume   = {17},
year     = {2025},
number   = {9},
pages    = {1533},
numpages = {16},
doi      = {10.3390/sym17091533},
url      = {https://www.mdpi.com/2073-8994/17/9/1533}
}

\end{document}